\documentclass{jfm}
\usepackage{graphicx}
\usepackage{epstopdf, epsfig}

\usepackage[dvipsnames]{xcolor} 
\usepackage{cancel}
\usepackage{color,soul}
\usepackage{amsmath}
\usepackage{amssymb} 
\usepackage{subcaption}
\usepackage{algorithmic}
\usepackage{algorithm2e}
\usepackage{booktabs} 
\usepackage{array} 
\usepackage{paralist} 
\usepackage{verbatim} 

\newcommand{\BJ}[1]{{\color{black}#1}}

\newcommand{\cmnt}[1]{\ignorespaces}

\shorttitle{Transition in Atmospheric Boundary Layer Turbulence Structure}
\shortauthor{Jayaraman and Brasseur}

\title{The Surprising Transition in Atmospheric Boundary Layer Turbulence Structure from Neutral to Moderately Convective Stability States and Mechanisms Underlying Large-scale Rolls}

\author{Balaji Jayaraman\aff{1}
  \corresp{\email{balaji.jayaraman@okstate.edu}}
  \thanks{Formerly Department of Mechanical Engineering, Pennsylvania State University.}
 \and James G. Brasseur \aff{2}
 \thanks{Affiliate Scientist, National Center for Atmospheric Research; Affiliate Faculty, Atmospheric and Oceanic Sciences, University of Colorado; Professor Emeritus and Adjunct Professor of Mechanical Engineering, Pennsylvania State University.}
}

\affiliation{\aff{1}School of Mechanical and Aerospace Engineering, Oklahoma State University,
Stillwater, OK 74078, USA
\aff{2}Ann and H.J. Smead Aerospace Engineering Sciences, University of
Colorado, Boulder, CO 80309, USA}

\begin{document}

\maketitle

\begin{abstract}
The vectoral wind structure of daytime atmospheric boundary layer (ABL) turbulence is strongly dependent on the balance between shear-driven turbulence production of horizontal fluctuations (driven by winds at the mesoscale), and buoyancy-driven turbulence production of vertical velocity fluctuations (driven by solar heating), characterized by the global instability state parameter $-z_i/L > 0$. In the fully shear-driven neutral limit $-z_i/L \rightarrow 0$, the surface layer is dominated by coherent streamwise-elongated concentrations of negative streamwise fluctuating velocity (``low-speed streaks''), while in the moderately convective state ($z_i/L \sim 10$ ) buoyancy generates streamwise-elongated thermal updraft “sheets” of concentrated vertical velocity fluctuations. Using large-eddy simulation (LES), we study the transition between the neutral and moderately convective states by quantifying correlations and integral scales as a function of $-z_i/L$ created by systematically increasing surface heat flux with fixed geostrophic wind. We discover a surprising sudden transition in ABL turbulence structure at $-z_i/L \approx 0.40$ with dramatic enhancement of streamwise coherence, particularly in the mixed layer, and a sudden change in ABL dynamic response to further increase in surface heating. In the supercritical ABL, continued increase in surface heat flux leads to a “maximal coherence state” at ($z_i/L \sim 1.0-1.5$)  associated with helical large-scale roll structure and exceptionally coherent thermal updrafts, a process driven by two key dynamical effects: (a) a surprising continual increase in streamwise coherence of streamwise velocity fluctuations and shear-driven low-speed streaks and (b) increasing spatial correlation between the coherent low-speed streaks in the surface layer below and in the coherent thermal updrafts with in the mixed layer above.
\end{abstract}


\section{Introduction}\label{sec:introduction}
The dominant energy-containing eddies that characterize the turbulence structure of the daytime atmospheric boundary layer (ABL) and define its essential non-steady and mixing properties are directly correlated with the global stability state of the boundary layer, reflecting the balance between convectively-driven and shear-driven turbulence production rate. We consider the clear-air fully developed daytime ABL capped by an inversion layer at height $z_i$ in regions where the boundary layer is well approximated as statistically homogeneous in the horizontal and during periods where the ABL can be reasonably modeled as quasi-steady and in quasi-equilibrium. This ``canonical'' ABL typically occurs over relatively flat uniformly rough terrain for a few tens of kilometers in the horizontal, and for a few hours in the early afternoon when the sun is beyond peak apogee, and for periods without significant nonsteady forcing from weather events at the mesoscale. In the canonical equilibrium state the statistical structure of the turbulence eddying motions in this rough-surface high Reynolds number boundary layer capped by an inversion layer are solely a function of the boundary layer capping inversion height $z_i$, and the magnitude of the Obukhov length scale $L$ (sufficiently far from the roughness elements).

The Obukhov length scale $L$ characterizes the balance between turbulence production by mean shear-rate vs. by buoyancy driven by surface heating:
\begin{equation}
L=-\frac{u^3_*}{\kappa \left(g/\theta_0 \right)Q_0}
\end{equation}
where $u_*$ and $Q_0$  are surface friction velocity and temperature flux into the fluid, respectively, $\theta_0$ is the reference potential temperature (taken to be the initial surface temperature), $g$ is gravitational acceleration and $\kappa = 0.4$ is a representative von Karman constant in the equilibrium state. The daytime ABL is unstable with $L < 0$, so that the level of global instability of the daytime ABL is parameterized by the ratio of boundary layer depth $z_i$ to negative of the Obukhov length $L$. The instability parameter  $-z_i/L > 0$ is an order-of-magnitude estimate of the relative global rate of turbulent kinetic energy production in the buoyancy-driven vertical motions from solar heating of the ground to the production rate of horizontal turbulence fluctuations by mean shear-rate driven by horizontal winds at the mesoscale. Shear production dominates near the ground and $-L$ is an order-of-magnitude estimate of the distance above the ground where buoyancy takes over as the larger turbulence production mechanism. Both mechanisms, however, originate near the surface, where shear-rate is highest and solar heating generates a flux of heat from the ground into the fluid above. 

In the ``neutral'' limit $-z_i/L \rightarrow 0$, shear-rate is the only source of turbulence production in the boundary layer so that $-L \gg z_i$, while in the ``fully convective'' limit $-z_i/L$  , buoyancy dominates turbulence production throughout the boundary layer and $-L \ll z_i$. These two limits have been well characterized in both the geophysical and engineering fluid dynamics communities~\citep{kaimal1976turbulence,wilczak1980three,schmidt1989coherent,lee1990structure,robinson1991coherent,panofsky1974atmospheric,moeng1994comparison,khanna1997analysis,bodenschatz2000recent,jimenez2004turbulent,lohse2010small}.

Whereas neutral boundary layer (NBL) turbulence is characterized by strong coherence in streamwise-elongated turbulent velocity fluctuations near the surface from the interaction between mean shear with turbulence \citep{robinson1991coherent,lee1990structure}, the fully convective atmospheric boundary layer (CBL) is characterized by highly coherent vertical updrafts within turbulent ``Rayleigh-Bernard'' cells (e.g., \cite{bodenschatz2000recent}) between the heated surface and a local temperature inversion that caps the boundary layer \citep{wilczak1980three,schmidt1989coherent}. The intermediate ``moderately convective'' atmospheric boundary layer (MCBL) range, with instability states $-z_i/L \sim O(1-10)$, is characterized by strong interaction between buoyancy-generated and shear-generated turbulence dynamics leading to large-scale roll-like turbulence eddies that span the ABL \citep{moeng1994comparison,khanna1998three,weckwerth1997horizontal}.

The basic structural characteristics of the canonical NBL vs. CBL vs. MCBL, as currently understood, are summarized in  \S~\ref{sec:contrastNBLMCBL}. Unclear, however, is characterization of the transition in turbulence eddy structure and dynamics as the balance between shear and bouyancy progressively shifts and the global instability state systematically transitions from  neutral to moderately convective ($-z_i/L \sim 0 \rightarrow  1$), as occurs, for example, in the morning hours of the diurnal cycle. We analyze this transition in detail in the current study for the canonical ABL in quasi-stationary equilibrium using large-eddy simulation (LES) with systematic increases in surface heat flux. 

The analysis is motivated, in part, by micrometeorological interest in the role of shear-rate and buoyancy in the mixing of the lower troposphere associated with specific turbulence structure and in the conditions underlying the existence and properties of ``large-scale atmospheric rolls''. A second motivation, however, is the potentially important role that ABL stability state plays in power capture and reliability of commercial wind turbines in particular, the correlation between deleterious nonsteady loading transients on wind turbine blades and shafts, and the characteristics of turbulent fluctuations in the daytime winds from the passage of stability-state-dependent eddying structures.

The structure of the MCBL, studied with field data \citep{etling1993roll,weckwerth1997horizontal,weckwerth1999observational} and large-eddy simulation \citep{sykes1989large,moeng1994comparison,khanna1998three}, is associated with the existence of ``large-scale rolls,'' roll-like convectively and advectively driven turbulence structures that span the atmospheric boundary layer. There has been a long history in the observation, characterization and definition of these structures (see \cite{etling1993roll} and \cite{weckwerth1997horizontal} for reviews) and a certain amount of confusion. Our use of the term ``large-scale roll" specifically refers to turbulent helical motions that span the thickness of the boundary layer due to the capping inversion and are highly elongated in the streamwise direction 	due to mean shear, driven by mesoscale winds above a capping inversion. These structures are sometimes indirectly visualized by assuming that observable cloud streets are directly correlated with underlying updrafts in the convergence zones between rolls. The updrafts drive moisture vertically between the rolls to the capping inversion, giving the subjective impression that the streamwise coherence in the large-scale rolls extend over tens, even hundreds, of boundary layer depths. Quantification of coherence in fluctuating velocity (see section \ref{sec:TransitionNBLtoMCBL}) or RADAR reflectivity indicate that large-scale rolls are associated with elongational aspect ratios of order $10$ or more. In the current study we address the transition in ABL structure as surface heating is systematically added to an otherwise neutral boundary layer. We find that the transitional characteristics are relevant to the process of large-scale roll formation and we define the instability state bounds associated with the existence and coherence of rolls in the equilibrium MCBL.

We characterize the canonical ABL, an equilibrium state with no residual dependence on initial conditions, no external forcing, and characterized statistically by non-dimensional ratios. The canonical daytime ABL is characterized by a global Reynolds number and global instability parameter $-z_i/L$. However, because the outer Reynolds number is extremely high and because the ABL surface is hydrodynamically rough, the entire boundary layer is inertia-dominated and the large-eddy structure is approximately Reynolds number independent. Therefore, the canonical ABL is characterized primarily by $-z_i/L$. Relative to the true ABL, the canonical ABL is best represented by the daytime ABL over flat terrain in the early afternoon with uniform winds, no cloud cover and a capping inversion with nearly constant $z_i$. The well-formed surface roughness elements must be statistically homogeneous and characterized by a single characteristic roughness scale $z_0$. For roughness not to disrupt surface-layer scaling, $z_0$ must be small relative to thickness of the  surface layer. The capping inversion grows slowly in time and the Coriolis time scale is typically an order of magnitude larger than largest-eddy turbulence time scales, so the approximately canonical ABL is, in practice, quasi-stationary and in quasi-equilibrium. The requirements for the daytime canonical ABL have been approximately realized in the field, for example, in the Kansas and Minnesota experiments of 1968 and 1973 \citep{kaimal1990kansas}.

The paper is organized as follows. In \S~\ref{sec:simanal} we describe the design of our large-eddy simulation (LES) experiments of the canonical daytime ABL, describe the range of instability states studied, provide basic ABL quantifications, and describe the analytical methods used to quantify turbulence eddying structure statistically in the transition from NBL to MCBL. To put the new research discoveries into perspective, section ~\ref{sec:contrastNBLMCBL}  provides a brief overview of current understanding of the three canonical ABL states: neutral (NBL), fully convective (CBL), and moderately convective (MCBL). We use current simulations in these descriptions and also summarize the current knowledge of large-scale roll structure from field experiments. \BJ{Sections \ref{sec:TransitionNBLtoMCBL}-\ref{sec:TemporalTransitionDynamics} present the results of our extensive analysis of the transition in ABL coherent structure as heat flux from the surface is systematically increased under an initially neutral boundary layer, until the ABL  can be described as ``moderately convective'' (see section \ref{sec:contrastNBLMCBL})}. 
\BJ{The changes in turbulent coherent structure is quantified statistically using key integral scales in vertical and horizontal turbulent motions (as discussed in  \S~\ref{sec:TransitionNBLtoMCBL}), with visualizations used to develop understanding of the structure underlying coherence.} 
These results lead naturally to descriptions of the process and requirements underlying the formation of large-scale rolls.
\BJ{We observe that this roll formation is triggered by a ``critical'' transition (see section~\ref{sec:CriticalTransition}) characterized by rapid changes in coherent structure.}
 In addition, we have discovered that the large-scale coherence has its own dynamics surrounding a ``max coherence'' ABL instability state as described in \S~\ref{sec:MaxCoherenceTransition}. 
\BJ{The well-known near-neutral and moderately convective instability states sandwich the two earlier regimes and are described in \S~\ref{sec:NearNeutralAndMCBL}.  
}  
We conclude in section \ref{sec:DiscussionAndConclusions} with a summary of essential new knowledge from this study related both to the overall transition in ABL structure and to the criteria for formation and characteristics of large-scale rolls in the equilibrium daytime ABL.


\section{Simulation and Analysis Methods\label{sec:simanal}}
\subsection{Large Eddy Simulations of the Atmospheric Boundary Layer\label{subsec:LESABL}}
The ABL large-eddy simulations were designed to allow for highly controlled systematic adjustments to the global ABL stablity state parameter $-z_i/L$ from neutral ($z_i/L = 0$) to moderately convective ($z_i/L \sim 1-10$). We do this by systematically increasing surface heat flux in the lower boundary condition while keeping the imposed geostrophic wind vector fixed. The computational domain was designed to capture 5 - 10 horizontal integral scales to minimize influence of the periodic boundary conditions in the horizontal. Similarly, the domain height was sufficiently large relative to the capping inversion depth $z_i$ to insure minimal influence of the upper boundary conditions on the simulated ABL below the capping inversion during the quasi-equilibrium period over which computational data were collected. 

The filtered Navier-Stokes equation, energy (potential temperature) equation, and continuity equation upon which LES of the ABL is developed, are:
\begin{equation}
\frac{\p \boldsymbol{u}^r}{\p t}+\bnabla \bcdot \left( \boldsymbol{u}^r\boldsymbol{u}^r \right)^r=-\frac{1}{\rho_0}\bnabla p^*-\bnabla \bcdot {\boldsymbol \tau}_a^{sfs}+\frac{\boldsymbol{g}}{\theta_0}\left( \theta^r-\theta_0\right)+\boldsymbol{f}\times \left( \boldsymbol{V}_g - \boldsymbol{u}^r\right),
\label{eq:NS1}
\end{equation}
\begin{equation}
\frac{\p \theta^r}{\p t}+\bnabla \bcdot \left( \boldsymbol{u}^r \theta^r \right)^r=-\bnabla \bcdot {\boldsymbol \tau}_{\theta}^{sfs},
\label{eq:NS2}
\end{equation}
\begin{equation}
\bnabla\bcdot\boldsymbol{u}^r = 0,
\label{eq:NS3}
\end{equation}
where LES resolved-scale (RS) variables are indicated with superscript $r$: $\boldsymbol{u}^r$ is resolved velocity, $p^*$ is effective resolved pressure (below) and $\theta^r$ is resolved potential temperature. The subfilter-scale (SFS) ``stress'' tensor ${\boldsymbol \tau}^{sfs}=\left( \boldsymbol{u}^r\boldsymbol{u}^s+ \boldsymbol{u}^s\boldsymbol{u}^r+\boldsymbol{u}^s\boldsymbol{u}^s\right)^r$ and heat flux vector  ${\boldsymbol \tau}^{sfs}_{\theta}=\left( \boldsymbol{\theta}^r\boldsymbol{u}^s+ \boldsymbol{\theta}^s\boldsymbol{u}^r+\boldsymbol{\theta}^s\boldsymbol{u}^s\right)^r$ must be modeled. ${\boldsymbol \tau}^{sfs}$ is separated into its isotropic and deviatoric parts, its isotropic part included in $p^*$ (below) and its deviatoric part, ${\boldsymbol \tau}^{sfs}_a$, modeled using the one-equation eddy viscosity closure described in \cite{moeng1984large} and originating in \cite{deardorff1980stratocumulus} (see also \cite{sullivan1996grid}  and \cite{khanna1997analysis}):
\begin{equation}
\boldsymbol{\tau}_a^{sfs}=-2\nu_T\mathsfbi{S}^r\textrm{,      }\nu_T=C_k\Delta\sqrt{e}\textrm{,      }\Delta=\left( \Delta_x,\Delta_y,\Delta_z \right)^{1/3}\textrm{,      }C_k=0.1,
\label{eq:uSFSdef}
\end{equation}
where $\mathsfbi{S}^r$ is the resolved strain-rate tensor and $\Delta_{x,y,z}$ are the grid cell dimensions for the uniform grid. The prognostic equation for the ``SFS turbulent kinetic energy” $e$ is given in \cite{moeng1984large}. 
The SFS heat flux vector $\boldsymbol{\tau}_{\theta}^{sfs}$ is also modeled with an eddy diffusivity closure:
\begin{equation}
\begin{split}
{\boldsymbol \tau}_{\theta}^{sfs}&=\nu_{\theta}\bnabla\theta^r\textrm{,      }\nu_{\theta}=\left[ 1+ \frac{2l_s}{\Delta}\right]\textrm{,      }\\
l_{s} &= \left\{
\begin{array}{ll}
\Delta & ,\ \p\theta/\p z < 0 \textrm{  (unstable)} \\[2pt]
0.76e^{1/2}\left( \frac{g}{\theta_0}\frac{\p\theta}{\p z} \right)^{-1/2} & ,\ \p\theta/\p z > 0 \textrm{  (stable)}
\end{array} \right.
\end{split}
\label{eq:thetaSFSdef}
\end{equation}
Over most of the ABL the turbulent Prandtl number is $1/3$. The lower expression for $l_s$ is active in the capping inversion where stable stratification suppresses mixing and reduces the mixing length scale \citep{deardorff1980stratocumulus,moeng1984large}. This adjustment to the eddy viscosity is necessary to maintain stability in the capping inversion.

The momentum eqn.\eqref{eq:NS1} uses the low Mach number Boussinesq approximation where buoyancy force is included for small variations in potential temperature around the background state ($\rho_0$ and $\theta_0$, taken as the surface values) with local resolved pressure $p^r=p^{r'}+P$   is separated into ensemble mean ($P$) and fluctuating parts. The canonical daytime ABL is driven by a uniform mean horizontal pressure gradient written in terms of a specified ``geostrophic wind'' vector $\boldsymbol{V}_g$, defined through the Ekman balance between pressure and Coriolis force:
\begin{equation}
\bnabla_h P\equiv \rho_0 \boldsymbol{f}\times \boldsymbol{V}_g,
\label{eq:geostrophic}
\end{equation}
where $\boldsymbol{f}=f\hat{e}_z$ is twice the projected Earth rotation rate vector. In our simulations, $f=|\boldsymbol{f}|=10^{-4}s^{-1}$, corresponding to mid latitudes. Thus, the specified geostrophic wind vector is perpendicular to the mean force driving the mean winds which, from horizontal homogeneity, are strictly horizontal. An impact of the Coriolis term in the momentum equation is the spiraling of the mean wind velocity vector as a function of height, $z$ \citep{wyngaard2010turbulence}. The specification of $\boldsymbol{V}_g$ in concert with surface flux establishes the stability state of the boundary layer after reaching quasi-equilibrium. The isotropic part of $\boldsymbol{\tau}^{sfs}$ is incorporated into an ``effective'' fluctuating resolved pressure, $p^*=p^{r'}+\frac{\rho_0}{3}\ \textrm{tr}\left\{ \boldsymbol{\tau}^{sfs} \right\}$ which is given by the solution to the Poisson equation produced when the divergence of eqn.~\eqref{eq:NS1} and \eqref{eq:NS3}. The boundary condition for $p^*$  at the surface is a derived Neumann condition (see \cite{moeng1984large,sullivan1996grid}).

The ABL is capped by a stabilizing inversion layer in potential temperature, which is specified as part of an initial one-dimensional potential temperature field as constant up to initial capping inversion height and then increasing with constant slope (lapse rate). The rate at which the capping inversion depth $z_i$ grows with time ($dz_i/dt$) increases with increasing surface heat flux $Q_0$, leading to concerns with the influences of nonstationarity on equilibrium (which requires quasi-stationarity) and deviations from the canonical state. In addition, to minimize boundary influences it could become necessary to increase the vertical domain height and grid with increasing $Q_0$, which introduces concern with potential grid influences on the predicted transitions in ABL structure with stability, $-z_i/L$ (see \cite{brasseur2010designing}).

\begin{figure}
	\begin{center}
	\centerline{\includegraphics[width=0.51\columnwidth]{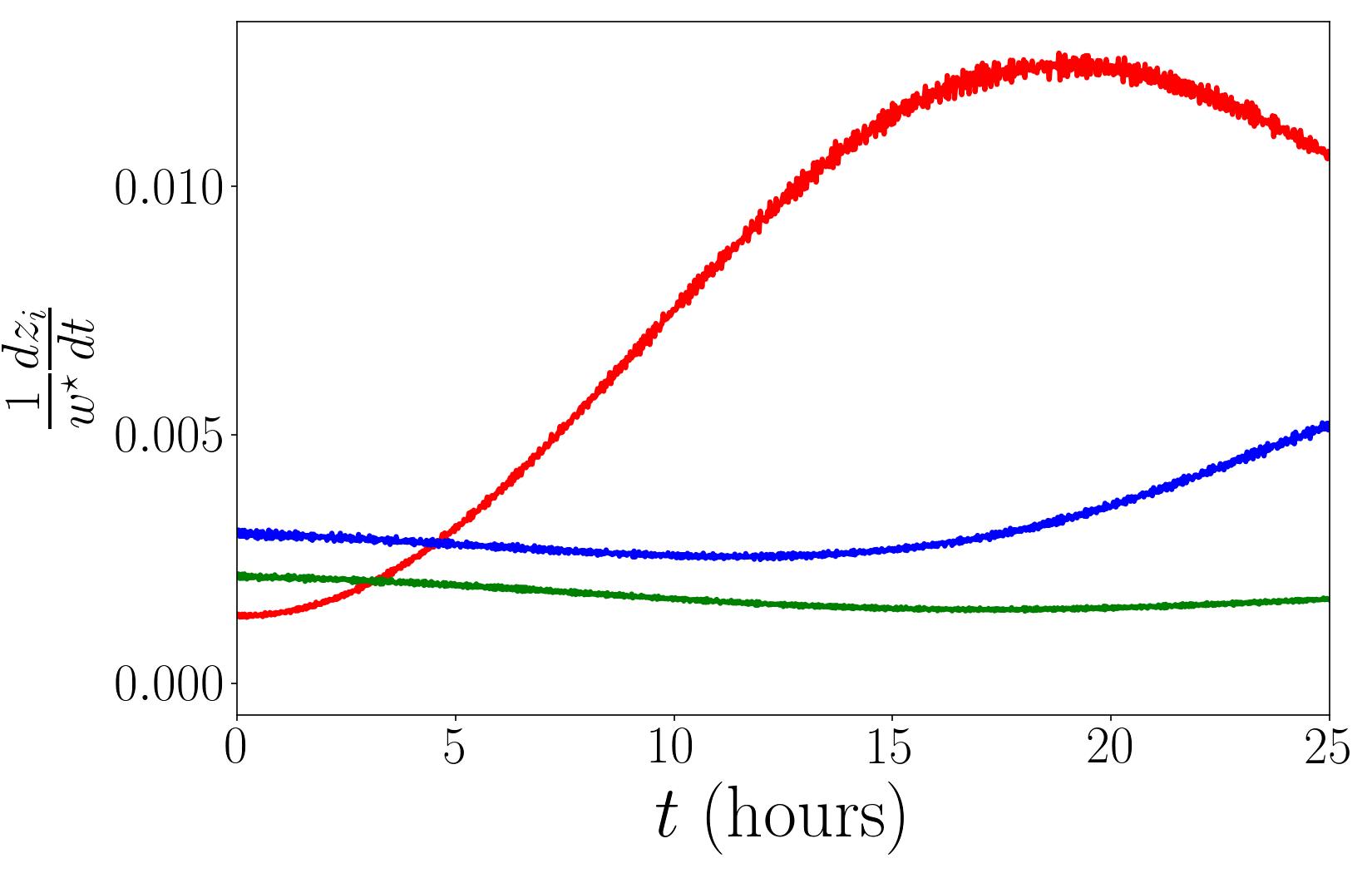}}
	\caption{\label{fig:cappinginversion} Time evolution of the ratio of capping-inversion-growth velocity to the buoyancy-driven mixed-layer velocity scale for three different capping inversion strengths. red: $[d\theta/dz]_{CI}=0.064$K/m, blue: $[d\theta/dz]_{CI}=0.128$K/m, green: $[d\theta/dz]_{CI}=0.192$K/m.}
	\end{center}
\end{figure}

To avoid these concerns, the initial capping inversion strength, quantified by maximum vertical gradient in mean potential temperature in the capping inversion, was adjusted to reduce the rate of growth in $z_i$ at higher. In figure~\ref{fig:cappinginversion} we plot the ratio of capping-inversion growth velocity to buoyancy-driven mixed-layer velocity scale $w_*$ for three different capping inversion strengths. Equilibrium requires that this ratio be small compared to $1$ and constant in time. From this result we chose the highest of the three capping inversion strengths \BJ{which is strength stronger than those typically observed}. This statement is based on our analysis of $13$ plots of potential temperature vs. height obtained in part from the open literature~\citep{sorbjan1996effects,sullivan1998structure,hennemuth2006determination}, and in part from a database collected from aircraft flights in Indiana \citep{Davis2017personal}. Whereas the depths of the capping inversions in our large-eddy simulations were in the range of those analyzed from the available field data (see also \cite{grabon2010airborne}), the strengths of our LES capping inversions were, on average, $7.5$ times stronger so as to limit $dz_i/dt$ in our simulated boundary layers in order to maintain quasi-stationarity and eliminate the need for grid adjustment. By quantifying the statistics of the ABL with varying capping inversion strength in large-eddy simulations of the ABL, \cite{sorbjan1996effects} showed that the statistics of the ABL are insensitive to the strength of the capping inversion except near the \BJ{inversion layer} itself. Since our focus is on changes in boundary layer dynamics near the surface in the equilibrium state, we argue that this approach of \BJ{employing a strong capping inversion is preferred and justified.} 

The dynamical system was discretized within a rectangular computational domain $5 \textrm{km} \times 5 \textrm{km} \times 2 \textrm{km}$ using uniformly spaced grid points on a pre-dealiased grid of $192 \times 192 \times 128$ grid points. The pseudo-spectral algorithm is applied in the horizontal with periodic boundary conditions to minimize numerical dissipation, with second-order finite differencing applied in the vertical on a staggered grid that resolves well the surface layer. Dealiasing is applied in the spectrally-resolved horizontal using the 2/3 rule, creating an effective grid of $128 \times 128 \times 128$ points on which the resolved scales evolve. A series of $11$ simulations were carried out to equilibrium with systematically varying surface heat flux for fixed geostrophic wind velocity of $10$ m/s in fixed horizontal direction aligned with x). As shown in Table \ref{tab:simparams}, at the time of the collection of data the vertical extent of the capping inversion is $2.5-3.1\ z_i$, so that the boundary layer is resolved in the vertical with $41-47$ points and the surface layer with $8-9$ points (as suggested by \cite{brasseur2010designing}). The exceptionally large distance between $z_i$ and the top of the computational domain ensures negligible influence from the upper boundary conditions on the predictions.The horizontal length of the domain is $6.5 - 7.8\ z_i$ to minimize influence of the periodic boundary conditions on predicted ABL dynamics. (Later we learn that the peak integral scales in the horizontal are $1.3-1.7\ z_i$, indicating good resolution of the energy-containing atmospheric turbulence eddies.) Time advance is performed with third order Runge-Kutta \BJ{integration of the spatially discretized system of equations}.

For each simulation, an initial potential temperature distribution was defined as constant in $z$ at up to an initial capping inversion layer of thickness $156 m$ and temperature jump of $30K$ ($d\theta/dz=0.192K/m$) above which we specify a lapse rate (slope) of $0.003K/m$ . The initial capping inversion height was $600 m$. On the upper boundary mean temperature gradient is specified at the initial lapse rate, mean resolved velocity was set to match the horizontal geostrophic wind vector $\boldsymbol{V}_g$, vertical fluctuating velocity was set to zero and zero vertical gradient was applied to transverse velocity fluctuations and SFS viscosity. On the lower surface, the temperature flux $Q_0=\langle w\theta\rangle_0$   was prescribed at the 11 values given  in Table \ref{tab:simparams}.

A lower boundary condition on surface fluctuating shear stress vector was applied in the standard way \citep{moeng1984large,khanna1997analysis} using the roughness parameter $z_0$ in a relationship with specified ratio of mean (horizontal) velocity magnitude at the first grid level, $U_1$  to the friction velocity $u_*=|\langle \boldsymbol{u}_hw \rangle|^{1/2}_0$   based on a specified continuous function of $z/z_0$ . As is typical, we use the following modified logarithmic form that includes the impact of buoyancy force in the transition from neutral to moderately convective ABL:
\begin{equation}
\frac{U(z_1)}{u_*}\equiv \frac{U_1}{u_*}=\kappa^{-1}\left[ \ln \left( \frac{z_1}{z_0}\right)-\psi_1(z_1/L) \right],
\label{eq:wallmodel}
\end{equation}
where $\psi_1(z_1/L)=2\ln\left( \frac{1+x_1}{2} \right)+\ln \left( \frac{1+x^2_1}{2} \right)-2\tan^{-1}x_1+\frac{\pi}{2}$, with $x_1=\left( 1-15{z_1/L} \right)^{1/4}$.
If $z_1$ were replaced by the continuous coordinate $z$, eqn.~\eqref{eq:wallmodel} would describe the empirical form of Monin-Obukhov scaling for the surface layer, $U(z)/u_*=f(z/L,z/z_0)$, developed by \cite{paulson1970mathematical}. However, because Eq. (8) is applied at only one point in z, this form does not force the simulation to satisfy law-of-the-wall in the surface layer \citep{brasseur2017role}.

\begin{table}
	\begin{center}
		\def~{\hphantom{0}}
		\setlength{\tabcolsep}{12pt}
		\begin{tabular}{c|c|c|c|c|c}
			${-z_i/L}$  & ${Q_0}$(K.m/s)   &   ${z_i}$(m) & ${|L|}$(m) & ${u_*}$(m/s) & ${w^*}$(m/s)\\[7pt]
			0.0   & 0.0 & 645 & $\infty$ & 0.44 & 0.0\\
			0.09   & 0.001 & 650 & 7139 & 0.45 & 0.27\\
			0.21  & 0.0025 & 656 & 3072 & 0.46 & 0.37\\
			0.29   & 0.0035 & 663 & 2295 & 0.47 & 0.42\\
			0.40 & 0.0050 & 670 & 1699 & 0.48 & 0.47 \\
			0.73 & 0.010 & 692 & 946 & 0.50 & 0.60 \\
			1.08 & 0.015 & 726 & 688 & 0.51 & 0.70 \\
			1.39 & 0.020 & 729 & 547 & 0.52 & 0.77 \\
			1.95 & 0.030 & 766 & 402 & 0.54 & 0.90 \\
			2.46 & 0.040 & 754 & 316 & 0.55 & 0.99 \\
			2.95 & 0.050 & 789 & 267 & 0.56 & 1.08 \\
		\end{tabular}
		\caption{Simulation parameters over the quasi-stationary analysis periods.}
		\label{tab:simparams}
	\end{center}
\end{table}

An issue inherent to large-eddy simulation of the rough-surface atmospheric boundary layer is the problem of the near-surface ``overshoot,'' an over-prediction of mean velocity gradient at the first several LES grid cells adjacent to the surface that appears as a near-surface peak in the law-of-the-wall-normalized mean gradient variation in z. This long-standing and well-known problem with LES prediction has been diagnosed by (\cite{brasseur2010designing}, BW10) for the neutral ABL. It was found there that the overshoot results from the interaction of model, algorithm and grid-dependent numerical diffusion and the horizontal vs. vertical grid resolution. It was also shown how the grid can be designed together with the parameterizations of the models for the SFS terms in the momentum and potential temperature equations. However there are two issues, one technical and one practical, that precludes the use of the BW10 method in our study. The technical issue is that nearly all our simulations are designed to mix buoyancy with shear driven motion, with corresponding buoyancy-driven additions to the model for the fluctuating surface stress vector embodied in eqn.~\eqref{eq:wallmodel} above, while the BW10 method has only been tested on the buoyancy-free neutral boundary layer. As importantly, BW10 found that the friction in the model and numerical algorithm that underlies the existence of the overshoot smooths over a fundamental problem with the universally-applied surface-exchange model for the surface stress vector. The practical issue with applying the BW10 method is the excessive computational cost and time required to explore the grid refinements and additional parameter variations that are not currently understood for mixed buoyancy-shear driven boundary layers. There has not yet been enough research to understand and develop practical solutions to the overshoot problem in the moderately convective boundary layer. As a result, our simulations do contain overshoots which we have analyzed in detail to determine if the overshoot could have a potential role in our key findings. We summarize our analysis in the Appendix~\ref{appOvershoot} with the conclusion that it is \emph{unlikely} that the existence of an overshoot in our simulations plays a significant role in our key results.

\begin{figure}
	\centering
	\begin{subfigure}[ht!]{0.46\columnwidth}
		\includegraphics[width=\columnwidth]{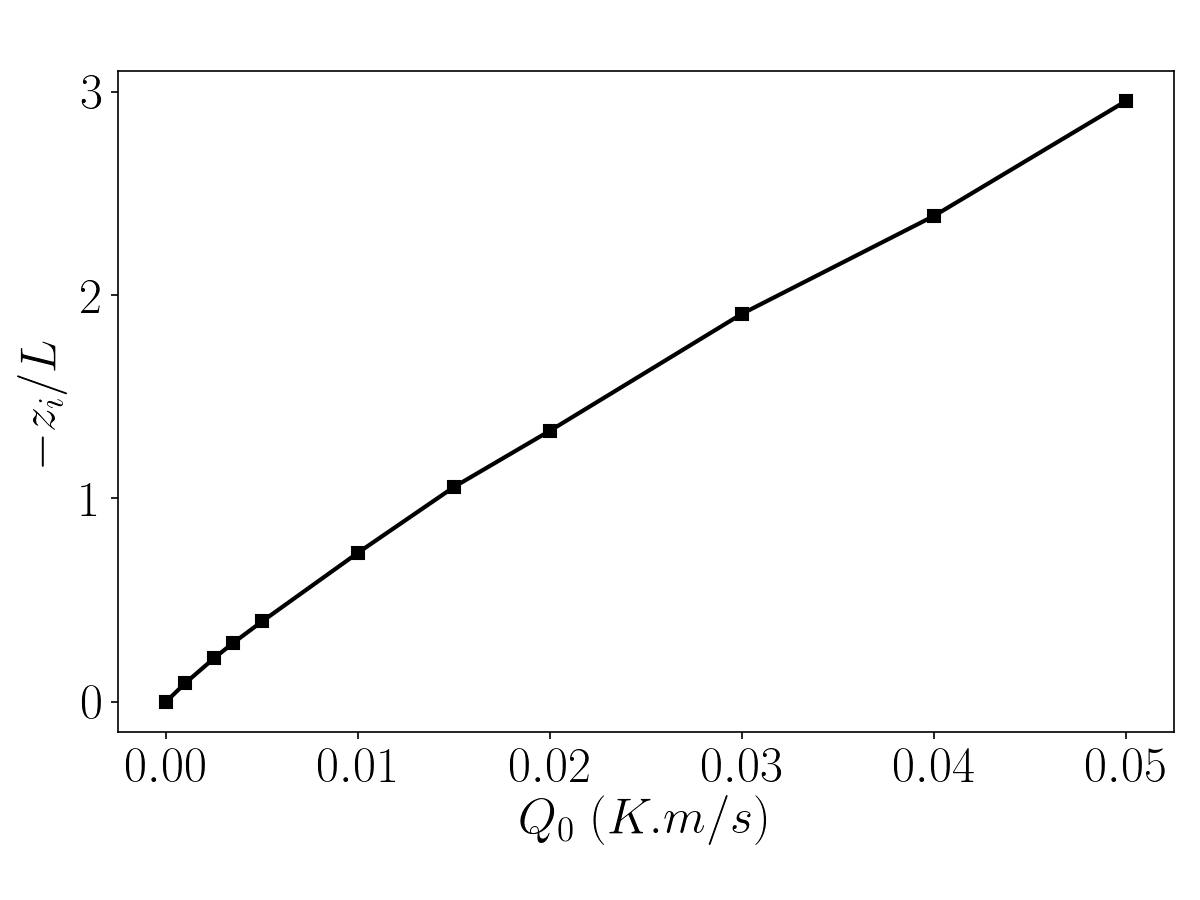}	
		\caption{\label{fig:zibylvsq0} }
	\end{subfigure}
	\begin{subfigure}[ht!]{0.48\columnwidth}
		\includegraphics[width=\columnwidth]{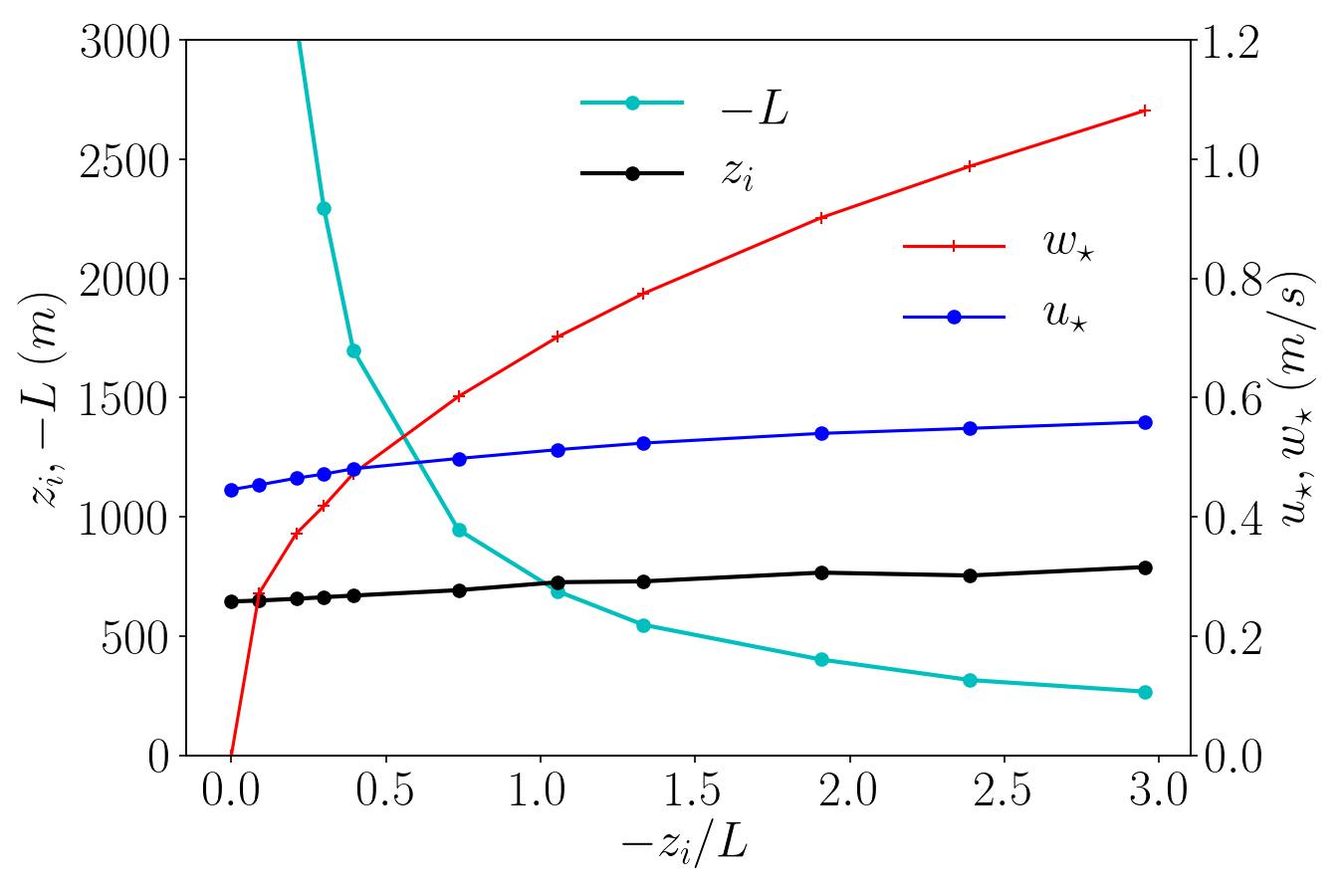}
		\caption{\label{fig:zibylvsutauwstar} }
	\end{subfigure}
\caption{\label{fig:zibylvssimparams}(a) Variation of instability parameter with surface heat flux; (b) Variation of simulation parameters with instability state, $–z_i/L$. }
\end{figure}

As shown in fig.~\ref{fig:zibylvsq0} and Table \ref{tab:simparams}, 11 simulations were carried out to quasi-steady-state and analyzed with instability state parameter $-z_i/L$ from $0$ to $2.95$ over a change in $Q_0$ from $0$ to $0.050$Km/s with a fixed geostrophic wind speed of $10$ m/s. fig.~\ref{fig:zibylvsutauwstar} and Table \ref{tab:simparams} show the variation of boundary layer height $z_i$, Obukhov length $L$, mixed layer buoyancy-driven characteristic velocity $w_*\equiv \left( gQ_0z_i/\theta_0 \right))$ and surface friction velocity $u_*$. The surface friction velocity $u_*$, a measure of shear-driven vertical flux of turbulence-driven momentum driven from the rough surface, varies relatively little with stability state. The Obukhov length $L$, on the other hand, decreases rapidly with increasing $–z_i/L$ as buoyancy production of turbulent kinetic energy progressively increases relative to shear production. Given the relatively mild increase in $u_*$, $L$ varies approximately as $1/Q_0$ while, due to the strong capping inversion as discussed previously, $z_i$ increases relatively little with increasing $-z_i/L$. For this reason, $w_*$  increases approximately as $Q_0^{1/3}$ (fig.~\ref{fig:zibylvsutauwstar}).

\subsection{Analysis Methods\label{subsec:AnalysisMethods}}
The data for analysis were extracted after ensuring that the ABL from LES has reached quasi-equilibrium over multiple large-eddy turnover timescales. We use horizontal homogeneity to average over horizontal planes together with temporal averaging to realize high-quality converged statistics so as to accurately quantify subtle changes in the statistical structure of ABL turbulence with small changes in stability state. The use of temporal averaging requires quasi stationarity, ensured by our treatement that minimizes growth of the capping inversion. Converged statistics were obtained by averaging over roughly $60$ eddy turnover times, $\tau_u=z_i/u_*$, at each instability state. During this period $z_i$ increased by $\approx 10\%$ for the most convective instability state ($-z_i/L = 2.95$); all other boundary layers grew at slower rates.

To study transitions in ABL structure we quantify various two-point correlations and coherence lengths along with second and third-order moments to characterize structure, coherence and intensity. The two-point correlations of \emph{rotated} streamwise ($x_1 \equiv x_{rot} $) and transverse ($x_2 \equiv y_{rot} $) components of fluctuating velocity are given by: 
\begin{equation}
{R}_{u_iu_j}(\boldsymbol{r};z)=\langle u_i(\boldsymbol{x};z)u_j(\boldsymbol{x+r};z) \rangle_{h,t},
\label{eq:Rij}
\end{equation}
where $u_1 \equiv [{u^{r}}^{\prime}]_{rot}$, $u_2 \equiv [{v^{r}}^{\prime}]_{rot}$,$u_3 \equiv [{w^{r}}^{\prime}]=w^r$   are the fluctuating components of resolved velocity in local coordinates $x_{1,2}$ that have been rotated horizontally into the direction of the mean velocity vector at that point. The subscript $h$ implies that the ensemble average is carried out over the horizontal plane at specified height $z$ and fixed time, and subscript $t$ implies averaging was also carried out over $60\tau_u$ during quasi-stationary state. The corresponding coherence lengths (integral scales) are given by:
\begin{equation}
{L}_{ijk}(z)\equiv{L}_{u_i,u_j,r_k}(z) =\int_0^{\infty} \frac{R_{u_iu_j}(r_k;z)}{R_{u_iu_j}(0;z)}dr_k.
\label{eq:Lij}
\end{equation}

It is important to note that because the mean velocity is rotated in the horizontal due to Coriolis acceleration, eqns.~\eqref{eq:Rij} and \eqref{eq:Lij} are written in coordinates locally rotated on the horizontal plane with $x$ in the direction of the ensemble mean velocity vector at that height (necessarily horizontal due to horizontal homogeneity). This is achieved by interpolating onto a mesh \BJ{of the same resolution} oriented in the local mean flow direction \BJ{after taking advantage of periodic boundary conditions}. Whereas in the homogeneous horizontal directions coherence length is independent of $\boldsymbol{x}_h$ in eqn.~\eqref{eq:Rij}, this is not true of coherence length in the vertical inhomogeneous direction, $x_3 = z$. In this study we estimate vertical correlations with respect to vertical level $z_{corr}/z_i=0.1$ in the positive $z$ direction (i.e., upwards). To interpret coherence and correlations we make use of 3D visualizations of isocontours and isosurfaces of resolved velocity and temperature fields. 

\section{Contrasting the Turbulence Structure of the Neutral and Moderately Convective Atmospheric Boundary Layers\label{sec:contrastNBLMCBL}}

In the Introduction we summarized the motivations for our study and briefly described the three canonical instability state classes⎯neutral, moderately convective and fully convective⎯ associated with the daytime ABL. In this section we summarize current understanding of these three states as reference points to relate to our new findings. Although most of the key points in this discussion may be found in \cite{khanna1998three,moeng1994comparison,salesky2017nature}, we use our current LES to illustrate the same.
\begin{figure}
	\centering
	\begin{subfigure}[ht!]{0.48\columnwidth}
		\includegraphics[width=\columnwidth]{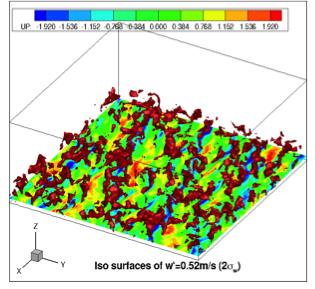}	
		\caption{\label{fig:NBL-wisosurf} }
	\end{subfigure}
	\begin{subfigure}[ht!]{0.48\columnwidth}
		\includegraphics[width=\columnwidth]{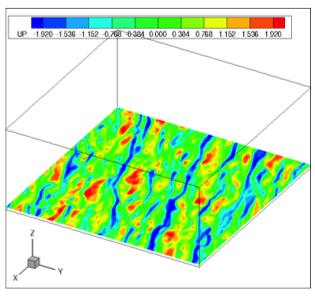}
		\caption{\label{fig:NBL-uisocont} }
	\end{subfigure}
	\caption{\label{fig:NBL-Baseline} (Color Online) The neutral boundary layer ($-z_i/L= 0$): (a) Isosurface of $w'$ (red, $2\sigma_w$) overlaying a plane of isocontours of $u'$ ($\pm2\sigma_u$) at $z = 0.1z_i$. (b) The plane of $u'$ isocontours at $z = 0.1z_i$ in (a). }
\end{figure}

\begin{figure}
	\centering
	\begin{subfigure}[ht!]{0.48\columnwidth}
		\includegraphics[width=\columnwidth]{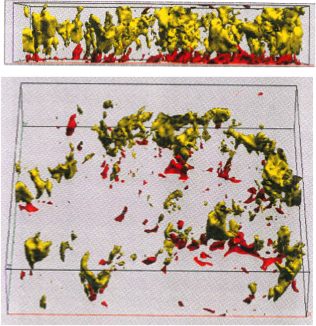}	
		\caption{\label{fig:CBL-wiso} }
	\end{subfigure}
	\begin{subfigure}[ht!]{0.48\columnwidth}
		\includegraphics[width=\columnwidth]{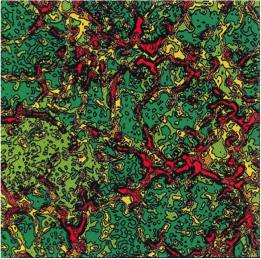}
		\caption{\label{fig:CBL-uisocont} }
	\end{subfigure}
	\caption{\label{fig:CBL-Baseline} (Color Online) The dominantly convective boundary layer ($-z_i/L = 730$): (a) Isosurfaces of $w'$ (yellow) and $\theta$ (red) with side view in top image to show spatial displacement between concentrations of temperature and vertical velocity;  (b) Isocontours of $\theta$ on a plane near the ground, at $z=0.05z_i$ ($\pm2\theta^*$ relative to base temperature). The figures are taken from \cite{khanna1998three}.
	}
\end{figure}

Consider the neutral atmospheric boundary layer (NBL) limit $-z_i/L\rightarrow 0$  shown in figure~\ref{fig:NBL-Baseline} in contrast with the fully convective boundary layer (CBL) state $-z_i/L \gg 1$ in figure~\ref{fig:CBL-Baseline}. The neutral ABL has negligible surface heating with mean shear driven by the large-scale horizontal wind above the capping inversion (in the quasi-stationary equilibrium state this is the geostrophic wind). Because the NBL is fully shear-dominated, turbulence production arises from the distortion of Reynolds shear stress by mean shear-rate, dominant adjacent to the surface. In the mean, shear-generated turbulence enters the fluctuating velocity in the direction of the local mean velocity before being distributed to other directions via pressure-strain-rate correlation. Thus, in the NBL the coherent structure of the streamwise fluctuations are of special interest, especially near the surface.

It is well known that the lower-than-mean streamwise fluctuations concentrate within highly coherent turbulence structures elongated in the direction of the local mean velocity, which is angulated relative to the geostrophic wind vector due to Coriolis acceleration (see figure~\ref{fig:NBL-Baseline}) and with average horizontal separation that scales on the boundary layer depth, $z_i$ \citep{khanna1998three}. ``Low-speed streaks'' are a ubiquitous response to turbulence production by mean shear \citep{lee1990structure} and are consequently the dominant energy-containing eddy structure of turbulent boundary layer flows, in the ABL both in the surface layer and above \citep{khanna1998three}. In contrast, concentrations of ``high-speed'' (i.e., higher $u$ than average) turbulence fluctuations are less coherent and much less elongated in comparison to the low-speed (i.e., lower $u$ than average) streaks. As shown visually in figure~\ref{fig:NBL-wisosurf}, vertical velocity fluctuations tend to be concentrated within smaller less coherent structures that are relatively unaffected by the horizontally extended streamwise coherence of the low-speed streaks \BJ{easily identifiable in figure~\ref{fig:NBL-uisocont}}.

In contrast to the shear-driven NBL, the fully convective boundary layer (CBL), illustrated in figure~\ref{fig:CBL-Baseline}, is driven by buoyancy forces that are generated from surface heat flux due to solar heating of the ground in the presence of negligible horizontal mean wind above the capping inversion, and correspondingly negligible turbulence ($u'$) production near the ground from mean shear. Consequently, turbulence fluctuations are produced first in the vertical before enhancing horizontal fluctuating velocity components through pressure-strain-rate correlation. Thus, whereas the dominant eddying features of the NBL are streamwise-elongated low-speed streaks, in the CBL the dominant eddy structures are coherent atmospheric updrafts surrounding turbulent Rayleigh-Bernard cells with less coherent downward motions within the cells. Whereas the horizontal dimension of the updrafts scale, in the surface layer, on the distance from the surface, $z$, the horizontal cell dimensions (i.e., between the updrafts) scale on the capping inversion height, $z_i$. These structures are illustrated in the lower panel of figure~\ref{fig:CBL-wiso}. However, as illustrated in the upper panel of figure~\ref{fig:CBL-wiso} together with figure~\ref{fig:CBL-uisocont}, the ``seeds'' of the updrafts are concentrations of high-temperature fluid near the surface underlying the vertical updrafts. The spatial separation between the high-temperature fluid near the surface and the vertical velocity fluctuations above reflects damping of vertical motions in the surface layer by the impermeable surface below.

\begin{figure}
	\centering
	\begin{subfigure}[ht!]{0.48\columnwidth}
		\includegraphics[width=\columnwidth]{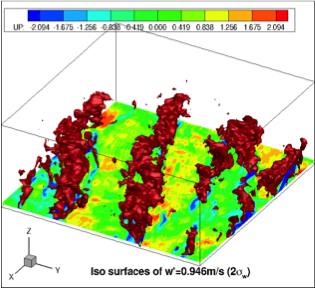}	
		\caption{\label{fig:MCBL-wiso} }
	\end{subfigure}
	\begin{subfigure}[ht!]{0.48\columnwidth}
		\includegraphics[width=\columnwidth]{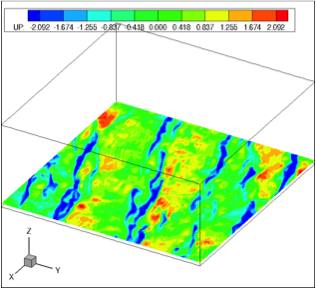}
		\caption{\label{fig:MCBL-uisocont} }
	\end{subfigure}
	\caption{\label{fig:MCBL-Baseline} (Color Online) The moderately convective boundary layer ($-z_i/L= 2.95$): (a) Isosurface of $w'$ ($2\sigma_w$) overlaying isocontours of $u'$ ($\pm2\sigma_u$) at $z = 0.1z_i$; (b) Isocontours of $u'$ at $z = 0.1z_i$ (range $\sigma_u= \pm2.1$). The vertical updrafts in (a) are angulated relative to the geostrophic wind vector (aligned with x) due to Coriolis turning of the low-speed streaks shown in (b).
	}
\end{figure}

\cite{khanna1998three} proposed that the contrasting mechanisms in integral-scale eddy production discussed above for the NBL and CBL interact in the production of the fundamental large-eddy turbulence structure of the canonical ``moderately convective'' atmospheric boundary layer (MCBL). The MCBL is driven from above by horizontal winds of sufficient magnitude to cause significant shear production of turbulence velocity fluctuations near the ground relative to simultaneous generation of buoyancy-driven turbulent motions by solar heating of the ground. This ABL is characterized by the instability parameter $-z_i/L$ between roughly $O(1)$ and roughly $O(10)$. Figure~\ref{fig:MCBL-Baseline} shows an isosurface of vertical fluctuating velocity (figure~\ref{fig:MCBL-wiso}) together with isocontours of horizontal fluctuating velocity on a plane near the ground (figure~\ref{fig:MCBL-uisocont}) for an equilibrium MCBL at $-z_i/L ≈ 2.95$. Figure~\ref{fig:MCBL-uisocont} shows that, even in the presence of strong buoyancy force, mean shear near the ground generates horizontal turbulence fluctuations within highly coherent elongated low-speed streaks. The transition from the moderately convective to the  fully convective ABL equilibrium states was studied by \cite{salesky2017nature}. Here we study the transition from the neutral to moderately convective equilibrium states.

It is well known that fluctuations of advected scalars such as temperature concentrate within the coherent low-speed streaks. \cite{khanna1998three} argue that, as a result of the strong correlation between high temperature and low-speed turbulent fluctuations (relative to the local mean), high-temperature fluctuations concentrate within the coherent low-speed streaks, thus localizing buoyancy force to the streaks, particularly in the high-shear region near the surface where coherence in the low-speed horizontal fluctuations is strong and where vertical fluctuations can be generated with strong surface influence, as shown in figure~\ref{fig:MCBL-Baseline}. Since, like Rayleigh Bernard convection in figure~\ref{fig:CBL-Baseline}, vertical fluctuations are less suppressed farther from the surface, it is visually evident in figure~\ref{fig:MCBL-Baseline} that strong updrafts concentrate away from the surface, but these updrafts are now spatially correlated with the buoyancy ``seeds'' in the coherent low-speed streaks below. Thus mean shear near the surface generated by winds at the mesoscale above the capping inversion, and buoyancy force near the surface generated by solar heating of the surface, work together to produce strong energy-dominant quasi-two-dimensional updrafts in the mixed layer overlaying streamwise elongated coherent low-speed steaks in the surface layer below, as illustrated in figure~\ref{fig:MCBL-Baseline}. The coupling of near-surface shear-driven streaks with the structure of the updrafts in the mixed layer will play an important role in the analyses of the current study (e.g., \S~\ref{sec:CriticalTransition}).

The \cite{khanna1998three} arguments for the generation of sheet-like thermal updrafts over low-speed streaks in the presence of the mean horizontal winds of the moderately convective ABL imply fluid particle trajectories that are helical at the largest vertical scale, moving warmer fluid near the surface to the cooler potential temperatures near the capping inversion, where the flow is redirected toward to the surface. When sufficiently strong, the helical fluid particle motions collectively define "large-scale atmospheric rolls," a class of coherent large-scale boundary layer motions that connect the lower and upper boundary layers over large horizontal distances. These large-scale rolls aligned with the updrafts of figure~\ref{fig:MCBL-wiso} are often observed visually from above as lines of clouds that mark condensations of humidity-rich upward flow over tens of kilometers. These cloud-visualized structures have been studied extensively in the field by the micro-meteorology community \citep{weckwerth1997horizontal,weckwerth1999observational,lemone1973structure,lemone1976relationship}.

The principle aim of the current study is to quantify the transition from the neutral atmospheric boundary layer state in (figure~\ref{fig:NBL-Baseline}) to the moderately convective ABL state (figure~\ref{fig:MCBL-Baseline})⎯which, we discover, contains interesting unexpected transitional features. In doing so, this study also advances understanding of the genesis of large-scale atmospheric rolls.

\section{Transition in Atmospheric Boundary Layer Turbulence Structure from Neutral to Moderately Convective Stability States\label{sec:TransitionNBLtoMCBL}}
The changes in structure of energy-dominant atmospheric turbulence eddies in the canonical daytime ABL as a function of instability parameter $-z_i/L$ are difficult to quantify systematically in the field, especially when the data used must be restricted to quasi-equilibrium periods. A major advantage of large-eddy simulation (LES) is the ability to approximate the canonical equilibrium ABL state with statistical structure that is dependent only on $z_i/L$, and the ability to simultaneously quantify and visualize details of 4D large-eddy motions with highly controlled systematic variation in atmospheric stability. Of particular interest here are the subtle (and not so subtle) changes in horizontal and vertical characteristic coherence length scales associated with progressive increases in surface heating from the neutral ABL state with fixed geostrophic wind. As was discussed in \S~\ref{subsec:LESABL}, we are careful to maintain quasi equilibrium with increased surface heating by special treatment of the capping inversion designed to be always well within the computational domain and to grow with time scale much larger than than the turbulence time scales.

In this section, we describe our analysis that presents a rich transition in the energy-dominant coherent structure of the ABL with increasingly instability from neutral to a roll-dominated moderately convective state in the instability range $-z_i/L\sim 0 - 3$. In subsequent sections we elaborate on specific characteristics in turbulence structure that define ``regimes of transition,'' including  a ``critical stability regime'' where small changes in near neutral  stability create major changes in ABL structure (see \S~\ref{sec:CriticalTransition}) and a ``peak coherence stability state'' that is associated with the highly coherent large-scale roll state that has been described from observation in the field (see \S~\ref{sec:MaxCoherenceTransition}). These transitions occur between the basic neutral and moderately convective ABL states (\S~\ref{sec:NearNeutralAndMCBL}) described in the literature and reviewed in \S~\ref{sec:contrastNBLMCBL}. 

The production of streamwise velocity fluctuations originates statistically from the interaction between mean shear and Reynolds shear stress. Furthermore, stability theory \citep{lilly1966instability,brown1980longitudinal} and studies of homogeneous turbulent shear flow~\citep{lee1990structure,brasseur1992structural,brasseur2005kinematics} show that streamwise coherence is created by the influence of mean shear on underlying turbulence fluctuations, an effect that strengthens with normalized shear-rate. Therefore, one anticipates that the streamwise coherence length of streamwise velocity fluctuations ($L_{11,1}$) in the surface layer will be highest in the purely shear-dominated neutral state ($-z_i/L = 0$) and that the addition of any level of surface heating ($-z_i/L > 0$) interferes with shear production of horizontal turbulence fluctuations and leads to reduction in streamwise coherence of horizontal velocity fluctuations, $u'$. If this is the case, then the streamwise coherence length of streamwise velocity fluctuations, $L_{11,1}$, should decrease monotonically with increasing $-z_i/L$ from the neutral state.

\begin{figure}
	\centering
	\begin{subfigure}[ht!]{0.48\columnwidth}
		\includegraphics[width=\columnwidth]{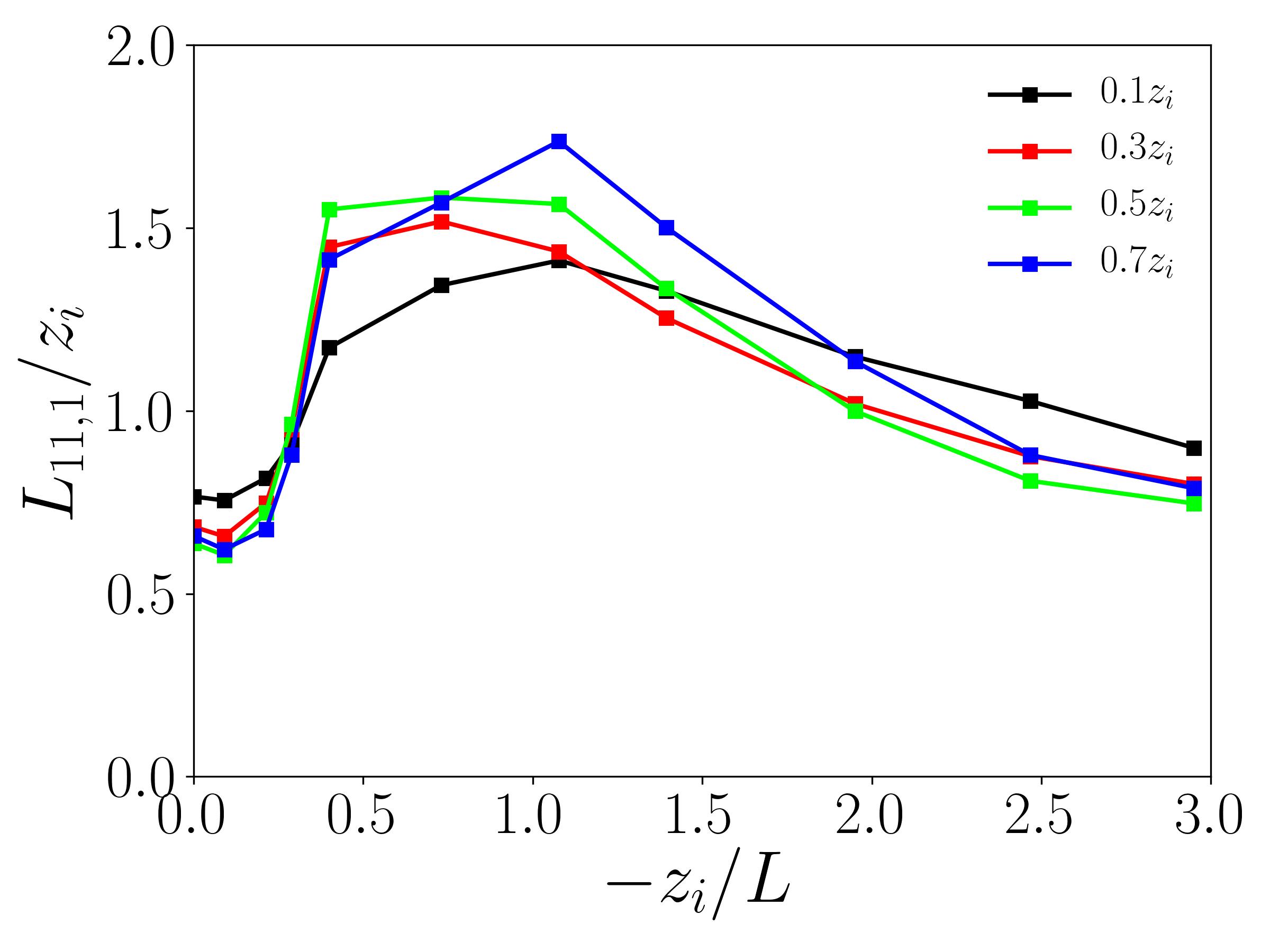}	
		\caption{\label{fig:L111vszibyL} }
	\end{subfigure}
	\begin{subfigure}[ht!]{0.48\columnwidth}
		\includegraphics[width=\columnwidth]{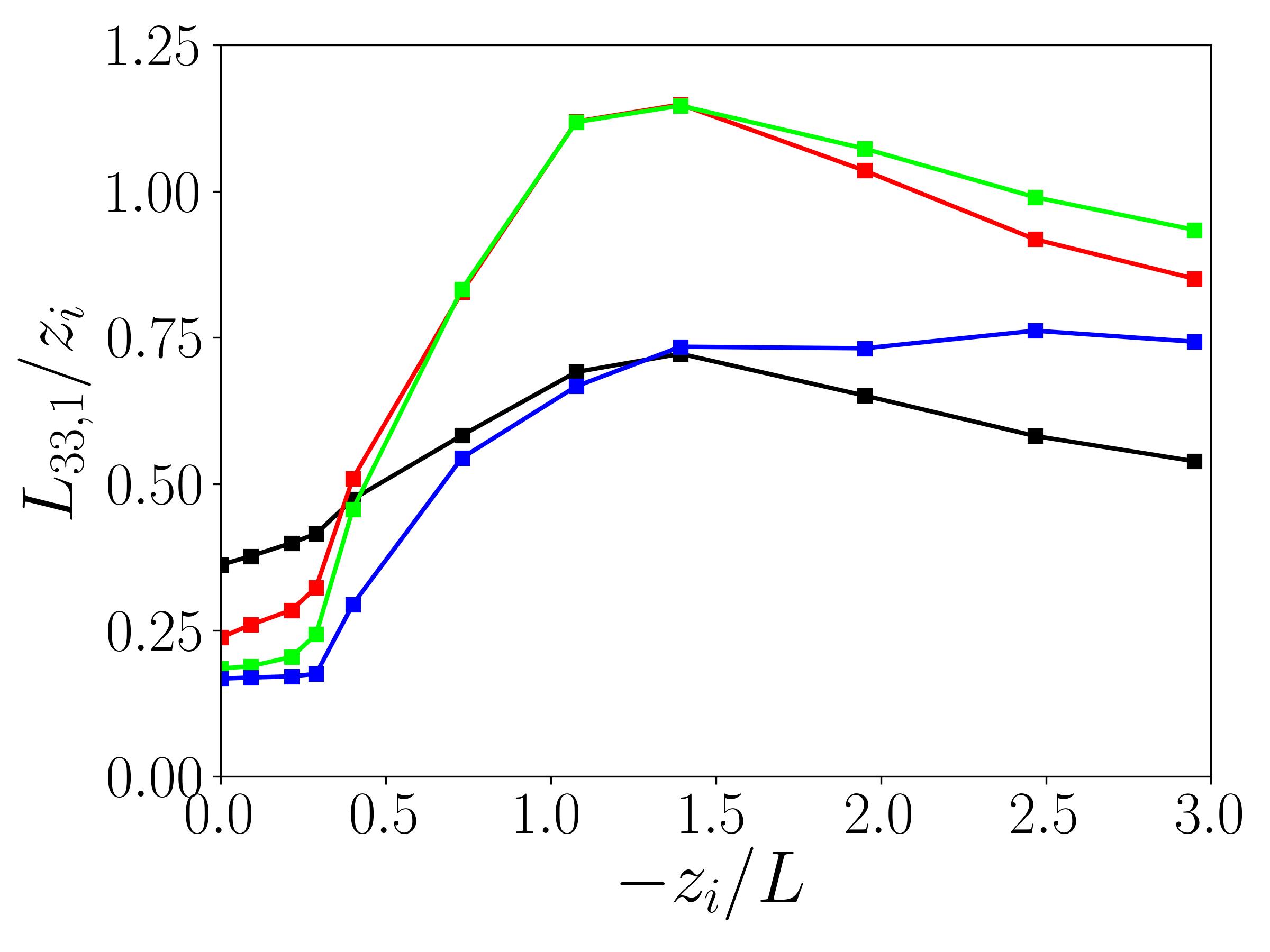}
		\caption{\label{fig:L331vszibyL} }
	\end{subfigure}
	\caption{\label{fig:Lxx1vszibyL} (Color Online) Variation of streamwise coherence lengths of (a) streamwise velocity fluctuations ($L_{11,1}$) and (b) vertical velocity fluctuations ($L_{33,1}$) with the global instability parameter $–z_i/L$. Each curve is as fixed height within the boundary layer as indicated.
	}
\end{figure}

\begin{figure}
	\centering
	\begin{subfigure}[ht!]{0.36\columnwidth}
		\includegraphics[width=\columnwidth]{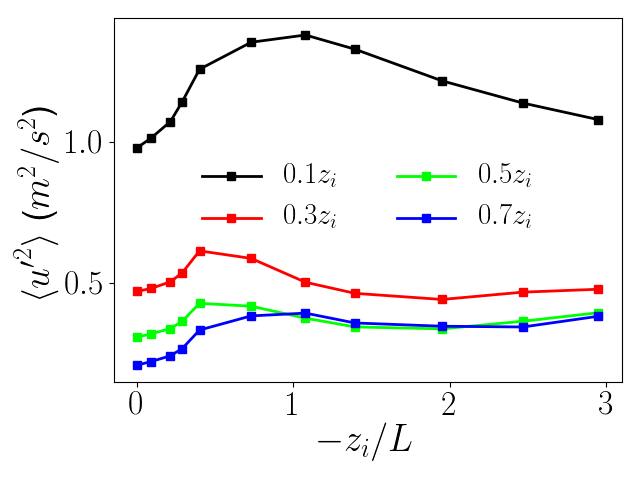}	
		\caption{\label{fig:uvarvszibyL} $\langle u'^2 \rangle$}
	\end{subfigure}
	\begin{subfigure}[ht!]{0.36\columnwidth}
		\includegraphics[width=\columnwidth]{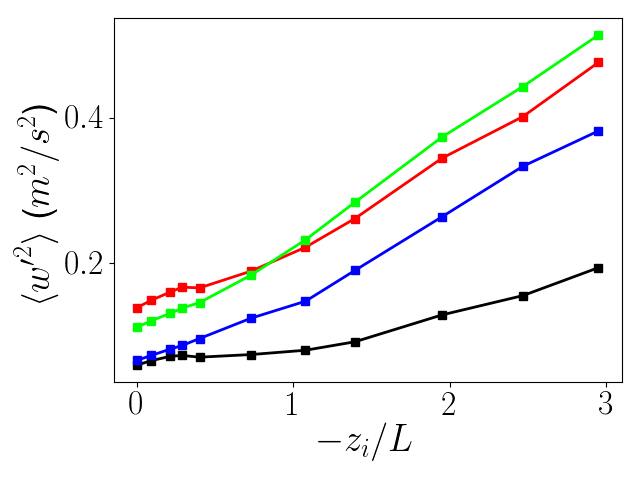}
		\caption{\label{fig:wvarvszibyL} $\langle w'^2 \rangle$}
	\end{subfigure}
	\begin{subfigure}[ht!]{0.36\columnwidth}
	\includegraphics[width=\columnwidth]{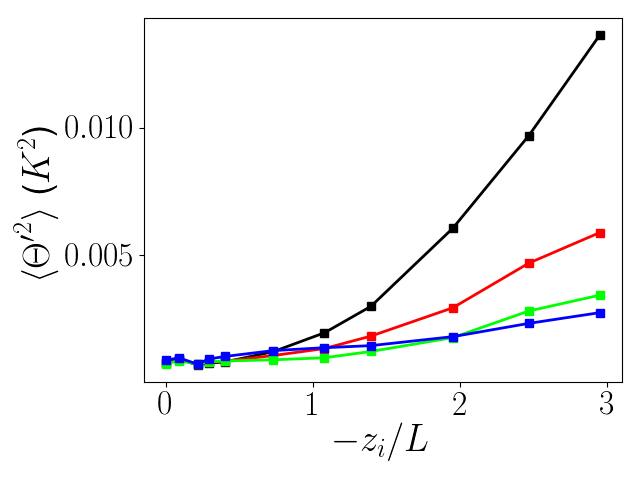}	
	\caption{\label{fig:tvarvszibyL} $\langle \theta'^2 \rangle$}
\end{subfigure}
\begin{subfigure}[ht!]{0.36\columnwidth}
	\includegraphics[width=\columnwidth]{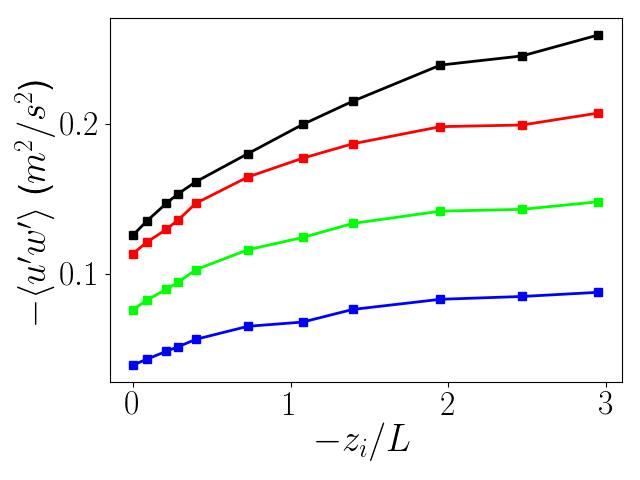}
	\caption{\label{fig:uwcovarvszibyL} $\langle u'w' \rangle$}
\end{subfigure}
	\begin{subfigure}[ht!]{0.36\columnwidth}
	\includegraphics[width=\columnwidth]{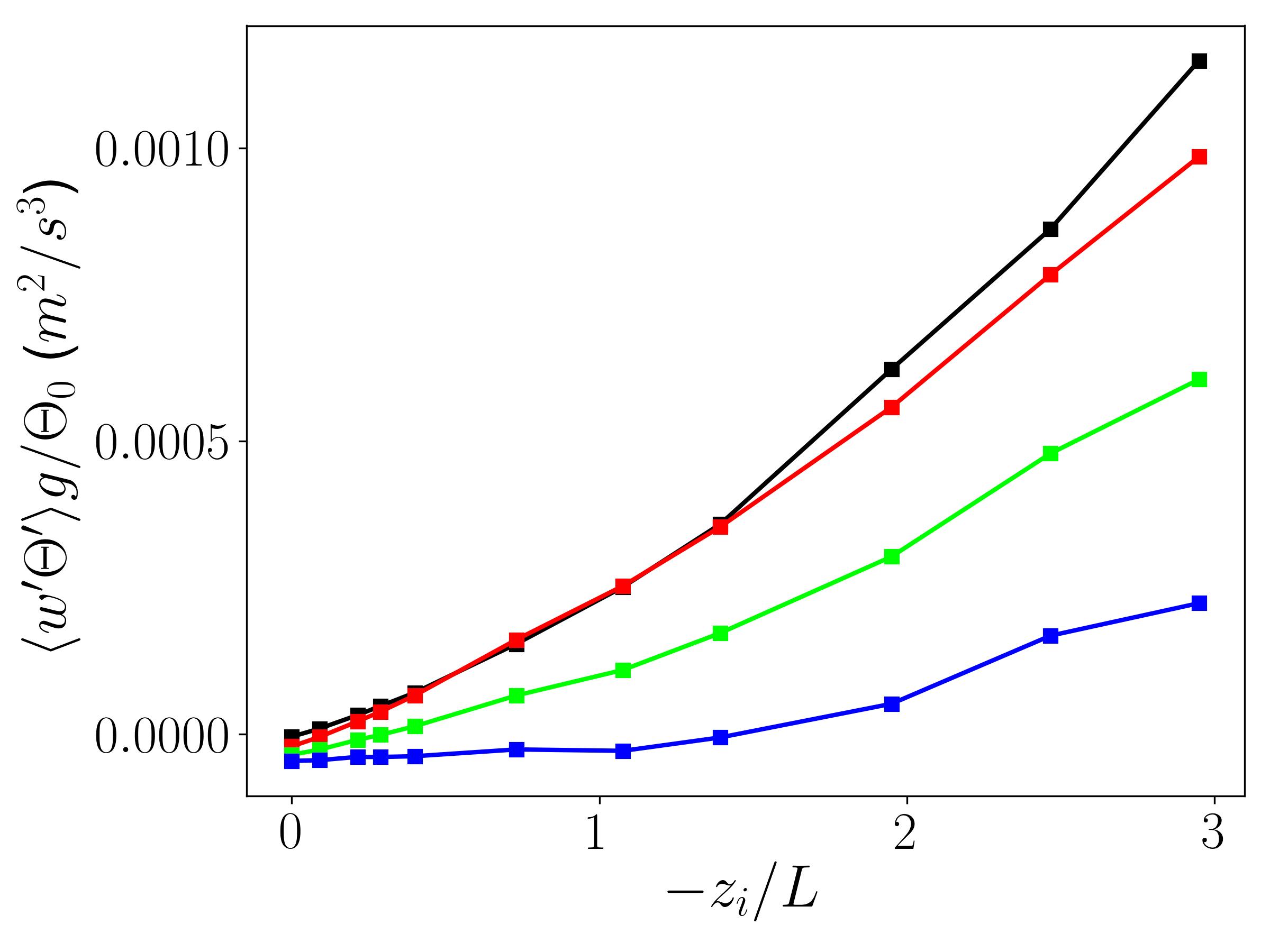}	
	\caption{\label{fig:wtcovarvszibyL} $\langle w'\theta' \rangle g/\theta_0$}
\end{subfigure}
\begin{subfigure}[ht!]{0.36\columnwidth}
	\includegraphics[width=\columnwidth]{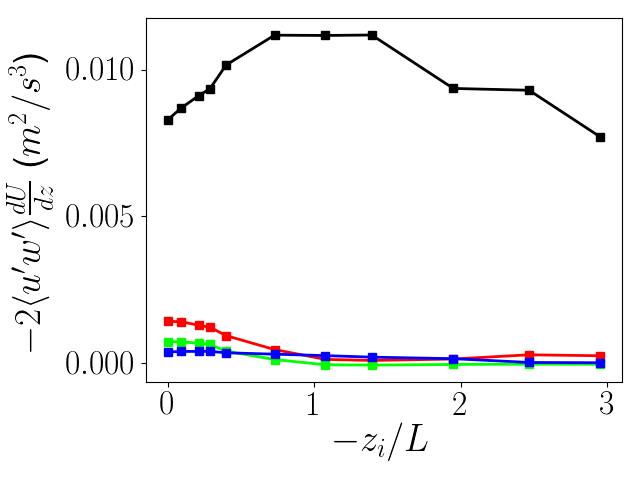}
	\caption{\label{fig:prodvszibyL} $-2\langle u'w' \rangle \frac{dU}{dz}$}
\end{subfigure}
	\begin{subfigure}[ht!]{0.36\columnwidth}
	\includegraphics[width=\columnwidth]{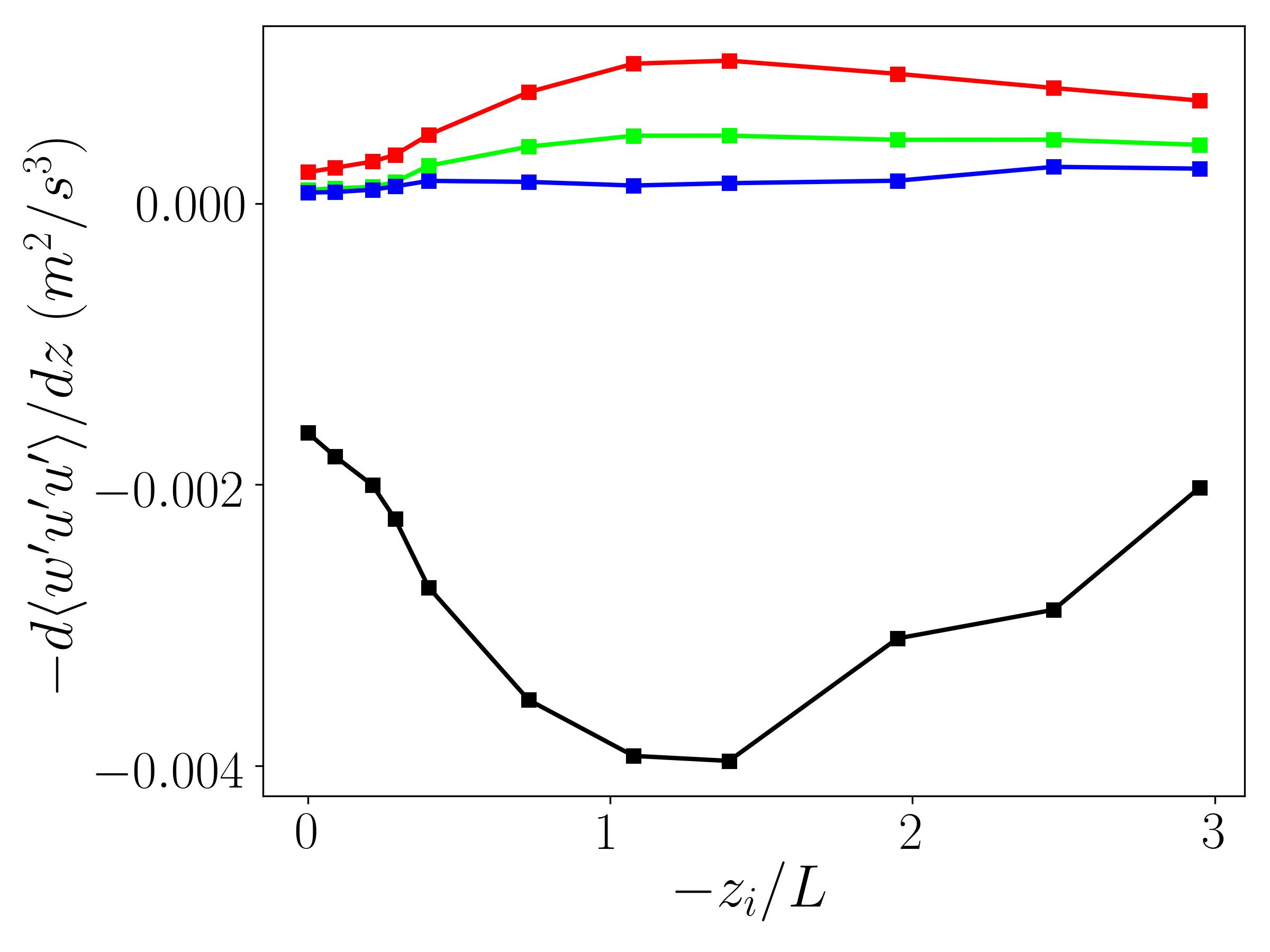}	
	\caption{\label{fig:transvszibyL} $-d\langle w'u'u' \rangle/dz$}
\end{subfigure}
\begin{subfigure}[ht!]{0.36\columnwidth}
	\includegraphics[width=\columnwidth]{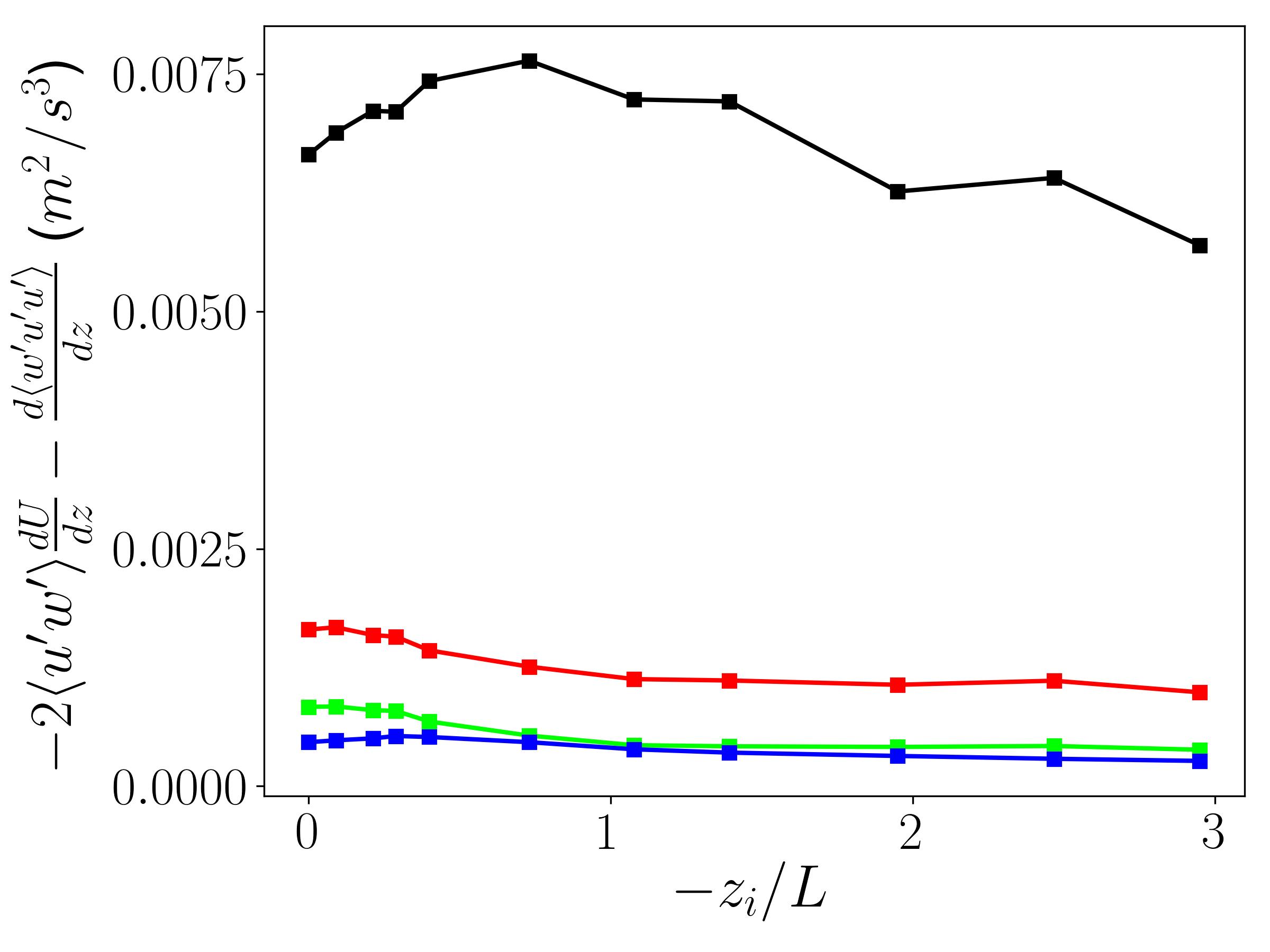}
	\caption{\label{fig:prodtransvszibyL} $-2\langle u'w' \rangle {dU}/{dz}-{d\langle w'u'u' \rangle}/{dz}$}
\end{subfigure}
	\caption{\label{fig:StatsvszibyL} (Color Online) Transition of turbulence statistics with the global instability parameter $-z_i/L$ at fixed vertical locations in the ABL, as specified with the line colors defined on figure (a). (a) Streamwise velocity variance; (b) vertical velocity variance; (c) potential temperature variance; (d) vertical turbulent momentum flux (i.e., shear component of Reynolds stress); (e) rate of production of vertical velocity variance by buoyancy force (proportional to vertical potential temperature flux); (f) shear production-rate of streamwise velocity variance (g) time rate of change of horizontal velocity variance due to vertical turbulent transport (h) net time rate of change of horizontal velocity variance from shear production and vertical turbulence transport. Note that the ($u',v^{\prime},w'$) are the fluctuating parts of resolved velocity in a local frame of reference with $x$ rotated into the direction of the mean velocity vector at that point.
	}
\end{figure}

Surprisingly, this turns out not to be the case, as shown in figure~\ref{fig:L111vszibyL} where $L_{11,1}$ is plotted against $–z_i/L$ at different distances from the surface. Our simulations predict that streamwise coherence lengths of streamwise ($u'$) fluctuations actually increase with increasing $–z_i/L$  as surface heat flux $Q_0$ is added to a previously unheated neutral canonical ABL. This response is strong and rapid with small increases in instability state, and occurs through the entire boundary layer from near the surface ($z/z_i=0.1$) to near the capping inversion ($z/z_i=0.7$). The anticipated reduction in streamwise coherence length with increase in $–z_i/L$ does occur, but not until $L_{11,1}$ has first increased by over a factor of two everywhere at $-z_i/L \approx 1-1.5$, after which $L_{11,1}$ does decrease monotonically. As will be discussed in detail in \S~\ref{sec:CriticalTransition}, the doubling of horizontal coherence length $L_{11,1}$  initiates suddenly at $-z_i/L \approx 0.21-0.29$, after the addition of very small levels of surface heating (from figure~\ref{fig:zibylvsq0} and Table~\ref{tab:simparams}, $Q_0 \approx 0.003$ Km at $-z_i/L \approx 0.21-0.29$). The sudden increase in $L_{11,1}$ from $-z_i/L \approx  0.21-0.29$ to $0.40$ is followed by a more gradual increase in streamwise coherence length with increasing $-z_i/L$ before initiating the expected decrease with increasing surface heat flux after peaking at $-z_i/L \sim O(1)$.  

This general trend is also observed in the streamwise coherence length of vertical velocity fluctuations $w'$, $L_{33,1}$, however with significant differences (figure~\ref{fig:L331vszibyL}). For example, whereas a sudden increase in coherence length at $-z_i/L\approx0.21-0.29$ occurs in $L_{11,1}$ throughout the boundary layer, the corresponding increase in $L_{33,1}$  initiates at $-z_i/L\approx0.29$ and occurs everywhere except near the surface at $z_i/z=0.1$, where vertical velocity fluctuations are severely damped at all $-z_i/L$ (see figures~\ref{fig:wvarvszibyL}). At all other heights, a sudden increase in horizontal coherence length occurs at $z/z_i\approx0.29$ for both horizontal ($L_{11,1}$) and vertical ($L_{33,1}$) fluctuations. However the horizontal coherence lengths of vertical fluctuations are roughly a factor of two smaller than horizontal fluctuations up to peak coherence lengths that occur at slightly higher values of $-z_i/L$ for $L_{33,1}$ than for $L_{11,1}$. Furthermore the sudden increases at $z/z_i\approx0.29$ are less dramatic for $L_{33,1}$ than that for $L_{11,1}$ and the further increases (and decreases) with $-z_i/L$ are more lethargic, particularly from blockage in the vertical, severely at the surface and less severely at the capping inversion (figure~\ref{fig:wvarvszibyL}). 

Blockage in the vertical close to the capping inversion is likely responsible for the lack of a peak in $L_{33,1}$ at height $z/z_i=0.7$ (figure~\ref{fig:L331vszibyL}); instead, we observe an asymptotic limit of $L_{33,1}/z_i \approx 0.75$. It is interesting that the normalized horizontal coherence lengths at the highest values of $-z_i/L$ simulated are all roughly in the range $0.75-0.90$ for both horizontal and vertical turbulence fluctuations at all levels in the ABL (with the exception of $L_{33,1}$ at the highly blocked location $z/zi = 0.1$, which has an apparent asymptotic limit closer to $0.5$). Also note that whereas both $L_{11,1}$ and $L_{33,1}$ increase significantly up to a peak coherence length, the total relative increase is, overall, much larger for $L_{33,1}$ due to the smaller horizontal coherence lengths of fluctuating vertical velocity in the neutral state compared to streamwise fluctuating velocity. These observations will be discussed in more detail in \S~\ref{sec:CriticalTransition} where we analyze the ``critical'' instability regime $-z_i/L\sim0.29-0.40$. 

The complex transition in the coherence lengths of the energy-containing turbulent velocity fluctuations ($L_{11,1}$ and $L_{33,1}$) is a manifestation of an equally interesting and complex transition in the organization and overall structure of the energy-dominant turbulent eddies as surface heat flux increases relative to a fixed geostrophic wind that drives the ABL at the mesoscale. As will be analyzed in more detail in subsequent sections, we observe here the complex interplay between energy-containing buoyancy and shear-driven turbulent motions in ways that can both increase and reduce streamwise coherence lengths. 

Figure~\ref{fig:StatsvszibyL} shows the changes in key terms in the budget equations for turbulent kinetic energy components and potential temperature variance with systematic change in stability state. Whereas figure~\ref{fig:Lxx1vszibyL} shows that the relative changes in streamwise coherence lengths of turbulent fluctuating velocities are non-monotonic with increasing levels of instability $-z_i/L$, the same is not true, in general, of the variances and covariances. An important exception is the streamwise velocity variance (figure~\ref{fig:uvarvszibyL}) which tends to increase to a peak and then decrease with $-z_i/L$, qualitatively similar to the streamwise coherence length, $L_{11,1}$, but peaking at slightly smaller values of $-z_i/L$ than $L_{11,1}$, and displaying less dramatic changes outside the surface layer. All buoyancy-driven variables − vertical velocity variance (figure~\ref{fig:wvarvszibyL}), potential temperature variance (figure~\ref{fig:tvarvszibyL}) and vertical heat flux (figure~\ref{fig:wtcovarvszibyL}) - display strong monotonic increases with increasing surface heat flux at all levels in the atmosphere, although with a distinctly different form than either surface heat flux, which increases roughly linearly with $-z_i/L$ (figure~\ref{fig:zibylvsq0}), or $w_*$, which increases initially very rapidly and then much more slowly with increasing $-z_i/L$ (figure~\ref{fig:zibylvsutauwstar}). Although monotonic increases in buoyancy-driven variables with increased surface heating are expected, the non-monotonic trends in horizontal velocity variance (figure~\ref{fig:uvarvszibyL}) are, like the coherence lengths of figure~\ref{fig:Lxx1vszibyL}, unexpected and observed over the entire ABL, albeit most dramatically near the surface. Like streamwise coherence length, as the ABL becomes increasingly unstable from increasing surface heat flux, the levels of horizontal velocity fluctuations asymptote after their respective peaks at higher $-z_i/L$ as they enter the moderately convective state. These asymptotes initiate with peaks at lower $-z_i/L$ outside the surface layer than is the case with $L_{11,1}$ or $L_{33,1}$ (figure~\ref{fig:Lxx1vszibyL}), suggesting that changes in horizontal coherence with increasing $-z_i/L$ are tied to the surface layer region more strongly than are changes in horizontal velocity variance.

This trend is not reflected in the variation of Reynolds stress with $-z_i/L$ (figure~\ref{fig:uwcovarvszibyL}), which increases monotonically in magnitude. However, figure~\ref{fig:prodvszibyL} indicates that near the surface, at $z/z_i = 0.1$, production-rate of streamwise turbulent velocity variance by mean shear increases with $-z_i/L$ from the neutral state to a peak at $-z_i/L\approx0.7-1.2$ followed by a decrease into the MCBL regime. There is no suggestion of this trend at other heights however, where shear production rate is relatively low and roughly constant with increasing $-z_i/L$. The near-surface result is discussed further in \S~\ref{sec:TemporalTransitionDynamics}.

The shear production rates in figure~\ref{fig:prodvszibyL} at $z/z_i=0.1$ do not fully explain the changes in streamwise velocity variance   with increasing $-z_i/L$ observed in figure~\ref{fig:uvarvszibyL} outside the surface layer. Vertical turbulent transport shown in figure~\ref{fig:transvszibyL}, however, further explains some of these observations. Figure~\ref{fig:transvszibyL} shows that, for the entire range of instability states, streamwise velocity fluctuations are transported from the near-surface region ($z/z_i=0.1$) into the ABL regions outside the surface layer ($z/z_i\sim0.3 - 0.7$).  Furthermore, near the surface at $z/z_i=0.1$, there is a strong correlation between the rate of generation of $\langle u'^2 \rangle$  by mean shear (figure~\ref{fig:prodvszibyL}) and the rate of loss of $\langle u'^2 \rangle$ by vertical turbulent transport (figure~\ref{fig:transvszibyL}) with increasing $-z_i/L$ from the neutral state: the peak in $\langle u'^2 \rangle$  is coincident with the rate of production of $\langle u'^2 \rangle$  at $-z_i/L\sim1-1.5$. Away from the surface production of $\langle u'^2 \rangle$ does not explain the variation in $\langle u'^2 \rangle$ with $-z_i/L$. Figure~\ref{fig:transvszibyL} suggests that away from the surface, where production of $\langle u'^2 \rangle$  is negligible (figure~\ref{fig:prodvszibyL}), vertical transport of $\langle u'^2 \rangle$ partially explains the variation of $\langle u'^2 \rangle$  with $z/z_i$ (figure~\ref{fig:uvarvszibyL}).

Approaching the highest values of $-z_i/L$ predicted, shear production-rate, vertical transport and streamwise velocity variance all asymptote. Figure~\ref{fig:prodtransvszibyL} shows that the rate of production of$\langle u'^2 \rangle$ by shear is significantly higher than the rate of vertical transport (compare with figures~\ref{fig:prodvszibyL} and ~\ref{fig:transvszibyL}). We have not considered the contribution of redistribution of $\langle u'^2 \rangle$  into other TKE components from pressure-strain-rate correlations (accurate estimation of these terms is problematic near the surface).

These observations suggest that the most interesting responses to increasing $-z_i/L$ are in the changes in coherence length of streamwise velocity fluctuations (figure~\ref{fig:L111vszibyL}) correlated with changes in streamwise velocity variance (figure~\ref{fig:uvarvszibyL}) and streamwise coherence length of vertical velocity fluctuations (figure~\ref{fig:L331vszibyL}). 
We elaborate on these major observations in following sections. In \S~\ref{sec:CriticalTransition} we argue that the boundary layer undergoes a ``critical'' transition at $-z_i/L\sim0.29-0.40$ marked by a sudden increase in streamwise coherence length of $u'$ with a slight increase in surface heat flux. 
In \S~\ref{sec:MaxCoherenceTransition} we show that this transition to the supercritical state also initiates a global transition in ABL structure with increasing $-z_i/L$ to a peak coherence state around $-z_i/L\sim1-1.5$ that is associated with the boundary-layer-scale roll vortices (\S~\ref{sec:contrastNBLMCBL}) that couple lower and upper ABL regions. In \S~\ref{subsec:NearNeutralAndMCBL-NBL} we argue that this coupling of the lower- and upper-ABL does not initiate below critical, the true ``near-neutral'' limit. 
The transition from the peak coherence state to what we define as the ``moderately convective'' atmospheric stability state is described in \S~\ref{subsec:NearNeutralAndMCBL-MCBL}, with specific focus on the reduction in coherence in the large-scale roll vortices within the overall ABL structure. Whereas these discussions center on spatial coherence, in \S~\ref{sec:TemporalTransitionDynamics} we discuss temporal coherence and show a correlation between changes in spatial and temporal coherence. 

\section{A Critical Stability Regime ($-z_i/L \sim 0.29-0.4$)\label{sec:CriticalTransition}}
Figure~\ref{fig:Lxx1vszibyL} shows that as heat is added at low rates to the surface of a previously neutral ABL and the boundary layer begins to deviate from neutral, a dramatic change in ABL turbulence structure takes place with very minor solar heating. There is negligible response to surface heat flux up to $Q_0\approx0.003$ Km/s, which is only about 1\% of the average peak (mid-day) temperature flux of roughly $0.25$Km/s \citep{wyngaard2010turbulence} and corresponds to very low values of the stability state parameter $-z_i/L\approx 0.21-0.29$. At just above above these low values in $-z_i/L$, however, dramatic changes in integral-scale turbulence eddying structure take place. Figure~\ref{fig:Lxx1vszibyL} shows a sudden increase in the streamwise coherence lengths $L_{11,1}$  and $L_{33,1}$ from $-z_i/L$ of $0.29$ to $0.40$ across most of the ABL induced by an additional slight additional increase in $Q_0$ of only $0.0025$ Km/s (Table~\ref{tab:simparams}). Interestingly, figure~\ref{fig:zibylvssimparams} indicates that this critical transition state occurs when $u_*/w_*\sim 1$\BJ{, i.e., buoyancy driven motions match the shear driven turbulence as quantified using characteristic velocity scales}. In the current section, we analyze in detail this unanticipated sudden increase in coherence length both statistically (\S~\ref{subsec:CriticalTransition-existence}) and visually (\S~\ref{subsec:CriticalTransition-StreakRole}).
\subsection{The Existence of a Critical Transition in Statistical Coherence\label{subsec:CriticalTransition-existence}}

The strong sensitivity shown in figure~\ref{fig:Lxx1vszibyL} between small increases in heating rate and streamwise coherence length of large-eddy streamwise turbulence velocity fluctuations suggests \emph{critical transition dynamics}. The streamwise coherence lengths of both streamwise and vertical turbulence velocity fluctuations ($L_{11,1}$ and $L_{33,1}$) double with extremely small increases in instability state parameter at $-z_i/L$ values generally regarded as ``near neutral.'' $L_{11,1}$, in particular, increases from $60-75\%$ the boundary layer height to about $150\%$ $z_i$ in its change from subcritical to supercritical over a very small increase in $-z_i/L$ of only about 0.15.

In the neutral ABL, the streamwise correlation lengths of vertical velocity fluctuations are about a factor of two smaller than that for streamwise velocity fluctuations ($L_{11,1}\approx2L_{33,1})$, reflecting shear production dynamics whereby streamwise fluctuations are generated in the mean with vertical fluctuations modified through pressure-strain-rate correlations. However, the sudden doubling in coherence length during critical transition in both streamwise ($u'$) and vertical ($w'$) fluctuating components occurs with much smaller relative increases in component variances (figures~\ref{fig:uvarvszibyL} and \ref{fig:wvarvszibyL}). 

As discussed in \S~\ref{sec:contrastNBLMCBL}, \cite{khanna1998three} suggest a mechanism that couples the near-surface coherent structure of $u'$ fluctuations with that of $w'$. They argued that buoyancy-generated vertical motions originate from concentrations of high temperature fluctuations within shear-generated low-speed streaks in the surface layer. The sudden change in coherence from $-z_i/L\sim0.29-0.40$ may indicate that at this critical state a sudden new interaction between buoyancy-driven and shear-driven turbulent motions initiates which, as will be discussed in the next section, may be associated with a sudden local accumulation of warm fluid within low-speed streaks.

\begin{figure}
	\centering
	\begin{subfigure}[ht!]{0.48\columnwidth}
		\includegraphics[width=\columnwidth]{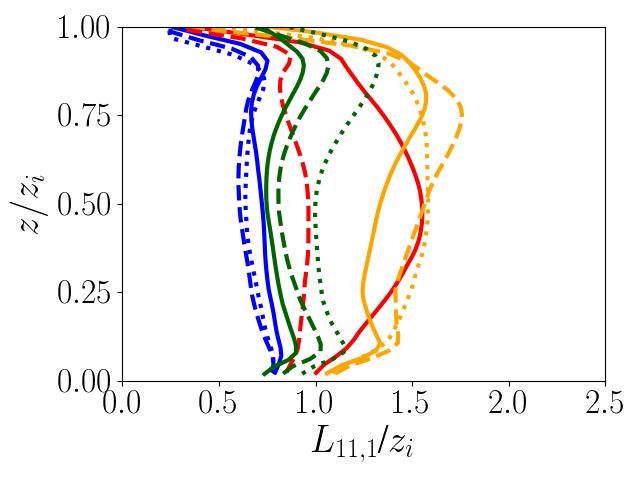}	
		\caption{\label{fig:L111vszAtzibyL} }
	\end{subfigure}
	\begin{subfigure}[ht!]{0.48\columnwidth}
		\includegraphics[width=\columnwidth]{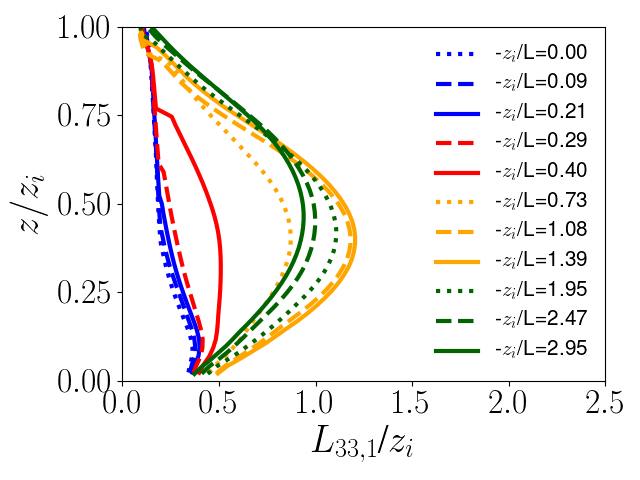}
		\caption{\label{fig:L331vszAtzibyL} }
	\end{subfigure}
	\caption{\label{fig:Lxx1vszAtzibyL} (Color Online) Transition of ABL turbulence coherence structure with the global instability parameter $-z_i/L$ over the \BJ{vertical extent of the} ABL. (a) Streamwise fluctuating velocity coherence length ($L_{11,1}$); (b) Vertical fluctuating velocity coherence length ($L_{33,1}$). 
	}
\end{figure}

However, the consequences of this critical phenomenon are not localized to the streaks in the surface layer, as shown clearly in figure~\ref{fig:Lxx1vszAtzibyL} where the coherence lengths $L_{11,1}$ and $L_{33,1}$ are plotted over the depth of the boundary layer at fixed $-z_i/L$. Over the small change in instability state between 
$-z_i/L\sim0.21-0.29$ and $0.40$ we observe a sudden abrupt change in coherence length at all $z/z_i$, implying a sudden change in the coherent structure of the entire boundary layer during the critical transition. The transition initiates in $L_{11,1}$ at slightly lower $-z_i/L$ than $L_{33,1}$. Furthermore, as previously noted, the increase in coherence length in the streamwise fluctuations occurs much more rapidly, with increasing $-z_i/L$, than the coherence length of vertical fluctuations. Note, in particular, the significant increase in $L_{11,1}$ from solid blue ($0.21$) to dashed red ($0.29$) followed by a massive $L_{11,1}$ increase to the solid red curve ($0.40$) in the mixed layer. By contrast, the increase in $L_{33,1}$ from $0.21$ to $0.29$ (solid blue to dashed red) is negligible while the increase from $0.29$ to $0.40$ (dashed to solid red) is much larger, but more gradual, than that for $L_{11,1}$ as the coherence lengths of w fluctuations continues to increase up to $-z_i/L\approx1.08$.

In fact, whereas figure~\ref{fig:Lxx1vszibyL} gave the impression that $L_{11,1}$ peaks near $-z_i/L\approx1$ and $L_{33,1}$ peaks nearer to $-z_i/L\approx1.3$, figure~\ref{fig:Lxx1vszAtzibyL} indicates that, in reality, $L_{11,1}$ peaks in the mixed layer right after the critical transition ($-z_i/L\approx0.40$) compared to the upper mixed layer where peak coherence length occurs at $-z_i/L=1.08$. Indeed, the critical transition in $L_{11,1}$ is immediate and much stronger in the mid mixing layer than near either the ground or the capping inversion. Furthermore, as $-z_i/L$ increases from the critical state ($0.40$), $L_{11,1}$ subsequently reduces in the mid mixing layer as the coherence lengths near the surface and below the capping inversion increase, ultimately exceeding the max coherence length in the mixed layer at $-z_i/L=1.08$. In contrast, the more gradual increases in horizontal coherence length of vertical velocity fluctuations ($L_{33,1}$) are clearly dominant at all times in the lower mixed layer, likely due to the suppression of vertical velocity at the surface and near the capping inversion.

The sub-critical states ($-z_i/L=0,0.09,0.21$) are characterized by streamwise coherence lengths ($L_{11,1}$ and $L_{33,1}$) in the surface layer which decrease into the mixed layer independently of stability state. Interestingly, the initial response of streamwise coherence length to the minimal surface heating is in the streamwise rather than vertical directions: at $-z_i/L=0.29$, $L_{11,1}$ has begun to change while $L_{33,1}$ had not. However between $-z_i/L\sim0.29 - 0.40$, the entire structure of the boundary layer suddenly and dramatically changes, with max coherence lengths shifting from near the surface to the mixed layer. Yet, while streamwise coherence lengths double, streamwise velocity variance increases by only $10-20\%$ and vertical velocity variance remains effectively unchanged (figures~\ref{fig:uvarvszibyL} and \ref{fig:wvarvszibyL}). The sudden change in ABL turbulence structure at $-z_i/L << 1$ is in its coherent structure rather than in its fluctuation levels.

\begin{figure}
	\centering
	\includegraphics[width=0.48\columnwidth]{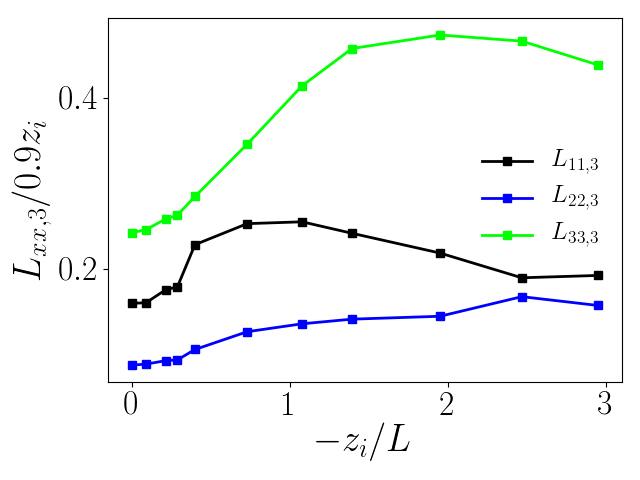}	
	\caption{\label{fig:Lxx3ForDiffx} (Color Online) Transitions of vertical coherence lengths with instability parameter ($-z_i/L$) calculated vertically from level $z=0.1z_i$ . The black, blue and green curves correspond to vertical coherence length of horizontal, spanwise and vertical velocity fluctuations, respectively. 
	}
\end{figure}

The sudden restructuring of the entire ABL at the critical state with minor changes in the velocity variances, suggests the initiation of a global instability from the interaction between shear-driven turbulent motions and low-level buoyancy-driven motions. In the transition from sub- to super-critical states, the streamwise extent of coherence changes from a maximum in the surface layer where shear is strongest, to a maximum in the mixed layer away from the surface − suggesting a critical instability process that suddenly links the lower and upper boundary layer. To test this hypothesis, we plot in figure~\ref{fig:Lxx3ForDiffx} the coherence lengths in the vertical direction from two-point correlations in $z$ referenced to $z/z_i=0.1$ (integral length scales are therefore normalized on $0.9z_i$). Peak vertical coherence lengths of horizontal motions are, as expected, well below the vertical coherence lengths of the buoyancy-driven vertical velocity fluctuations.The correlation of vertical velocity fluctuations extend to roughly $50\%$ of the capping inversion depth. In contrast, vertical coherence of horizontal fluctuations extend to roughly $25\%$ of $z_i$. Interestingly, when $L_{11,1}$ suddenly increases at the critical state, so does $L_{11,3}$ in such a way as to cap the increase in aspect ratio $L_{11,1}/L_{11,3}$ to roughly $5.6$ (as opposed to $\approx9.7$ if the vertical coherence length had not changed across the critical state).

Of all the vertical coherence lengths, $L_{11,3}$ is the only one to abruptly increase at the critical transition $-z_i/L\approx0.4$. The buoyancy driven correlation $L_{33,3}$ does increase mildly at critical transition, however it continues to increase with increasing heat flux to reach a maximum at $-z_i/L\sim1.4-2$, while $L_{11,3}$ decreases. The implication is that communication between the lower and upper boundary layer via vertical coherence increases with increasing surface heat flux above critical $-z_i/L$, suggesting that the dramatic restructuring of the entire boundary layer around the critical transition state is not due to a sudden creation of nonlocal communication between the lower and upper ABL.

As will be discussed in \S~\ref{sec:MaxCoherenceTransition}, the `critical’ transition at $-z_i/L\approx0.4$ initiates a change in ABL turbulence structure that continues to evolve towards a special peak coherence state characterizing the large-scale roll structure commonly described from cloud structure. It is surprising to find that this initiation of fundamental transition in ABL structure is observed most strongly in the coherence of streamwise fluctuations and only later in the coherence of vertical fluctuations. It is also surprising that the transition initiates at very low $-z_i/L$ previously assumed to be quasi neutral (e.g., \cite{khanna1997analysis,khanna1998three}).

\subsection{The role of Near-surface Streaks in the ‘Critical’ Transition in Turbulence Structure \label{subsec:CriticalTransition-StreakRole}}
We have described the apparent existence of a ``critical transition,'' or perhaps nonlinear ``bifurcation,'' that manifests as an abrupt restructuring of the entire ABL turbulence coherence structure without the creation of a strong linkage between the lower with the upper ABL. We show here that this critical transition initiates post-critical ABL states in which low-speed streaks align with concentrations of high temperature and vertical velocity fluctuations, a process that we shall learn in \S~\ref{sec:MaxCoherenceTransition} strengthens in the development of large-scale rolls and the strong linking of the upper and lower boundary layer.

\begin{figure}
	\centering
	\begin{subfigure}[ht!]{0.36\columnwidth}
		\includegraphics[width=\columnwidth]{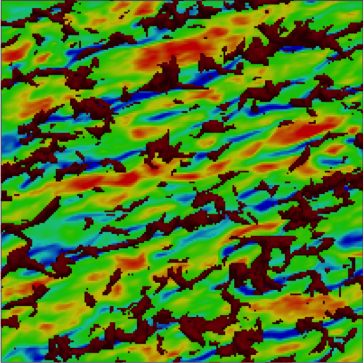}	
		\caption{\label{fig:CriticalTrans-uisozibyL-021} }
	\end{subfigure}
	\begin{subfigure}[ht!]{0.36\columnwidth}
		\includegraphics[width=\columnwidth]{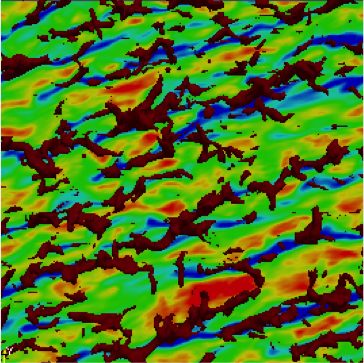}
		\caption{\label{fig:CriticalTrans-uisozibyL-029} }
	\end{subfigure}
	\begin{subfigure}[ht!]{0.36\columnwidth}
	\includegraphics[width=\columnwidth]{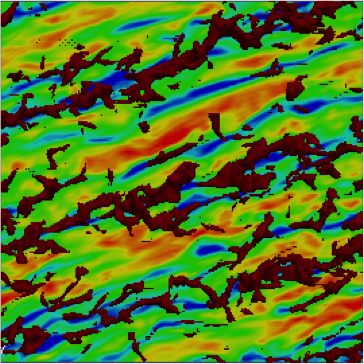}
	\caption{\label{fig:CriticalTrans-uisozibyL-040} }
\end{subfigure}
	\begin{subfigure}[ht!]{0.36\columnwidth}
	\includegraphics[width=\columnwidth]{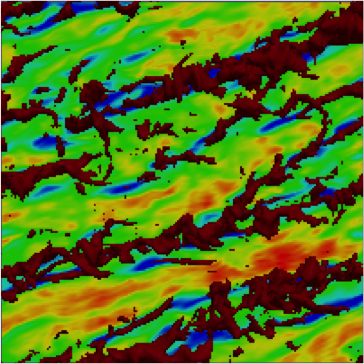}
	\caption{\label{fig:CriticalTrans-uisozibyL-073} }
\end{subfigure}
	\begin{subfigure}[ht!]{0.36\columnwidth}
	\includegraphics[width=\columnwidth]{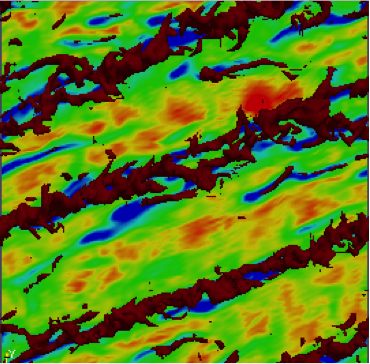}
	\caption{\label{fig:CriticalTrans-uisozibyL-108} }
\end{subfigure}
	\begin{subfigure}[ht!]{0.36\columnwidth}
	\includegraphics[width=\columnwidth]{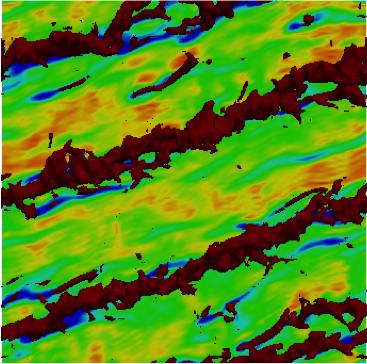}
	\caption{\label{fig:CriticalTrans-uisozibyL-139} }
\end{subfigure}
	\caption{\label{fig:CriticalTrans-uisozibyL} (Color Online) Simultaneous visualization of isocontours of x-component velocity ($u'$) on the $z/z_i=0.1$ plane (the underlying blue-green-red isocontours) and a single isosurface of vertical velocity (dark red) at a $w'$ level of $2\sigma_w$ (standard deviations). The isocontours of $u'$ vary from $-2\sigma_u$ (dark blue) to $+2\sigma_u$ (bright red); the solid green color is close to zero. The dark-blue isocontours ($u\approx -2\sigma_u$) identify the coherent ``low-speed-streaks'' that result from the action of mean shear-rate, dominant in the surface layer. The red isosurface identifies the structure of the strongest updraft motions, which occur spatially in the mixed layer. The relationship between the shear-driven low-speed-streaks below and thermal updrafts above are shown as a function of stability state. The transition between subcritical and supercritical occurs at $-z_i/L\approx0.4$ (\S~\ref{subsec:CriticalTransition-existence}\cmnt{\ref{sec:CriticalTransition}\ref{subsec:CriticalTransition-existence}}). The coherence length $L_{11,1}$ peaks at the supercritical state $-z_i/L\approx1.08$ (\S~\ref{subsec:CriticalTransition-StreakRole} and \S~\ref{sec:MaxCoherenceTransition}\cmnt{\ref{subsec:CriticalTransition-StreakRole}}). 
	}
\end{figure}

In figure~\ref{fig:CriticalTrans-uisozibyL} we superpose an isosurface of vertical velocity (at $+2\sigma_w$) over a near-surface plane ($z/z_i=0.1$) of isocontours of fluctuating velocity in the (unrotated) x direction. Figure~\ref{fig:CriticalTrans-uisozibyL} shows the view of the ABL from above over the 5 km × 5 km horizontal domain. Whereas the red isosurfaces show localized coherent thermal updrafts in the mixed layer, the plane of isocontours show the coherent shear-generated turbulence structures in the surface layer referred to as ``low-speed streaks'' in blue with the less coherent ``high-speed'' streaks in red/orange.

Figures~\ref{fig:CriticalTrans-uisozibyL-021} \& \ref{fig:CriticalTrans-uisozibyL-029} correspond to sub-critical instability states. At these states the regions of concentrated vertical velocity visually display much less coherence than do the low-speed streaks below, with relatively little alignment between the updrafts and streaks. By contrast, the turbulence at critical and super-critical instability states (figures \ref{fig:CriticalTrans-uisozibyL-040} \& \ref{fig:CriticalTrans-uisozibyL-073}) indicate higher levels of organization in the updrafts and significantly more alignment between the updrafts in the mixed layer and the underlying low-speed streaks in the surface layer. Indeed, comparison of the super-critical state (d) with the sub-critical state (a) shows clearly a strong change in the both the structure of the updrafts − they become more concentrated, well-defined and coherent − and the alignment of the updrafts with the strongest low-speed-streak structures below. 

The transition from sub-critical to super-critical is associated with both stronger more coherent updrafts in the mixed layer and enhanced alignment of the stronger mixed layer updrafts above with low-speed-streaks in the surface layer below. \cite{khanna1998three} argued that the increased alignment reflects the near-surface process of concentration of higher-temperature fluid within shear-generated, highly coherent, low-speed streaks, concentrating buoyancy force there and driving flow vertically within sheet-like structures (see \S~\ref{sec:contrastNBLMCBL}). In this way, the shear-induced low-speed-streaks become the ``seeds'' to updrafts. Observationally (figure~\ref{fig:CriticalTrans-uisozibyL}) and quantitatively (figure~\ref{fig:Lxx1vszibyL}), the coherence of the surface layer elongated streaks increase with increasing surface heating, but only after transition to the super-critical state. Correspondingly, continued increases in streak coherence within increasing heating in the post-critical state leads to increasingly strong concentrations of high-temperature fluid within thermal updrafts that increase in strength and coherence with increasing $-z_i/L$ in a manner directly tied to low-speed-streaks below, elongated in the mean flow direction and relatively narrow in the spanwise direction.

What is particularly fascinating is that once the ABL has transitioned to a discretely different dynamical state at very low $-z_i/L$, continued increases in surface heating and buoyancy somehow strengthen coherence of the shear-driven horizontal fluctuations within ``low-speed streaks''. This strengthening of streak correlation in turn causes strengthening of the concentration of temperature fluctuations within the streaks, further strengthening buoyancy force and vertical velocity fluctuations. However, this nonlinear interplay between surface-heating-induced buoyancy and shear-driven streak coherence does not manifest until the boundary layer has undergone a critical transition to super-critical, creating a dynamically altered ABL that responds differently to increases in surface heating than in the sub-critical ``near neutral'' state. 

\begin{figure}
	\centering
	\begin{subfigure}[ht!]{0.48\columnwidth}
		\includegraphics[width=\columnwidth]{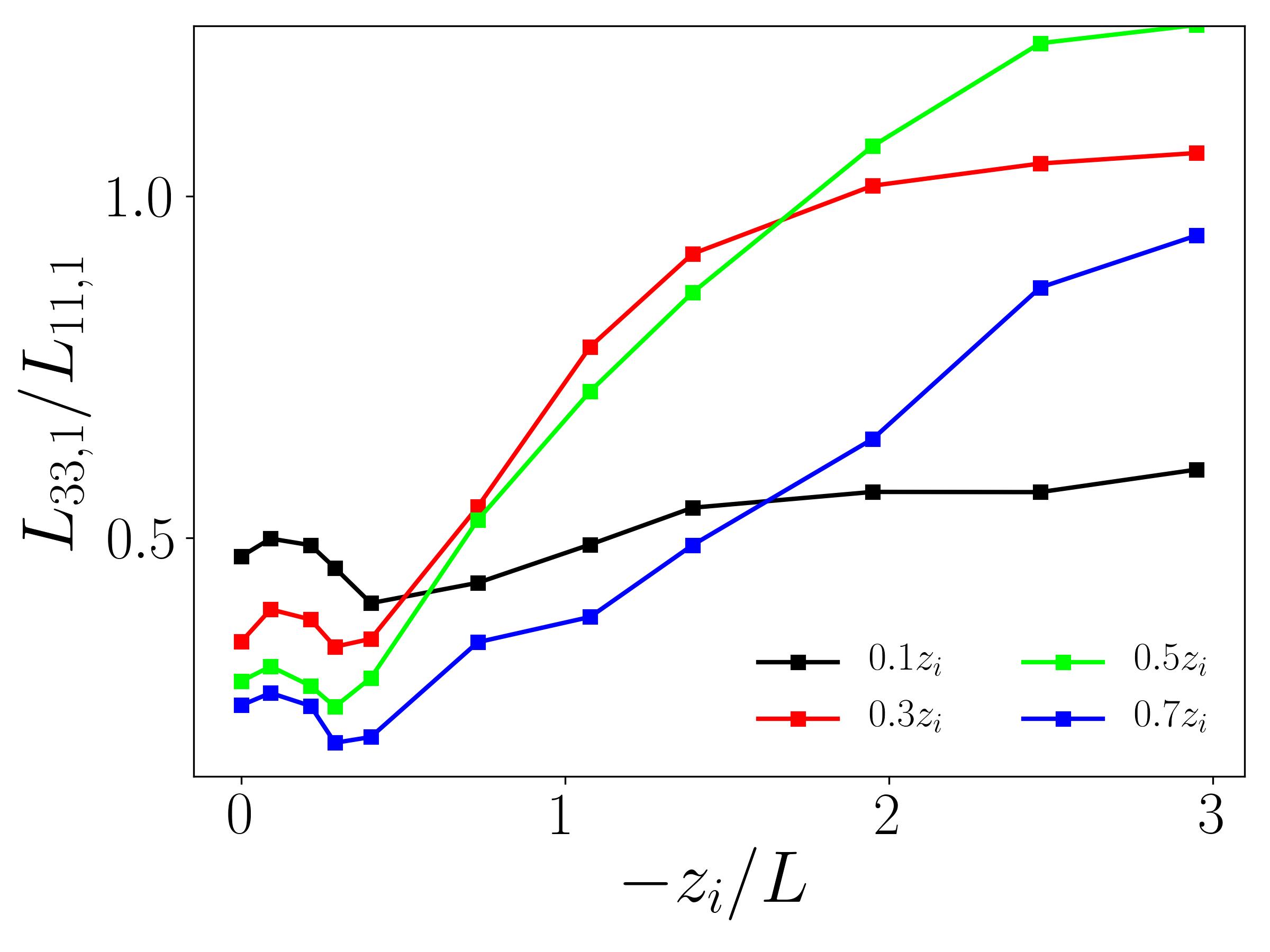}	
		\caption{\label{fig:Lyy1byLxx1vszibyL} }
	\end{subfigure}
	\begin{subfigure}[ht!]{0.48\columnwidth}
		\includegraphics[width=\columnwidth]{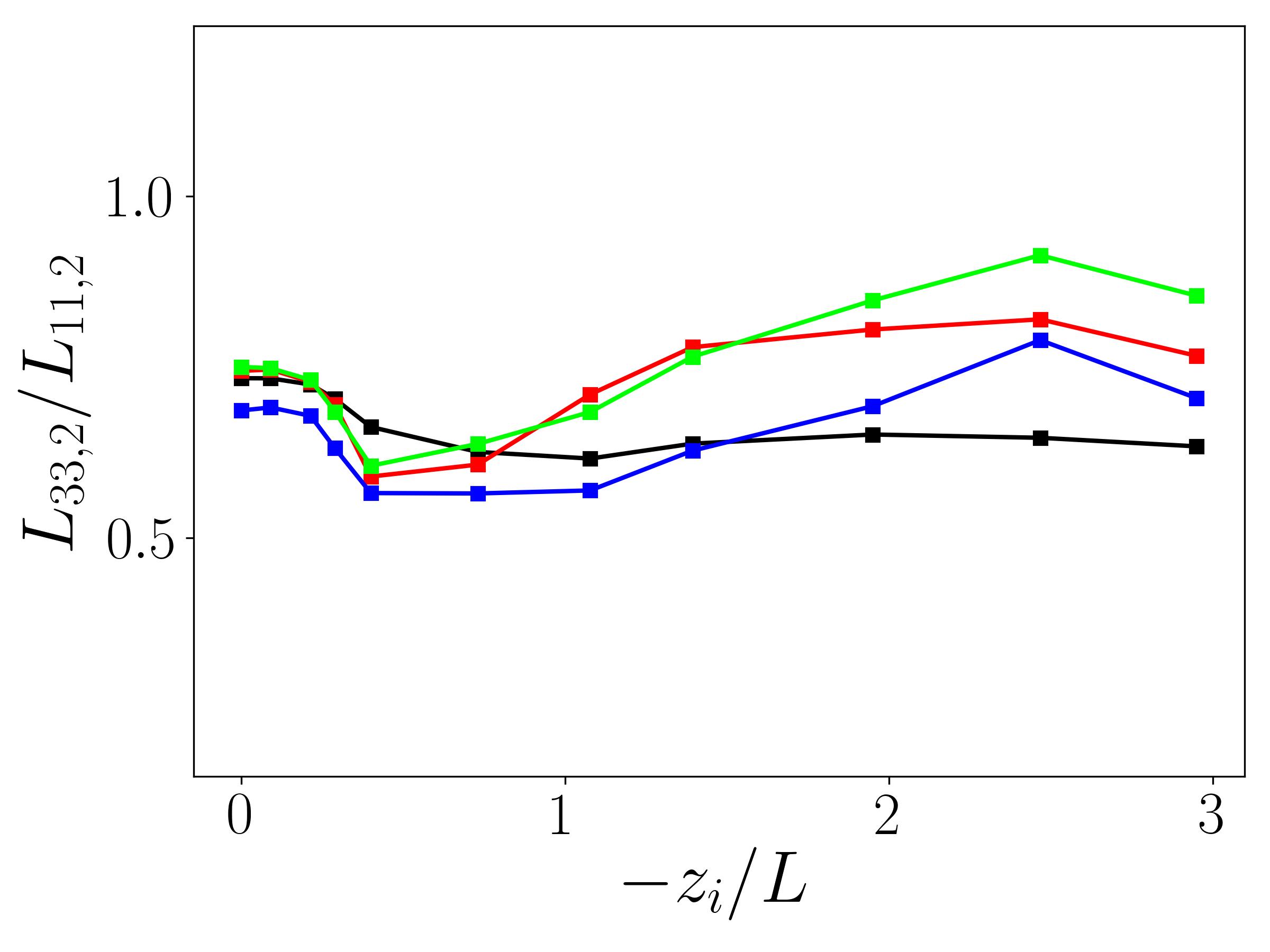}
		\caption{\label{fig:Lyy2byLxx2vszibyL} }
	\end{subfigure}
	\caption{\label{fig:LyykbyLxxkvszibyL} (Color Online) Transition in the ratio of coherence lengths of vertical to horizontal velocity fluctuations with stability state $-z_i/L$ at fixed heights: (a) streamwise coherence lengths; (b) transverse coherence lengths. 
	}
\end{figure}

\section{A Maximum Coherence Stability State and Development of “Large-scale Rolls”\label{sec:MaxCoherenceTransition}}
Figures \ref{fig:CriticalTrans-uisozibyL-108} \& \ref{fig:CriticalTrans-uisozibyL-139} show post-critical evolution to a maximum coherence state that we identify below with exceptionally well organized ABL-scale coherent roll eddies. We analyze this transition and the maximum coherence state in the current section.
\subsection{The Transition in Stability State from Super-critical to Maximum Coherence\label{subsec:MaxCoherenceTransition-SupCritToMaxCoh}}
Visually, figure~\ref{fig:CriticalTrans-uisozibyL} suggests that the average streamwise coherence length of the vertical velocity fluctuations (thermals) increases relative to the coherence length of the horizontal velocity fluctuations in the transition to the super-critical stability state and with increasing $-z_i/L$ when in the super-critical state. This qualitative observation is shown quantitatively in figure~\ref{fig:LyykbyLxxkvszibyL} where the ratio of the streamwise and spanwise coherence lengths of streamwise and vertical fluctuating velocities are plotted against $-z_i/L$. We observe from figure~\ref{fig:Lyy1byLxx1vszibyL} that the critical stability state ($-z_i/L\approx0.4$) represents the minimum in the ratio of the coherence lengths of vertical to streamwise velocity fluctuations ($L_{33,1}/L_{11,1}$). In fact, the transition to the supercritical state represents the initiation of a continuous growth in $L_{33,1}/L_{11,1}$ with increasing $-z_i/L$, from  $L_{33,1}/L_{11,1}\sim0.2$ to $L_{33,1}/L_{11,1}\sim1.0$ in the mixed layer, consistent with the visual impression given by figure~\ref{fig:CriticalTrans-uisozibyL}. This results is additional evidence that, as buoyancy force strengthens and increasingly connects the lower and upper regions of the boundary layer (figure~\ref{fig:Lxx3ForDiffx}), the change in the coherent structure of the thermal updrafts in the mixed layer with increasing $-z_i/L$ become progressively more connected to the change in the coherent structure of the low-speed-streaks in the surface layer. However this change initiates only after the ABL has transitioned to super-critical.

Figure~\ref{fig:Lyy2byLxx2vszibyL} shows that the spanwise coherence length of vertical fluctuations is consistently smaller than that for streamwise velocity fluctuations at all instability states. The decrease in $L_{33,2}/L_{11,2}$ at the critical transition ($-z_i/L\approx0.40$) is a result of a corresponding small increase in $L_{11,2}$ with minimal change in $L_{33,2}$. In fact, $L_{33,2}$ is roughly constant with increasing $-z_i/L$ (not shown). Figure~\ref{fig:Lyy1byLxx1vszibyL} shows that the streamwise coherence length of $u'$ exceeds that of $w'$ until $-z_i/L$ approaches $\sim1.4-2$, where streamwise coherence lengths of horizontal and vertical velocity fluctuations roughly coincide in the mixed layer. (In the surface layer and near the capping inversion, where vertical velocity is suppressed, the streamwise coherence length of $u'$ always exceeds that of $w'$.) In the vertical, however, the coherence length of vertical velocity always exceeds that of horizontal velocity fluctuations (figure~\ref{fig:Lxx3ForDiffx}). 

The existence of relatively broad maxima in ABL coherence parameters over the instability states $-z_i/L\sim1-2$ (figures~\ref{fig:Lxx1vszibyL},\ref{fig:Lxx3ForDiffx} and \ref{fig:LyykbyLxxkvszibyL}) indicates a smooth transition from increasing coherence with increasing surface heat flux, to the destruction of this enhanced coherence at higher surface heat rates. Thus, contrary to the expectation that streak coherence would immediately diminish upon the addition of any level of surface heating from the neutral state (see \S~\ref{sec:TransitionNBLtoMCBL}), the destruction of coherence is delayed to instability state parameters higher than $-z_i/L\sim1-1.5$, instability states far from neutral. Instead, streamwise coherence is suddenly enhanced with the introduction of very small levels of surface heating to an otherwise neutral ABL. It is not until surface heat flux exceeds roughly $10\%$ of the mid-day peak of $\sim0.25$ Km/s \citep{wyngaard2010turbulence} that destruction of streamwise coherence takes place. Noting the near-linear relationship between $Q_0$ and $-z_i/L$ (figure~\ref{fig:zibylvsq0}), figures ~\ref{fig:Lxx1vszibyL} and \ref{fig:Lxx3ForDiffx} indicate that the rate of increase  in coherence length with increasing $-z_i/L$ preceding peak coherence is much more rapid than its destruction by further increases in $Q_0$ after peak coherence. Clearly the maximum coherence state is a special stability state with unique properties.

\subsection{The Maximally Coherent Stability State: Large-scale Rolls\label{subsec:MaxCoherenceTransition-LargeScaleRolls}}
Although the increase in $u'$ coherence length is substantial throughout the boundary layer, it is generally concentrated near the surface, while the growth in coherence length of $w'$ manifests in the mixed layer away from the surface (figure~\ref{fig:Lxx1vszAtzibyL}). As the ABL approaches the maximal-coherence state, the updrafts concentrate more and more within coherent sheet-like thermals that extend vertically from the upper surface layer to the capping inversion with lateral coherence scale small compared to the streamwise coherence scale. The transition to the max coherence state is associated with ever-stronger coupling between the updrafts above and the increasing coherence of the low-speed streaks below, from where the updrafts originate.

Figure~\ref{fig:CriticalTrans-uisozibyL} visually illustrates this transition. At instability states $-z_i/L$ below the maximum coherence state ($-z_i/L\sim1.08-1.39$), the vertical updrafts in the supercritical states $-z_i/L$ between $0.4$ and $1.08$ are spread over a larger number of smaller thinner streamwise streaks as compared to the peak coherence state, where the correlation between low-speed streaks is visually maximal. In comparison with the lower $-z_i/L$ post critical states, the max coherence state is characterized by maximally extended low-speed streaks with maximal alignment with vertical updrafts that, themselves, are maximally coherent and extended in the streamwise direction. Visually it is clear from figures~\ref{fig:CriticalTrans-uisozibyL-108} and \ref{fig:CriticalTrans-uisozibyL-139} that the max coherence stated is characterized by exceptionally well organized and coherent updrafts tied to exceptionally coherent and extended low-speed streaks, with essentially no updraft or low-speed streaks in between. Figures~\ref{fig:CriticalTrans-uisozibyL-108} and \ref{fig:CriticalTrans-uisozibyL-139} indicate that fluid particles originating at the lower margins of the red updraft regions in the lower boundary layer will move vertically, on average, within the exceptionally strong coherent updrafts and longitudinally with the mean velocity. As the fluid particles arrive at the capping inversion they are blocked from vertical motion and forced laterally towards the centerline between neighboring highly coherent updraft sheets, where they are forced downward to conserve mass, creating ``large-scale atmospheric rolls.''

Because the lateral separations between the coherent thermal sheets scale on the boundary layer depth ($z_i$), the helical streamline patterns often extend tens of capping inversion scales ($O(10)z_i$) in the streamwise direction. It is visually clear from figure~\ref{fig:CriticalTrans-uisozibyL} that the quantitatively determined max coherence state is characterized by helical streamline patterns that connect exceptionally coherent updrafts with maximally coherent downdraft regions within high aspect ratio roll-like coherent structures that extend longitudinally over many boundary layer depths. 

Whereas the large-scale roll pattern is observable over a range on instability states, the maximally coherent stability state is the strongest most coherent manifestation of large-scale rolls, often described in the literature in relationship to parallel lines of clouds near the capping inversion created by the condensation of moisture carried aloft within the coherent sheet-like updrafts \citep{etling1993roll,weckwerth1997horizontal,weckwerth1999observational,lemone1973structure,lemone1976relationship}. Interestingly, while the streamwise coherence lengths of the velocity fluctuations peak at about $1.75z_i$ in the max coherence state, the streamwise extent of the coherent large-scale roll structure visually, in both figure~\ref{fig:CriticalTrans-uisozibyL} and in meterological images of cloud-topped large-scale rolls, shows large-scale coherence that extends well outside the simulation domain of $7-8 z_i$; cloud visualizations suggest the extension of coherent large-scale rolls often over tens of $z_i$.

A number of quantitative measures, in addition to peaks in coherence lengths, characterize the maximally coherent large-scale roll state. The distribution of streamwise coherence lengths within the boundary layer in the max coherence state must be seen in context with the transition process. Figure~\ref{fig:L111vszAtzibyL} shows that whereas the critical stability state is characterized by a sudden and dramatic enhancement of streamwise coherence in the mixed layer, at $-z_i/L$ below and above critical, the streamwise coherence length of streamwise fluctuations peaks near the lower surface and upper capping inversion, where blockage in the vertical forces the flow to be horizontal. What distinguishes the max coherence state is the dramatic increase in coherence length in the upper margins of the boundary layer adjacent to the capping inversion, where coherence length is locally maximal. \BJ{As in the lower near-wall region where the enhanced coherence length is associated with well-defined streak-like structures, one can expect the region near the capping inversion to also be heavily characterized by streaks.  }

The streamwise coherence length of vertical velocity fluctuations also peaks at the max coherence state, however this peak is squarely in the mixed layer. Thus, the large-scale roll state is distinguished by exceptionally strong vertical coherence of vertical velocity in the mixed layer in combination with exceptionally strong horizontal coherence of horizontal velocity fluctuations in the upper boundary layer, \emph{a cylindrical topology}. Figure~\ref{fig:StatsvszibyL} shows that the max coherence state is also associated with peaks in the variance of streamwise (but not vertical) velocity fluctuations, production rate of streamwise velocity variance, and vertical turbulent momentum flux. All of these variations are physically consistent with the first-order enhancement of coherence within the low-speed streaks that underlies the post-critical stability states up to max coherence.

\begin{figure}
	\centering
	\begin{subfigure}[ht!]{0.33\columnwidth}
		\includegraphics[width=\columnwidth]{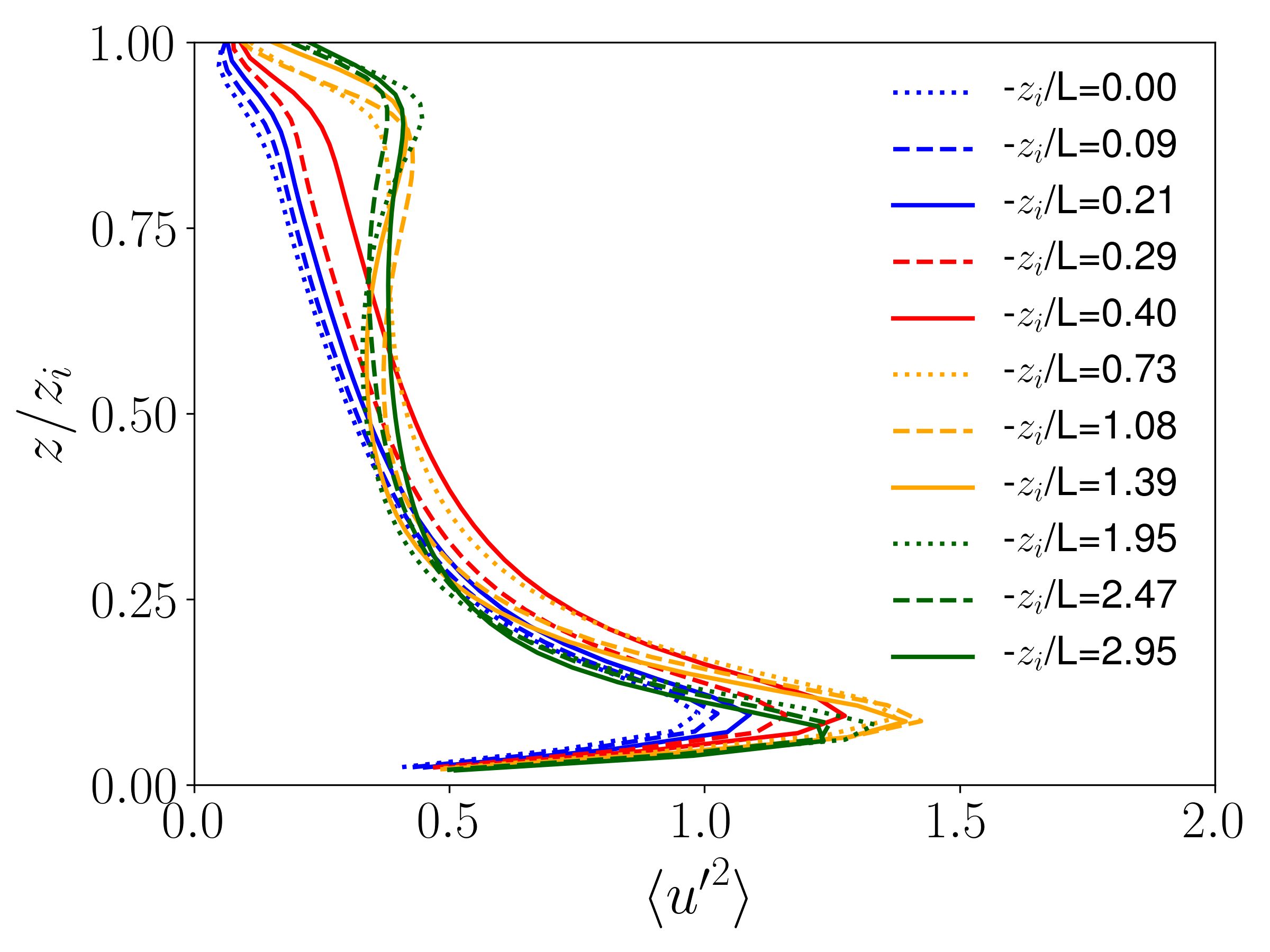}	
		\caption{\label{fig:uvarVszAtDiffzibyL} $\langle u'^2 \rangle$}
	\end{subfigure}
	\begin{subfigure}[ht!]{0.32\columnwidth}
		\includegraphics[width=\columnwidth]{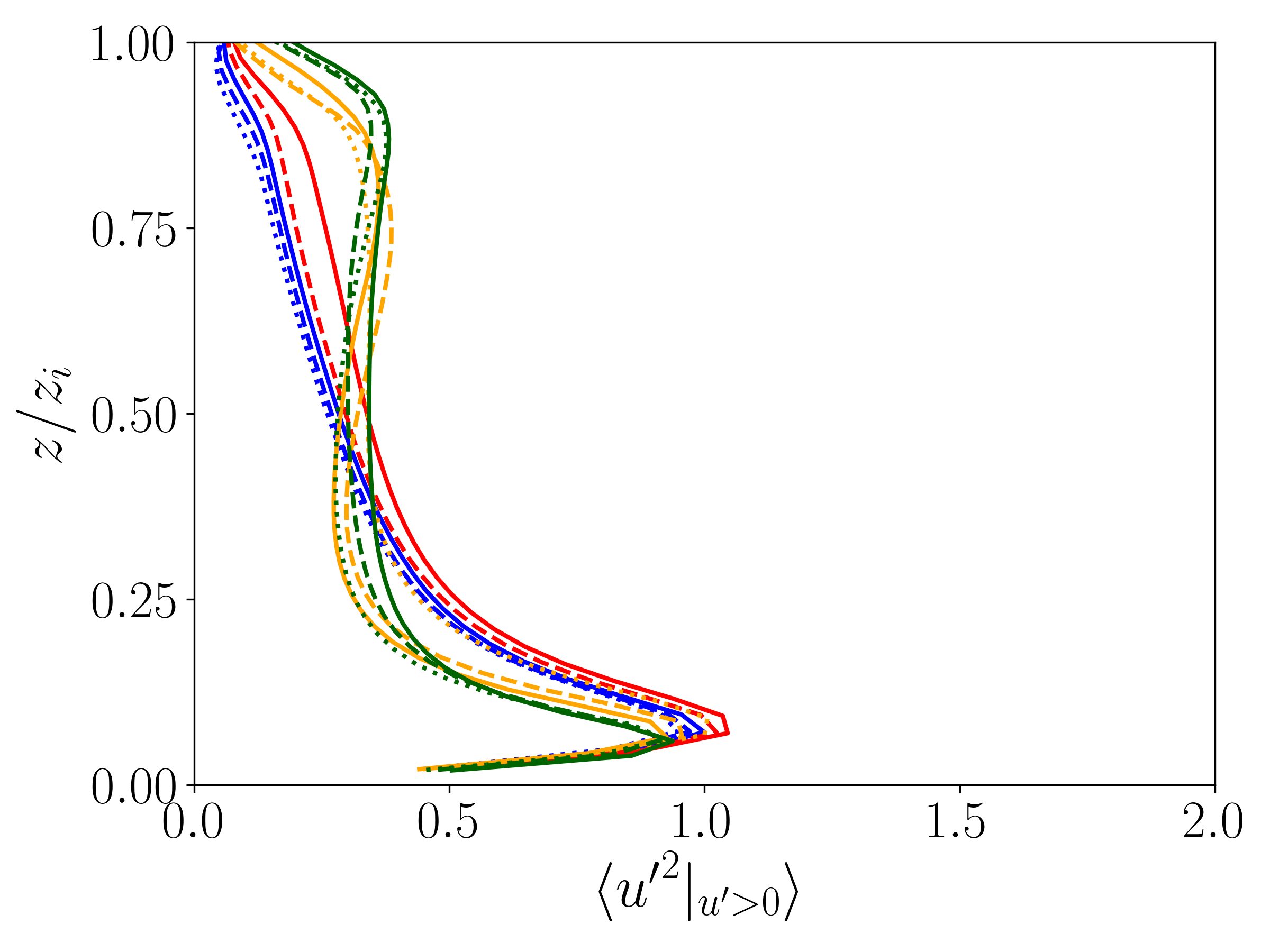}
		\caption{\label{fig:uvarHSVszAtDiffzibyL} $\langle u'^2 \rangle|_{u'>0}$}
	\end{subfigure}
	\begin{subfigure}[ht!]{0.32\columnwidth}
	\includegraphics[width=\columnwidth]{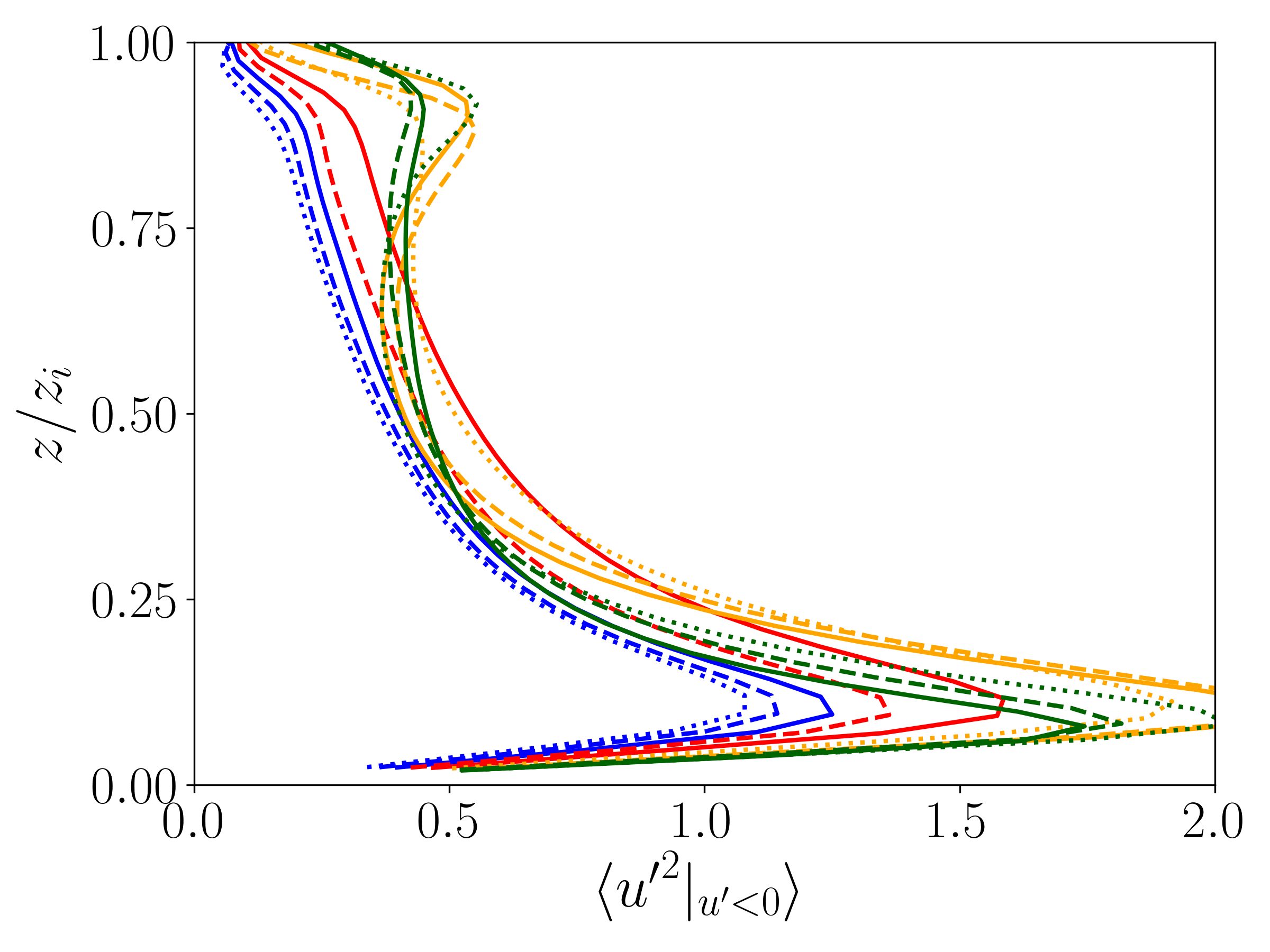}
	\caption{\label{fig:uvarLSVszAtDiffzibyL}$\langle u'^2 \rangle|_{u'<0}$ }
\end{subfigure}
	\begin{subfigure}[ht!]{0.32\columnwidth}
	\includegraphics[width=\columnwidth]{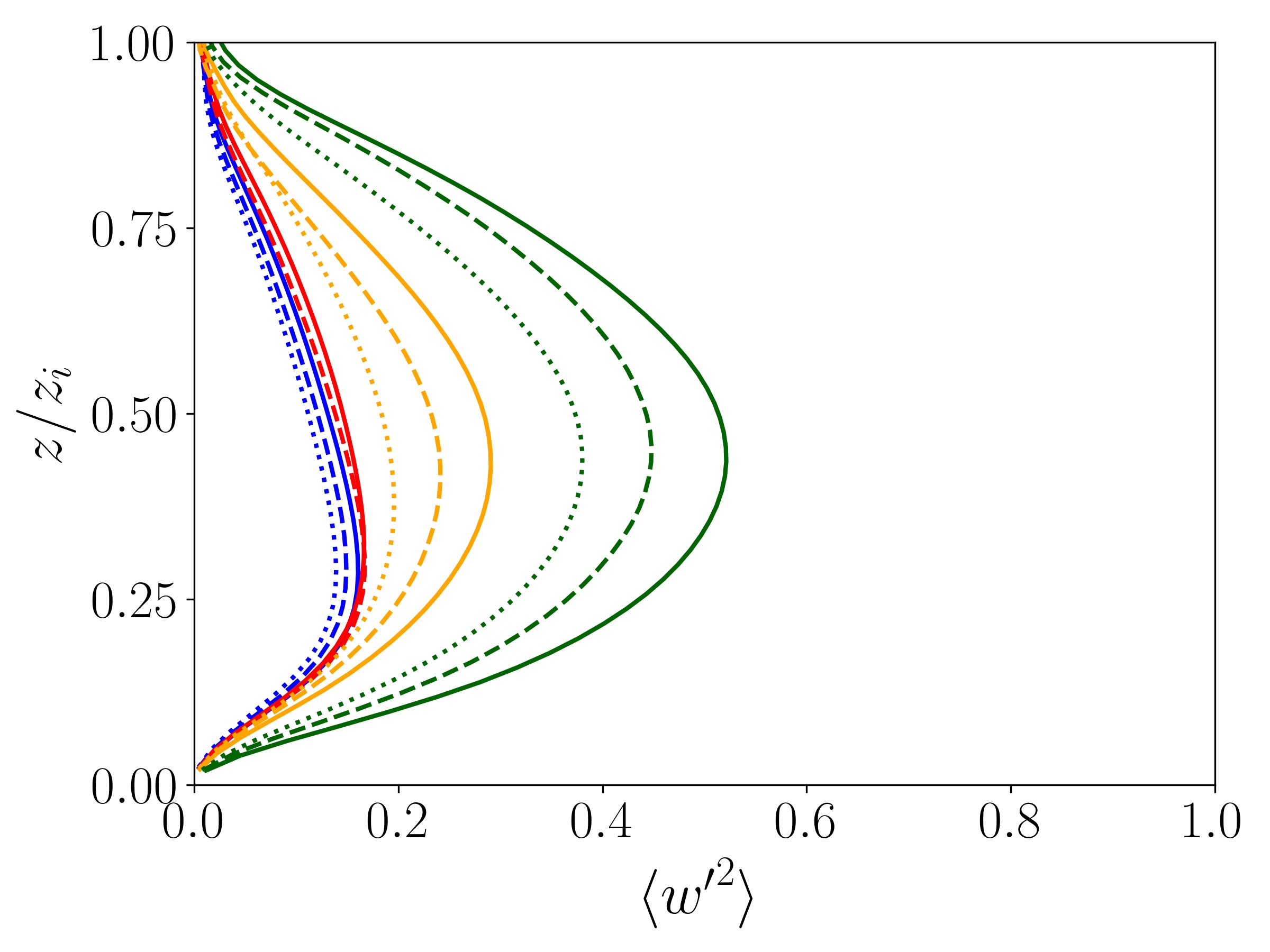}	
	\caption{\label{fig:wvarVszAtDiffzibyL} $\langle w'^2 \rangle$}
\end{subfigure}
\begin{subfigure}[ht!]{0.32\columnwidth}
	\includegraphics[width=\columnwidth]{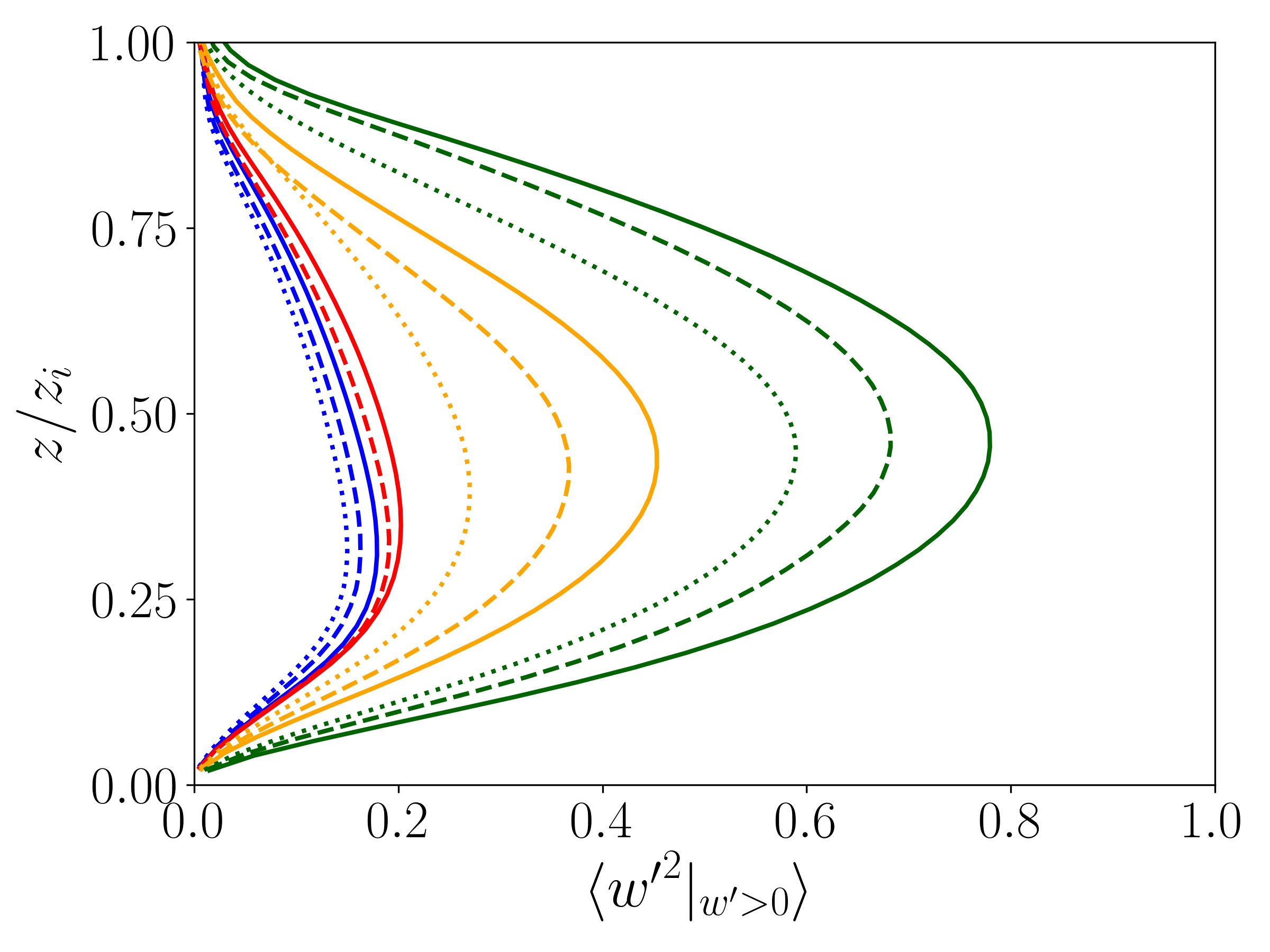}
	\caption{\label{fig:wvarHSVszAtDiffzibyL} $\langle w'^2 \rangle|_{w'>0}$}
\end{subfigure}
\begin{subfigure}[ht!]{0.32\columnwidth}
	\includegraphics[width=\columnwidth]{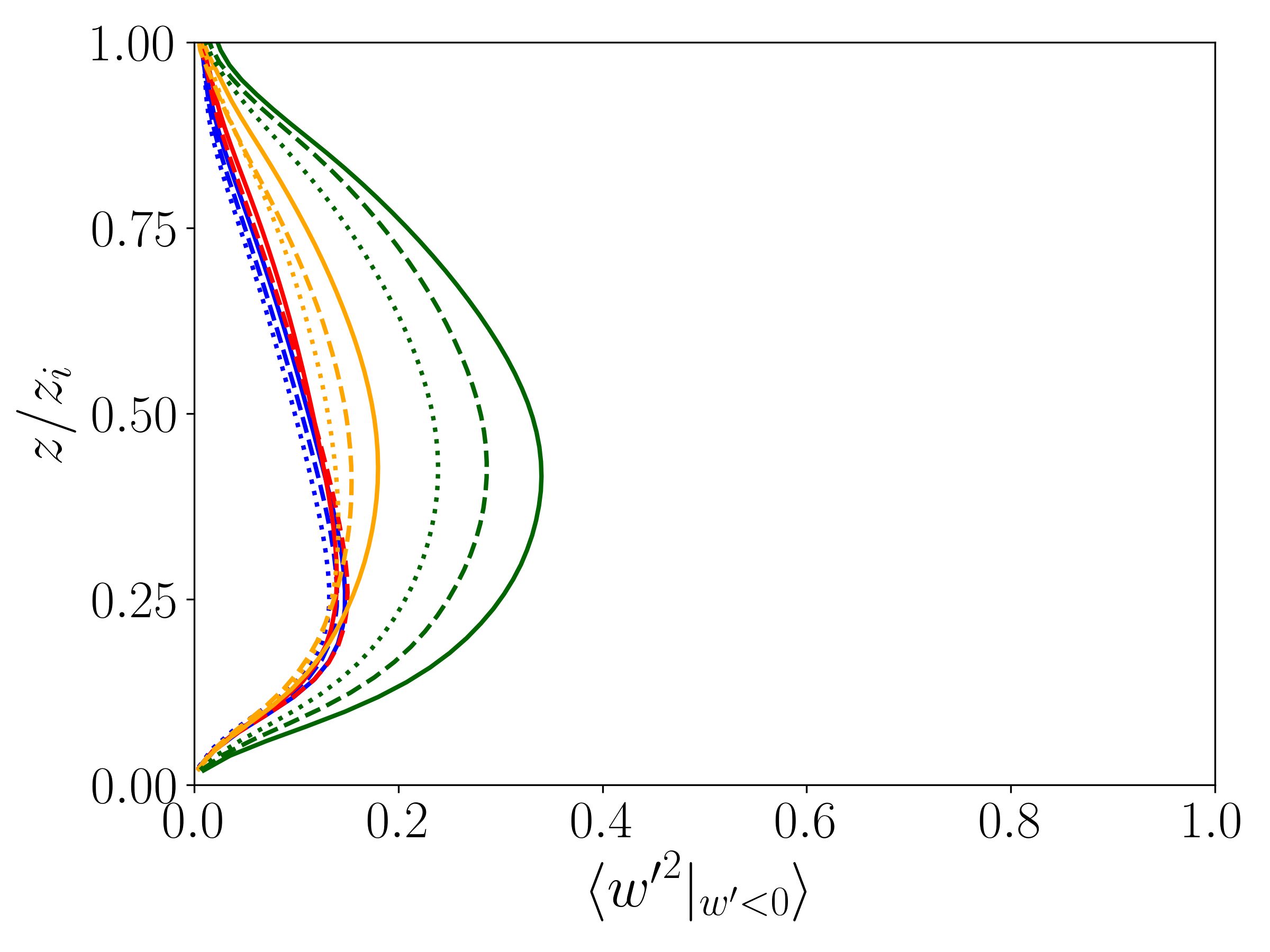}
	\caption{\label{fig:wvarLSVszAtDiffzibyL} $\langle w'^2 \rangle|_{w'<0}$}
\end{subfigure}
	\caption{\label{fig:VarVszAtDiffzibyL} (Color Online) Vertical variation of streamwise and vertical velocity variances as a function of stability state in comparison with conditionally sampled variances based on the sign of horizontal velocity fluctuations (low/high-speed streaks) and vertical velocity fluctuations (updrafts/thermals, downdrafts). All variances are in units of $(m/s)^2$. 
	}
\end{figure}
\begin{figure}
\begin{subfigure}[ht!]{0.48\columnwidth}
	\includegraphics[width=\columnwidth]{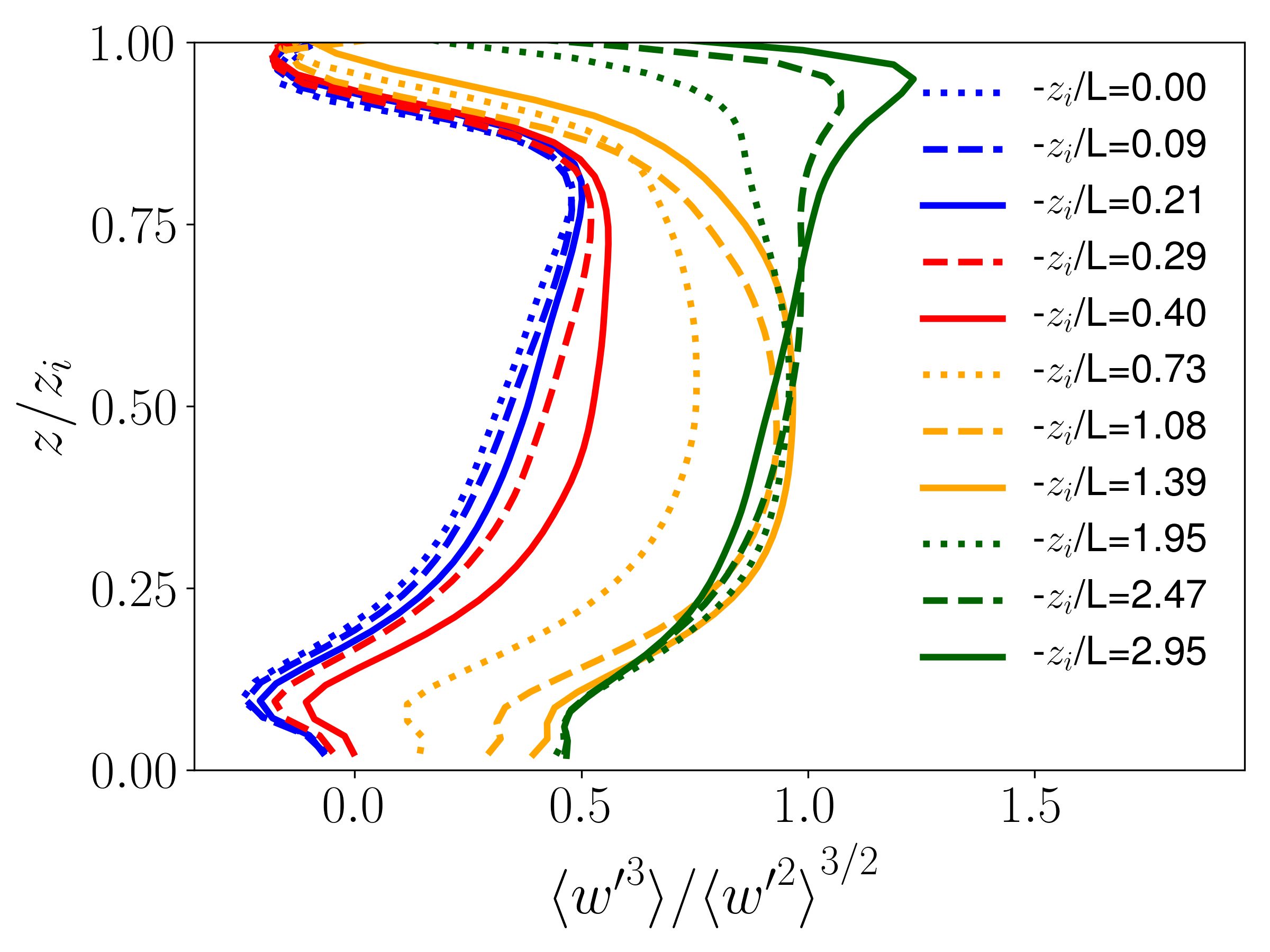}
	\caption{\label{fig:wskewVszAtDiffzibyL} $\langle w'^3 \rangle/{\langle w'^2 \rangle}^{3/2}$}
\end{subfigure}
\begin{subfigure}[ht!]{0.48\columnwidth}
	\includegraphics[width=\columnwidth]{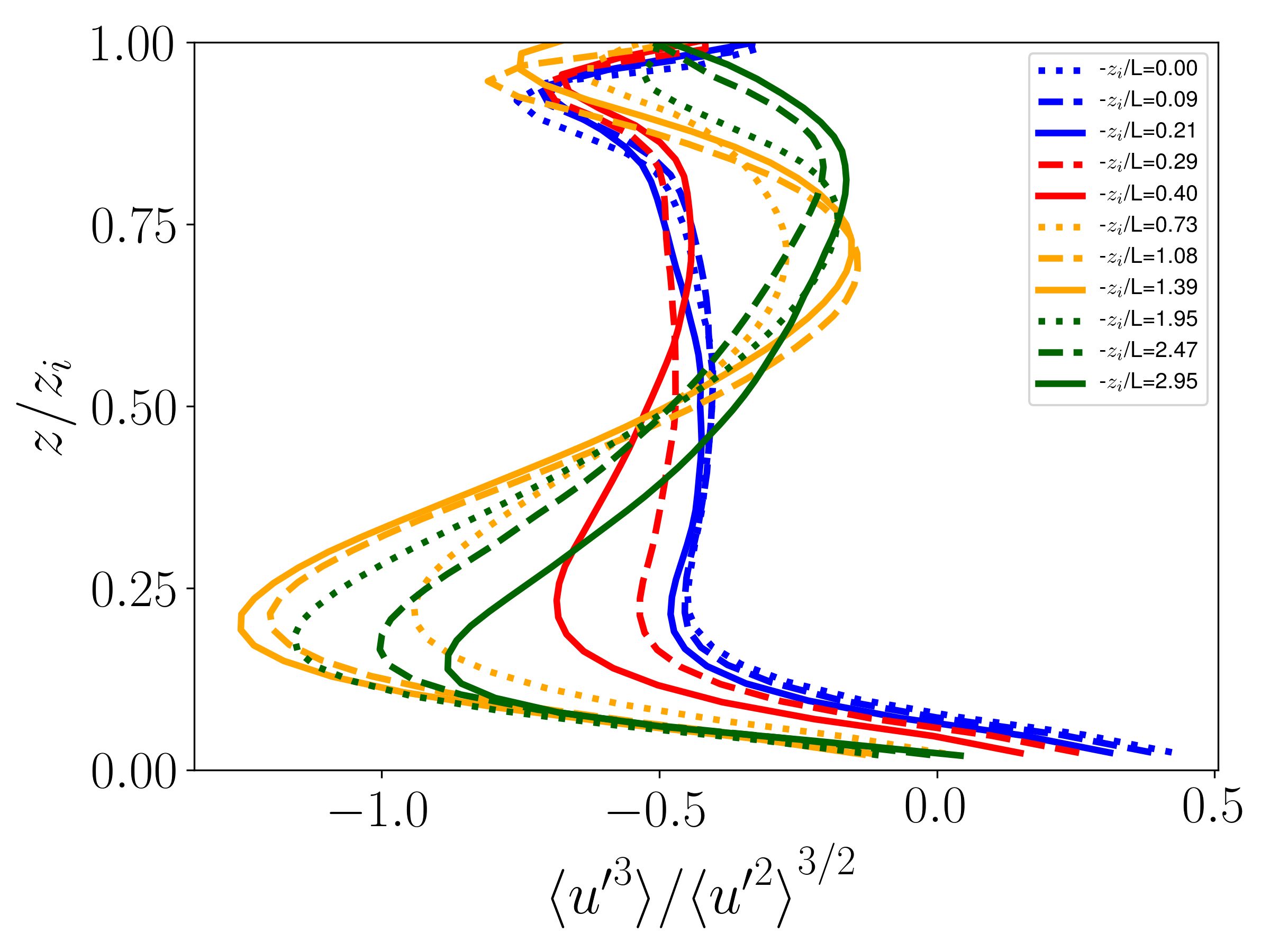}
	\caption{\label{fig:uskewVszAtDiffzibyL} $\langle u'^3 \rangle/{\langle u'^2 \rangle}^{3/2}$}
\end{subfigure}
\caption{\label{fig:SkewVszAtDiffzibyL} (Color Online) Transition in the spatial variation of skewness of (a) vertical and (b) streamwise velocity fluctuations as a function of stability state.
}
\end{figure}

The last statement is supported by figure~\ref{fig:VarVszAtDiffzibyL} where variances are conditionally sampled based on negative vs. positive fluctuating horizontal velocity (low/high-speed streaks), and on upward vs. downward vertical motions (updrafts/thermals vs. downdrafts). It is clear from figure~\ref{fig:VarVszAtDiffzibyL} that the dominant contributor to horizontal velocity variance is the low-speed streaks and the dominant contributor to vertical velocity variance is updrafts/thermals. The max coherence roll state is quite clearly characterized by strong response in the low-speed streaks in the horizontal (figure~\ref{fig:uvarLSVszAtDiffzibyL}). Furthermore, at the max coherence state horizontal velocity variance associated with negative fluctuations develops a pronounced local peak adjacent to the capping inversion. The latter is consistent with a strong maximum in negative skewness of horizontal velocity fluctuations in the lower mixed layer at the max coherence state, shown in figure~\ref{fig:uskewVszAtDiffzibyL}. Furthermore, figure~\ref{fig:wskewVszAtDiffzibyL} shows that the max coherence state is the state at which positive vertical velocity skewness reaches its maximum level in the mixed layer as vertical velocity variance continues to increase at higher $-z_i/L$ (figures~\ref{fig:wvarVszAtDiffzibyL}-\ref{fig:wvarLSVszAtDiffzibyL}). 

These results, taken together with observations drawn from figure~\ref{fig:VarVszAtDiffzibyL}, provide strong support for two overarching conclusions regarding the development of the max coherence state of large-scale rolls within the atmospheric boundary layer: (1) there exists strong sensitivity to the structure and dynamic enhancement of the low-speed streaks, first in the transition to the critical state at extremely low levels of surface heating, and then in the continued enhancement to a peak coherence that is central to the character of the max coherence roll state; and (2) there exists an exceptionally strong spatial and structural relationship between the maximally coherent low-speed streaks below and the maximally coherent vertical velocity fluctuations above associated with dominantly coherent narrow streamwise-extended sheet-like thermal updrafts that drive the production of helical roll structure in particle trajectories at the boundary layer scale.

Taken together, the directed coherence lengths of horizontal and vertical velocity fluctuation lead to the development of a special stability state where coherence lengths are maximal in the vicinity of  $-z_i/L\sim1.5$, where maximum horizontal coherence of horizontal fluctuations is achieved at $-z_i/L \approx 1.2-1.5$, and where maximum vertical coherence of vertical fluctuations is achieved at $-z_i/L \approx 1.5-2$. Large-scale rolls are strongest when $L_{33,1}$ approaches $L_{11,1}$ and the upper and lower boundary layer regions become exceptionally strongly coupled. However, the predominant mechanism for roll formation is associated with horizontal-wind-driven strong shear-rate near the surface with dynamics that leads to the formation of low-speed streaks, together with an \emph{enhancement mechanism} that involves an interesting interaction between buoyancy and shear, and a tightly coupled organization of updrafts along intensified low-speed streaks.

\section{The Near-neutral and Moderately Convective ABL \label{sec:NearNeutralAndMCBL}}
To this point in our analysis, we have discussed the dramatic transition in ABL structure that takes place across a critical ABL state close to neutral ($-z_i/L\approx0.3-0.4$), and the post-critical transition in ABL coherent structure to the max coherence roll state ($-z_i/L\approx1.2-2$). We complete the description in change in ABL structure with systematic increases in surface heat flux by describing the pre-critical and post max-coherence instability states.
\subsection{The Near-neutral Stability State ($0<-z_i/L< 0.3$) Regime\label{subsec:NearNeutralAndMCBL-NBL}}
Here we describe our understanding of what may be described as a ``near neutral'' stability state, an equilibrium ABL with finite levels of surface heat flux sufficiently small that the ABL has properties close to the neutral state. Apparently the near-neutral state is characterized by negligible impact of buoyancy force so that temperature fluctuations are effectively passive tracers. The transition to temperature as an active scalar occurs just above the critical stability state $-z_i/L\approx0.29$, where surface heat flux is $\approx0.0035$ K.m/s, only $1-2\%$ of peak mid day surface heat flux. The near-neutral character of the equilibrium ABL is dramatically destroyed at these extraordinarily small levels of surface heat flux (figure~\ref{fig:Lxx1vszibyL}). From neutral through $-z_i/L\approx0.21$, however, our simulations produce an ABL that is effectively neutral with temperature fluctuations \BJ{playing an almost passive role in the dynamics}.

Whereas figure~\ref{fig:StatsvszibyL} indicates the existence of small but measurable changes in the statistical structure of the near-neutral ABL below critical transition ( $z_i/L\lesssim0.29$), figure~\ref{fig:Lxx1vszAtzibyL} indicates negligible change in the coherence structure of the boundary layer. The first significant adjustment in ABL structure occurs above $-z_i/L\approx0.29$ where the horizontal integral scales of horizontal velocity fluctuations show their first clear deviation from the near-neutral state. However, at slightly higher $-z_i/L$, the entire coherence structure of the ABL has changed, driven by the interaction between buoyancy and shear-driven motions and the conversion in temperature from passive to active scalar. 

\begin{figure}
	\centering
	\begin{subfigure}[ht!]{0.36\columnwidth}
		\includegraphics[width=\columnwidth]{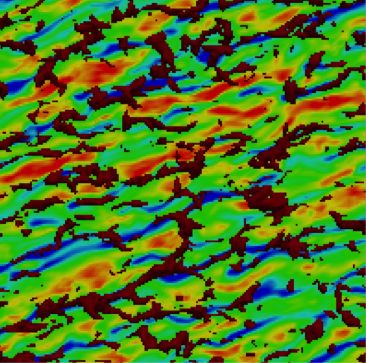}
		\caption{\label{fig:SubCriticalTrans-uisozibyL-000} $-z_i/L=0.00$ (neutral)}
	\end{subfigure}
	\begin{subfigure}[ht!]{0.36\columnwidth}
		\includegraphics[width=\columnwidth]{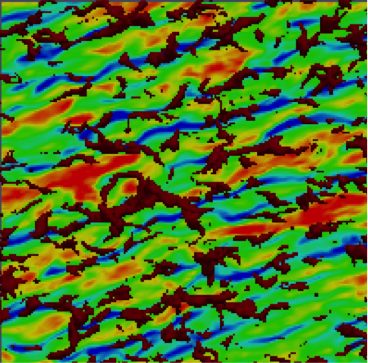}
		\caption{\label{fig:SubCriticalTrans-uisozibyL-009} $-z_i/L=0.09$ (near-neutral)}
	\end{subfigure}
	\begin{subfigure}[ht!]{0.36\columnwidth}
	\includegraphics[width=\columnwidth]{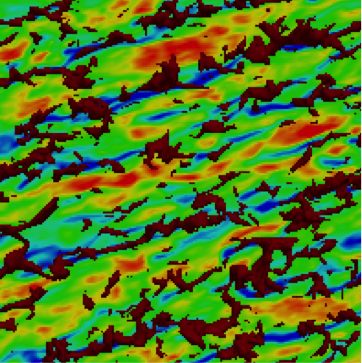}
	\caption{\label{fig:SubCriticalTrans-uisozibyL-021} $-z_i/L=0.21$ (sub-critical)}
\end{subfigure}
\begin{subfigure}[ht!]{0.36\columnwidth}
	\includegraphics[width=\columnwidth]{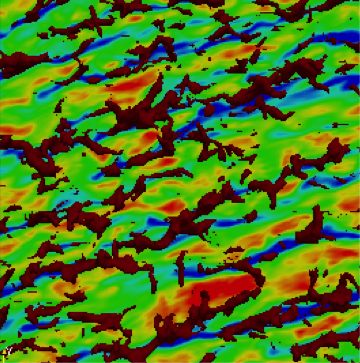}
	\caption{\label{fig:SubCriticalTrans-uisozibyL-029} $-z_i/L=0.29$ (near critical)}
\end{subfigure}
	\caption{\label{fig:SubCriticalTrans-uisozibyL} (Color Online) Transition in the instantaneous 3D structure of fluctuating resolved velocity with $-z_i/L$ across the near-neutral instability state regime. As in figure~\ref{fig:CriticalTrans-uisozibyL}, isocontours of fluctuating streamwise velocity $u'$ are shown on the plane $z/z_i=0.1$(surface layer) in the range $\pm 2\sigma_u$ together with a superposed isosurface of vertical fluctuating velocity at the threshold level $w'=2\sigma_w$, predominantly in the mixed layer. Low-speed streaks are given by the blue isocontour regions on the $z/z_i=0.1$ plane.}
\end{figure}

Figure~\ref{fig:SubCriticalTrans-uisozibyL} is an extension of figure~\ref{fig:CriticalTrans-uisozibyL} to include the near-neutral range of stability states. Visually, the instantaneous turbulence structure at stability states $-z_i/L = 0.09$ and $0.21$ look qualitatively the same as the neutral state: the existence of well-defined highly coherent low-speed streaks in the surface layer overlaid with relatively incoherent concentrations of vertical velocity roughly aligned globally with the low-speed streaks below. It is only at the near-critical stability state, and strongly at the critical state (figure~\ref{fig:CriticalTrans-uisozibyL-040}) that an observable strengthening, growth and alignment of the vertical velocity concentrations occurs together with a visually (as well as quantitatively) observable strengthening of the low-speed streaks below. 

\subsection{Transition to the ``Moderately Convective'' Stability State\label{subsec:NearNeutralAndMCBL-MCBL}}

\begin{figure}
	\centering
	\begin{subfigure}[ht!]{0.36\columnwidth}
		\includegraphics[width=\columnwidth]{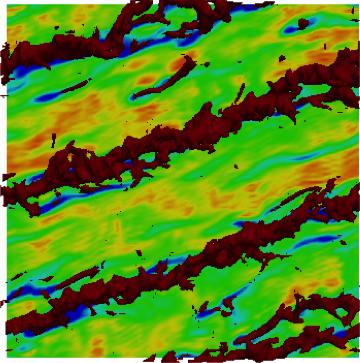}
		\caption{\label{fig:PostMaxCoherentTrans-uisozibyL-139} $-z_i/L=1.39$ (max. coherent)}
	\end{subfigure}
	\begin{subfigure}[ht!]{0.36\columnwidth}
		\includegraphics[width=\columnwidth]{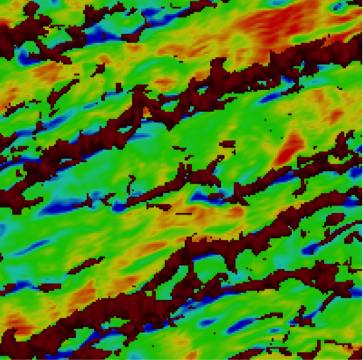}
		\caption{\label{fig:PostMaxCoherentTrans-uisozibyL-195} $-z_i/L=1.95$ (MCBL)}
	\end{subfigure}
	\begin{subfigure}[ht!]{0.36\columnwidth}
		\includegraphics[width=\columnwidth]{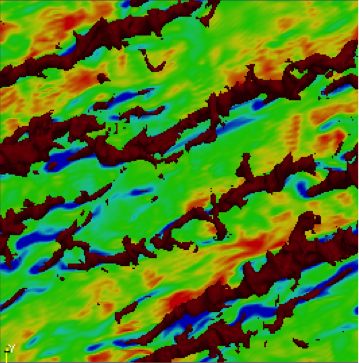}
		\caption{\label{fig:PostMaxCoherentTrans-uisozibyL-247} $-z_i/L=2.47$ (MCBL)}
	\end{subfigure}
	\begin{subfigure}[ht!]{0.36\columnwidth}
		\includegraphics[width=\columnwidth]{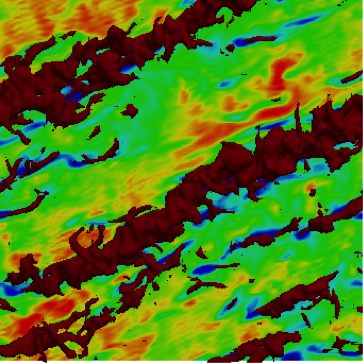}
		\caption{\label{fig:SubCriticalTrans-uisozibyL-295} $-z_i/L=2.95$ (MCBL)}
	\end{subfigure}
	\caption{\label{fig:PostMaxCoherentTrans-uisozibyL} (Color Online) Transition across moderately convective stability state regime in the instantaneous 3D large-eddy velocity structure with instability parameter  $(-z_i/L)$ at $z/z_i=0.1$. As in figure~\ref{fig:CriticalTrans-uisozibyL}, isocontours of fluctuating streamwise velocity $u'$ are shown on the plane $z/z_i=0.1$(surface layer) in the range $\pm 2\sigma_u$ together with a superposed isosurface of vertical fluctuating velocity at the threshold level $w'=2\sigma_w$, predominantly in the mixed layer. Low-speed streaks are given by the blue isocontour regions on the $z/z_i=0.1$ plane.}
\end{figure}

The existence of coherence measures that peak implies a reduction in coherence at the more unstable states when $-z_i/L$ exceeds the max coherence range $-z_i/L\approx1.2 - 2$ (figures~\ref{fig:Lxx1vszibyL} and \ref{fig:Lxx3ForDiffx}). This reduction in coherence in the structure of the ABL is apparent in figure~\ref{fig:PostMaxCoherentTrans-uisozibyL}, which continues the isocontour-isosorface images of figure~\ref{fig:CriticalTrans-uisozibyL} to $-z_i/L$ up to $2.95$. It is clear from the progression of visual representations of coherent structure with increasing $-z_i/L$ in figure~\ref{fig:CriticalTrans-uisozibyL} that the exceptionally strong coherence of the max coherence state begins to break down as global instability parameter $-z_i/L$ increases to $-z_i/L=2.95$. The exceptionally coherent sheets of vertical updraft become less so, and smaller-scale less-well-organized updraft features appear between the dominant updrafts associated with the large-scale rolls. Consequently, the large-scale roll structure loses coherence. Still, a large-scale roll structure continues to exist at $-z_i/L=2.95$.

Continued increase in the instability state towards higher values of $-z_i/L$ implies greater production of turbulence fluctuations from buoyancy-driven mechanisms relative to mechanisms associated with the interaction between mean shear-rate and Reynolds stress. Similarly, it implies that the shear-induced near-surface velocity scale $u_*$  is progressively dominated by the mixed-layer buoyancy-induced velocity scale $w_*$. Figures \ref{fig:Lxx1vszibyL}, \ref{fig:Lxx3ForDiffx} and \ref{fig:LyykbyLxxkvszibyL} suggest that the coherence structure of the ABL asymptotes to a less-coherent ``moderately convective'' state with increasing $-z_i/L$ and $w_*/u_*$. What distinguishes the ``moderately convective'' ABL is the continued importance of shear-rate, and correspondingly a roll-like structure, even as convection increasingly dominates and roll coherence diminishes. The current study has made clear the importance of shear-rate in the surface layer to the formation and coherence of thermal updrafts, dominantly in the mixed layer. It is this relationship that leads to the formation of rolls and it is this relationship that strengthens to peak dominance when the ABL is in the ``max coherence'' roll state. It is therefore also this relationship that breaks down as surface heating increases, the ABL becomes progressively more unstable globally, and 
\BJ{\cmnt{\st{the relative contribution of buoyancy to turbulence fluctuations to shear-rate increases (as quantified by $-z_i/L$ and $w_*/u_*$)}}the contribution of buoyancy to turbulence generation relative to shear-rate increases (as quantified by $-z_i/L$ and $w_*/u_*$).} 

Extrapolating figure~\ref{fig:zibylvsq0} to the average surface heat flux mid-day ($0.25$K.m/s according to data in \cite{wyngaard2010turbulence}), implies that the equilibrium ABL with geostrophic wind of $\approx10$ m/s has max $-z_i/L$ of roughly $12-14$ mid-way within the ``moderately convective'' range. For $-z_i/L$ to be higher, the wind speed above the capping inversion must be lower. When $-z_i/L$ is sufficiently high, we anticipate a transition from a less coherent helical roll-like structures to turbulent Rayleigh-Bernard convective cells that fill the volume of convection-dominated flow, as illustrated in figure~\ref{fig:CBL-Baseline} and discussed in \S\ref{sec:contrastNBLMCBL}. Large-scale rolls are highly extended in the streamwise direction as a result of strong coupling with the shear-generated low-speed streaks below. As the strength of the streaks diminishes sufficiently with increasing $-z_i/L$ at the end of the moderately convective stability state regime, the streaks become sufficiently weak and the correlation between low-speed streak and thermal updraft becomes sufficiently tenuous that the relationship between shear and buoyancy driven motions breaks down entirely. It is at this instability state that the high-aspect-ratio streamwise-elongated roll-like pattern transitions to a packed array of Rayleigh-Bernard convection cells. Whereas the large-scale rolls have a streamwise correlation scale that is coupled to the streamwise scale of the low-speed streaks below (figure~\ref{fig:Lyy1byLxx1vszibyL}) and decoupled from the ABL depth $z_i$, Rayleigh-Bernard cells are roughly axisymmetric around a vertical axis with horizontal dimensions that scale on the boundary layer depth, $z_i$. The transition from the ``moderately convective'' roll-based topological form to a ``fully convective'' Rayleigh-Bernard-cell-based form, the subject of a recent study by \cite{salesky2017nature}, likely occurs when $-z_i/L$ is multiple orders of magnitude greater than one~\citep{moeng1994comparison,khanna1998three,salesky2017nature}.

\section{Temporal Dynamics of ABL Turbulent Coherent Structure and Stability State\label{sec:TemporalTransitionDynamics}}

 The previous discussions were based on fully converged statistics with spatial averaging over $5 \textrm{km} \times 5 \textrm{km}$ horizontal planes and temporal averaging over time windows of $60$ eddy turnover times, $\tau_u$, where $\tau_u=z_i/u_*$.Whereas the horizontal domain is designed to resolve well the horizontal integral scales, the resulting domain averages have insufficient sample to produce fully converged statistics and residual temporal variability remains. Here we study temporal dynamics within the $60\tau_u$ averaging time window in relationship to the temporal changes in coherent ABL turbulence structure as a function of instability state by analyzing the temporal variability in the horizontal averages.
 
 We use the symbol $L'_{xx,y}=L'_{xx,y}(t)$  to describe horizontal averages over $5 \textrm{km} \times 5 \textrm{km}$ horizontal planes (i.e., $\langle\dots\rangle_h$ in \eqref{eq:Rij}), so that $L_{xx,y}=\langle L'_{xx,y}(t)\rangle_t$ is the fully converged mean coherence length. In Fig. 16 we plot normalized temporal standard deviations $\sigma_{xx,y}$ as a function of instability state $-z_i/L$, where $\sigma^2_{xx,y}=\langle \left( L'_{xx,y}(t)-L_{xx,y}\right)^2\rangle_t$ defines the temporal variances of $L'_{xx,y}(t)$. In each plot we superpose the mean values for $L_{11,1}$ and $L_{33,1}$ from figure~\ref{fig:Lxx1vszibyL} to make a number of interesting observations.
 
\begin{figure}
 	\begin{subfigure}[ht!]{0.48\columnwidth}
 		\includegraphics[width=\columnwidth]{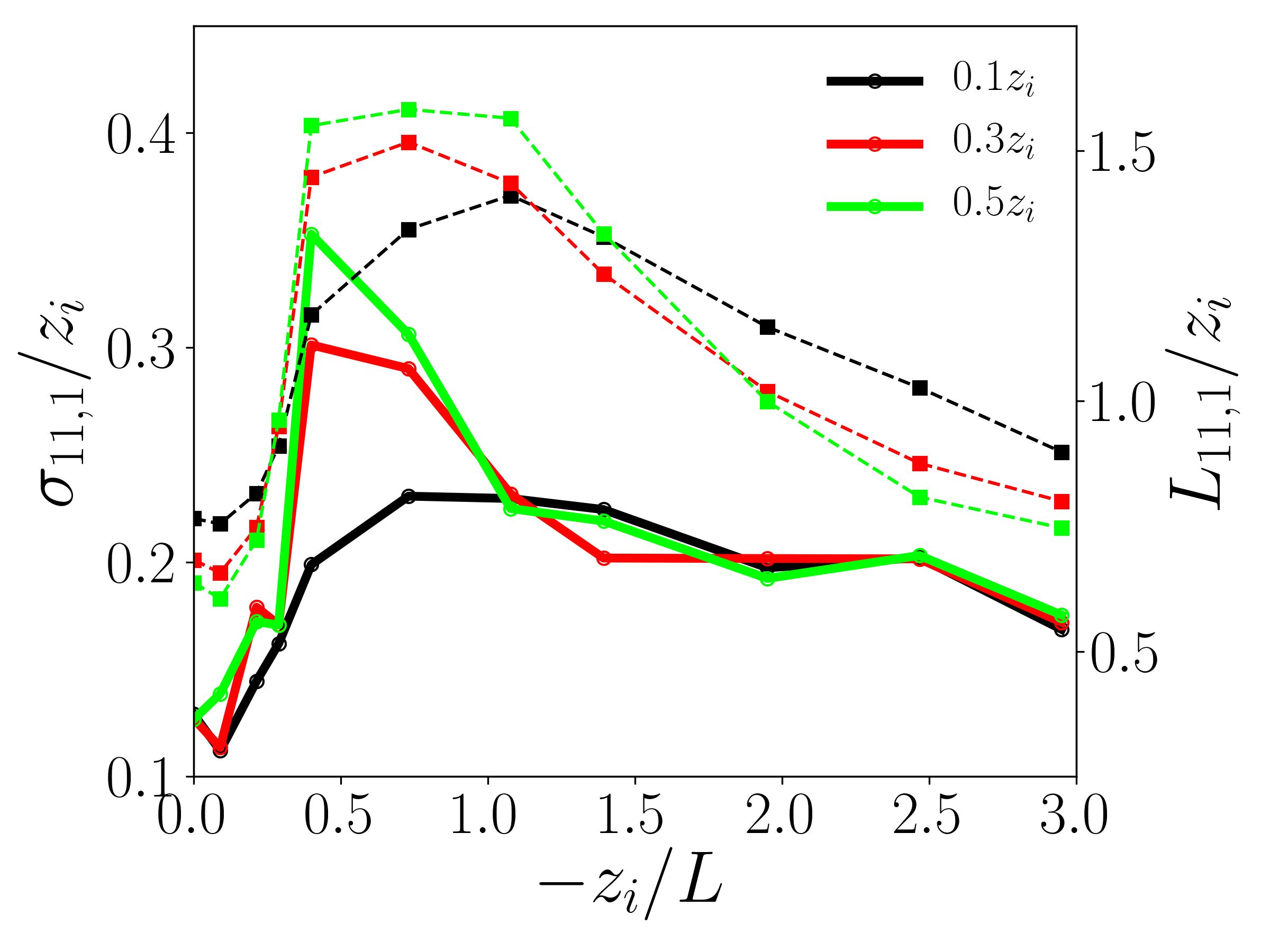}
 		\caption{\label{fig:L111VarVszibyL}}
 	\end{subfigure}
 	\begin{subfigure}[ht!]{0.49\columnwidth}
 		\includegraphics[width=\columnwidth]{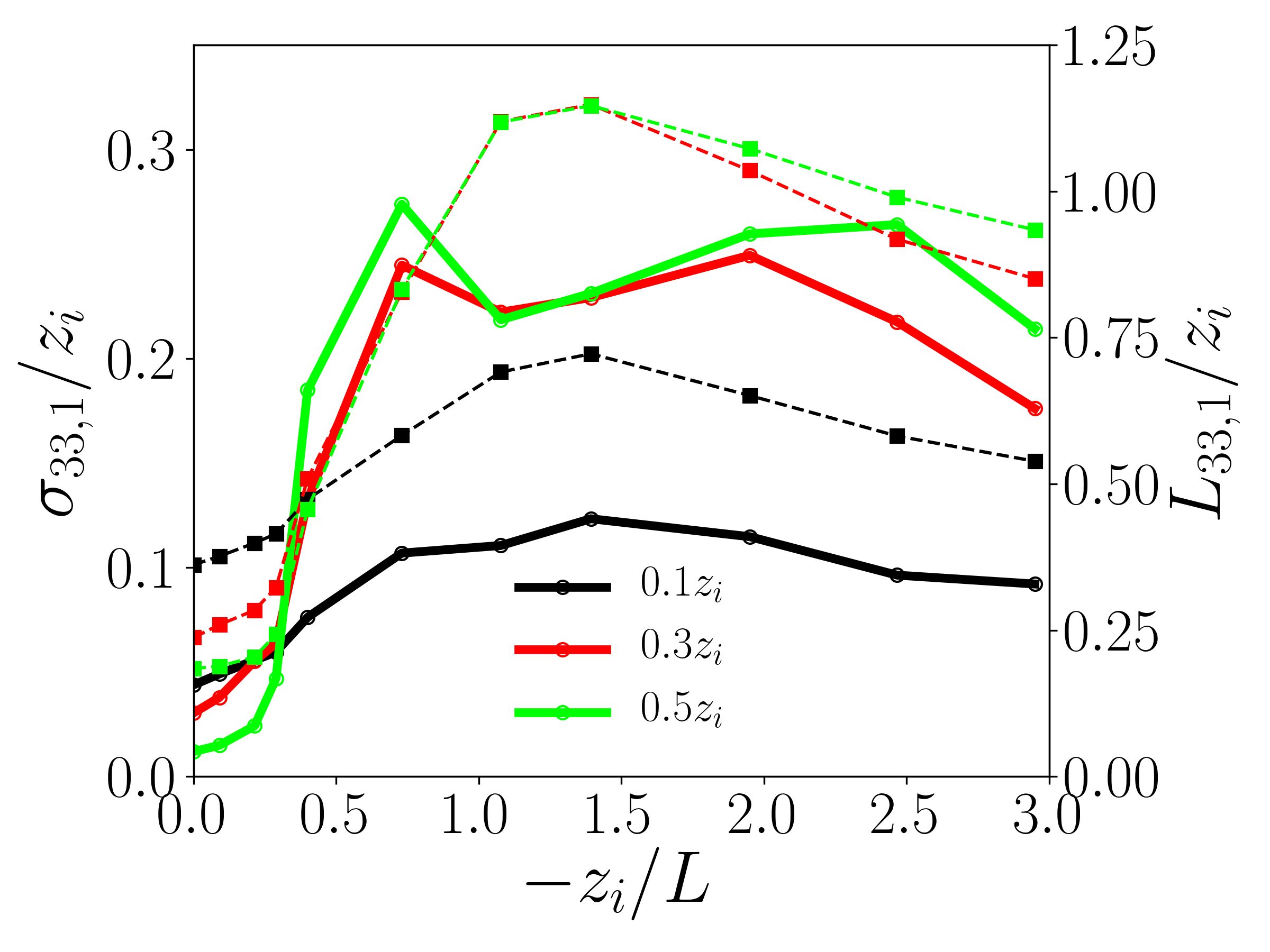}
 		\caption{\label{fig:L331VarVszibyL}}
 	\end{subfigure}
 	\caption{\label{fig:Lxx1VarVszibyL} (Color Online) The solid curves show the transition in temporal variance of streamwise coherence lengths normalized by $z_i$: (a) streamwise velocity fluctuations $u'$ ($\sigma_{11,1}/z_i$) and (b) vertical velocity fluctuations $w'$ ($\sigma_{33,1}/z_i$) across different stability states at $z/z_i= 0.1, 0.3 \textrm{ and } 0.5$. The dotted curves show mean streamwise coherence lengths of the same variables ($L_{11,1}/z_i, L_{33,1}/z_i)$ from figure~\ref{fig:Lxx1vszibyL}, for comparison.}
 \end{figure}

 Note that the sudden increases observed in the mean integral scales at the critical stability state $-z_i/L=0.40$ are even more apparent in the sudden jumps observed in figure~\ref{fig:Lxx1VarVszibyL} in the standard deviations of the temporal fluctuations, $\sigma_{xx,y}$, within the mixed layer. Indeed, figure~\ref{fig:L331VarVszibyL} shows that $\sigma_{33,1}$ displays a stronger jump in the mixed layer from $-z_i/L=0.29$ to $0.40$ than does $L_{33,1}$, indicating a sudden increase in temporal variability coincident with a slightly less sudden increase in mean coherence length. However, after the standard deviation of $L'_{33,1}$  suddenly increases at $-z_i/L=0.40$, it oscillates around a mean up to $-z_i/L\sim2-2.5$ while the mean $L_{33,1}$ continues to increase to peak broadly between $-z_i/L\sim1-1.5$. In contrast, figure~\ref{fig:L111VarVszibyL} shows a dramatic increase in $\sigma_{11,1}$ at $-z_i/L=0.40$, followed by a rapid reduction with increasing $-z_i/L$ after the critical stability state. At the max coherence length large-scale roll state, $-z_i/L\sim1-1.5$, the standard deviation of $L'_{11,1}$ approaches a minimum while the standard deviation of $L'_{33,1}$ is maximal. At higher stability states with $-z_i/L$ well above the max coherence large-scale roll state, $\sigma_{33,1}$ and $\sigma_{11,1}$ decrease, likely towards the ``moderately convective'' stability state. The changes in temporal variance are apparent in the mixed layer and less so near the surface where vertical fluctuations are damped more than horizontal fluctuations (figures~\ref{fig:L111VarVszibyL} and \ref{fig:L331VarVszibyL}).
 
 \begin{figure}
 	\centering
	\begin{subfigure}[ht!]{0.99\columnwidth}
		\includegraphics[width=\columnwidth]{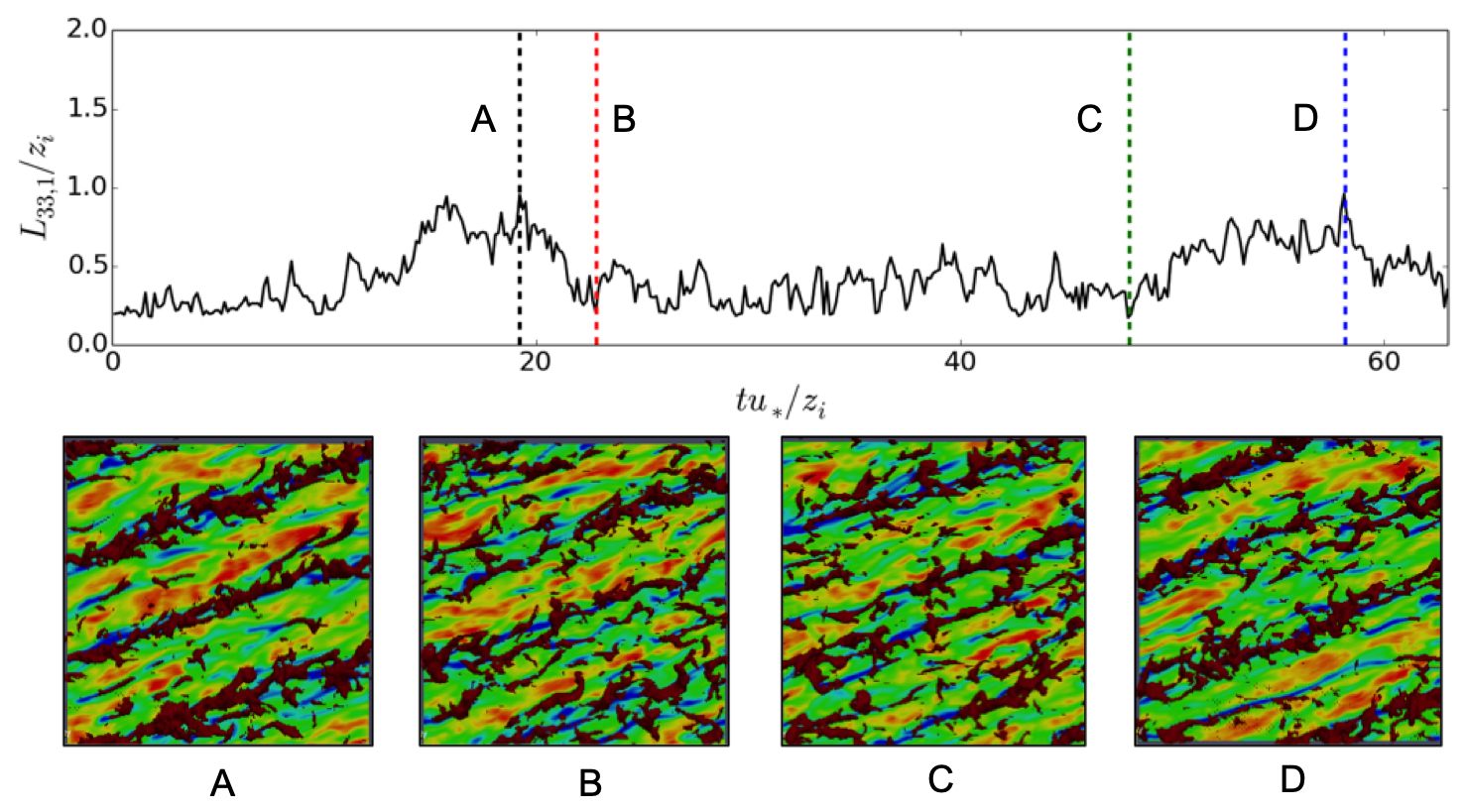}
		\caption{\label{fig:L331tDissection-zibyL040}}
	\end{subfigure}
	\begin{subfigure}[ht!]{0.99\columnwidth}
		\includegraphics[width=\columnwidth]{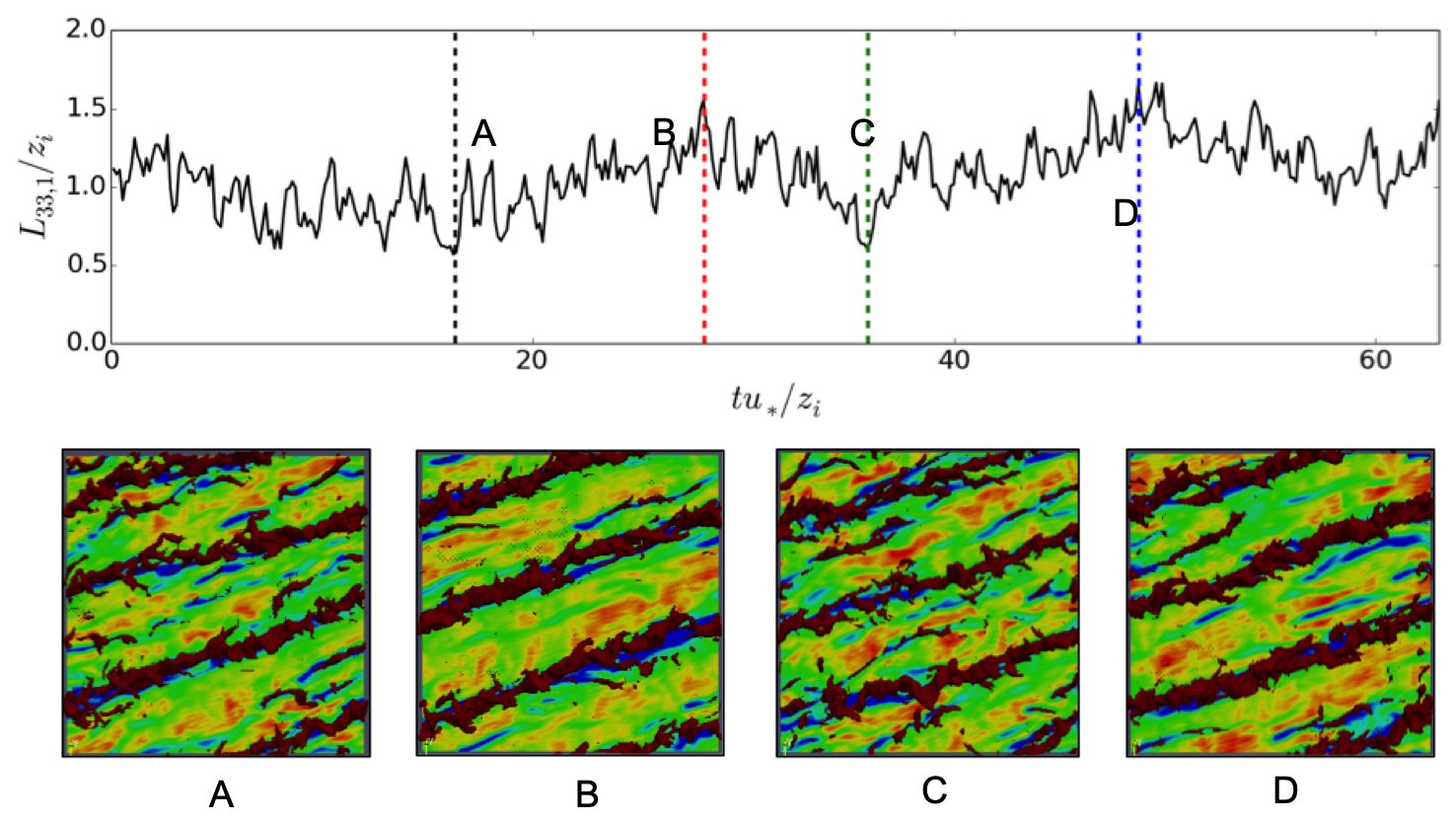}
		\caption{\label{fig:L331tDissection-zibyL108}}
	\end{subfigure}
	\caption{\label{fig:L331tDissection} (Color Online) In each pair of figures, the upper line plot shows the time variations in $L'_{33,1}(t)$ at $z/z_i=0.5$ over the averaging time period of $60$ eddy time scales $\tau_u=z_i/u_*$. In each group, the lower four figures show the instantaneous 3D large-eddy velocity structure at the four times in the time series plot marked A, B, C and D in the upper figure at the following stability states: (a) $-z_i/L= 0.40$, the critical stability state, (b) $-z_i/L=1.08$, the max coherence large-scale roll state, (c) $-z_i/L=2.95$, the moderately convective state (continued in the next page). The isocontours are of streamwise velocity fluctuations $u'$ with range $\pm2\sigma_u$ with blue isocontours showing the low-speed streaks and the red isocontours indicating the less coherent high-speed regions in between. The isosurface corresponds to positive (upward) vertical fluctuating velocity at $w'=2\sigma_w$.}
\end{figure}

 \begin{figure}
 	\ContinuedFloat
     \centering
	\begin{subfigure}[ht!]{0.99\columnwidth}
		\includegraphics[width=\columnwidth]{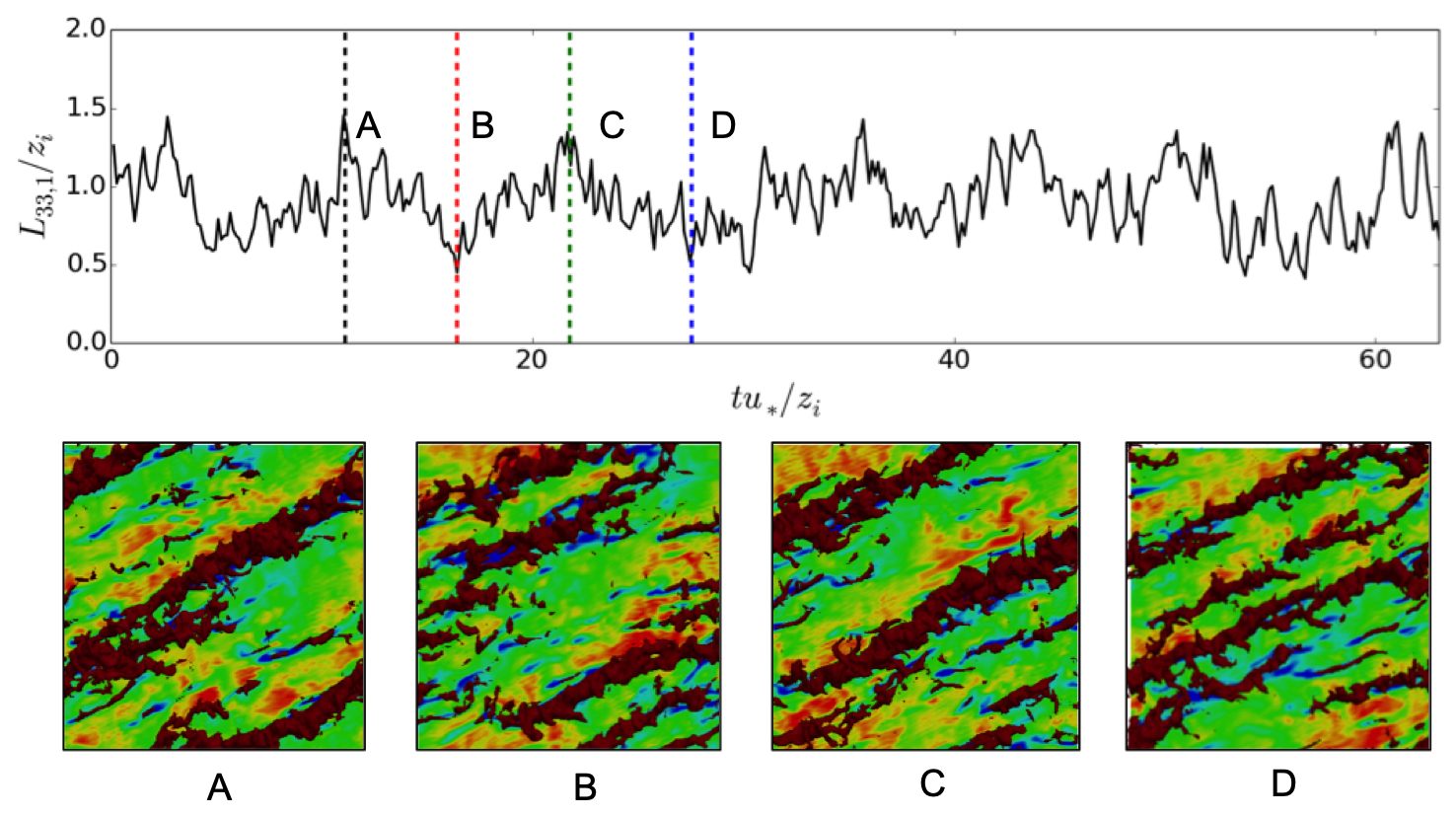}
		\caption{\label{fig:L331tDissection-zibyL295}}
	\end{subfigure}
	\caption{ Contd.}
\end{figure}

To interpret the changes in standard deviation shown in figure~\ref{fig:Lxx1VarVszibyL}, consider figures~\ref{fig:L331tDissection-zibyL040}, \ref{fig:L331tDissection-zibyL108} and \ref{fig:L331tDissection-zibyL295} where time variations of $L'_{33,1}(t)$ in the mixed layer are plotted along with visualizations of spatial eddy structure at key points in time in a repetitive cycle that changes with ABL instability state. We find that, whereas time series of $L'_{33,1}(t)$  at $-z_i/L$ below the critical stability state do not display significant temporal variability, the critical state $-z_i/L\approx0.40$ initiates an interesting large-time-scale feature that systematically changes with increasing $-z_i/L$. Figure~\ref{fig:L331tDissection-zibyL040} ($-z_i/L=0.40$) shows that the sudden increase in $\sigma_{33,1}$ at the critical state is associated with the sudden creation of a large-scale time variation in $L'_{33,1}(t)$ with a peak occurring repetitively with a period roughly $35$ times larger than the eddy turnover time $\tau_u=z_i/u_*$. The visualizations below the line plots indicate that the peaks in coherence lengths $L'_{33,1}(t)$  are associated with more organized and more highly concentrated sheets of vertical fluctuations that periodically break down into less coherent thermal updrafts with shorter flucutating streamwise correlation lengths. The low-coherence periods appear to last roughly $20-25\tau_u$ while roughly every $35\tau_u$, more coherent thermals form for a period of roughly $10\tau_u$. Careful comparison of isosurfaces of vertical updrafts with the blue low-speed streaks below suggests that the formation and breakdown of the more coherent periods of thermal updraft are tied to the formation and breakdown of the more coherent underlying low-speed streaks.

The transition from the critical to maximum coherence large-scale roll state is illustrated by comparing figures~\ref{fig:L331tDissection-zibyL040} and \ref{fig:L331tDissection-zibyL108}, where figure \ref{fig:L331tDissection-zibyL108} visualizes the updraft and low-speed streak structure at the max coherence state, $-z_i/L=1.08$. The time scale between periods of increase in $L'_{33,1}(t)$ is reduced from $35$ at critical $-z_i/L$ to roughly $20-25\tau_u$ at the max coherence state. However, at the max coherence state, the visual differences in updraft/streak structure between max vs. min $L'_{33,1}(t)$  periods (B, D vs. A, C) are minimal \BJ{(less variance $\sigma_{33,1}$)} and the vertical updraft structure is exceptionally coherent in both time and space. Interestingly, in the transition from the max coherence state ($-z_i/L = 1.08$) to the moderately convective state ($-z_i/L=2.95$), figures~\ref{fig:L331tDissection-zibyL108} to \ref{fig:L331tDissection-zibyL295}, another periodic transition in time occurs at $-z_i/L = 2.95$ between more coherent (A, C) to less coherent (B, D) updraft structures as $L'_{33,1}(t)$   oscillates between maxima and minima. Although the formation and breakdown process in the moderately convective state (figure~\ref{fig:L331tDissection-zibyL295}) is similar to the critical state (figure~\ref{fig:L331tDissection-zibyL040}), the moderately convective state is associated with more concentrated and streamwise coherent updraft elements than in the critical stability state, reflecting the existence of residual less-coherent large-scale rolls.

Ergodicity implies that in horizontally homogeneous stationary turbulent flows additional time averaging is statistically equivalent to horizontal averaging over expanded horizontal domains. Thus, the temporal variations in structure from A to D in figures~\ref{fig:L331tDissection-zibyL040}, \ref{fig:L331tDissection-zibyL108} and \ref{fig:L331tDissection-zibyL295} can be viewed equivalently as spatial variations at fixed time over larger horizontal domains, and increases in standard deviation in $L'_{33,1}(t)$ in time in the mixed layer can be interpreted equivalently as increases in concentration of vertical velocity within updrafts over larger horizontal planes, with correspondingly higher peak-to-peak variations in vertical velocity. This is the case, on average, with each stability state shown in the visualizations of figure~\ref{fig:L331tDissection}. However the structure of the more concentrated, most spatially coherent thermals, is strongly dependent on stability state. The max coherence roll state is characterized by exceptional local coherence in all sheets of concentrated vertical velocity that define the thermal updrafts⎯either over time over fixed horizontal planes of limited extent, or over extended horizontal planes at fixed time. Yet even at the max coherence roll state of figure~\ref{fig:L331tDissection-zibyL108}, careful examination of the four isosurface images suggests the existence of slight temporal, and therefore spatial, transitions between slightly greater vs. slightly lesser concentrations of vertical velocity within sheet-like structures.

Away from the max coherent roll instability states, the temporal (and spatial) variations between greater and lesser concentrations on local sheet-like thermals are much more apparent. However, figure~\ref{fig:L331tDissection} suggests that at each stability state the structure of the local updraft concentrations are spatially tied to the structure of the underlying low-speed streaks. As discussed in \S~\ref{subsec:CriticalTransition-StreakRole}, the critical transition corresponds to the initiation of a spatial correlation between low-speed streak structure below and vertical updraft structure above, a correlation that increases with increasing supercritical instability state parameter, $-z_i/L > 0.40$. Thus, as the streamwise lengths of coherent surface-layer streaks increase with increasing surface heating in the post-critical state, not only does the strength and degree of concentration of vertical velocity within the updrafts increase in response to increasingly strong buoyancy force, the updrafts become progressively more attached to the increasingly coherent streaks below.

Interestingly, whereas figure~\ref{fig:L331VarVszibyL} suggests that the variance in spatial coherence lengths of vertical updrafts is roughly the same from $-z_i/L\sim0.4 \textrm{ to } 2-2.5$ in the mixed layer, figure~\ref{fig:L331tDissection} indicates that the structure of the spatial variability in vertical updrafts is strikingly different. At the same time, the coherence lengths themselves continue to increase and peak broadly between $-z_i/L\sim1 - 1.5$, suggesting that, whereas the concentrations of vertical velocity within the thermal updrafts increase suddenly at the critical stability state, the horizontal coherence lengths of the thermals increase more gradually with increasing $-z_i/L$ as the updrafts couple progressively more strongly, in the streamwise direction, with the coherent low-speed streaks below. This coupling process continues until the streamwise coherence lengths of vertical updrafts and low-speed streaks match at the max coherence roll state (figure~\ref{fig:LyykbyLxxkvszibyL}).
Figure~\ref{fig:L111VarVszibyL}, on the other hand, shows a dramatic increase in $\sigma_{11,1}$ at the critical state followed by a rapid drop with increasing $-z_i/L$ to a minimum at the max coherence state. Thus, as the streamwise coherence lengths of horizontal and vertical velocity fluctuations peak at the max coherence large-scale roll state, the variability of $L'_{11,1}(t)$  is a minimum while that of $L'_{33,1}(t)$  is a maximum, consistent with increased attachment to underlying low-speed streaks below as the ``seeds'' to the formation of the more highly concentrated more highly correlated thermal updrafts above.

\section{Discussion and Conclusions\label{sec:DiscussionAndConclusions}}
A central conclusion from the current study is that in the transition in ABL structure associated with systematic increase in surface heating from a fully neutral ABL towards a moderately convective ABL state is that the coherent shear-driven low-speed streak structure near the surface and the coherent vertical velocity fluctuations concentrated within thermal “updrafts” in the mixed layer above are dynamically and structurally linked, and that the strength of the linkage increases with increasing levels of surface heating from the neutral state. Whereas the argument has been made that the near-surface motions associated with coherent near-surface streaks lead to the concentration of higher-temperature fluid within low-speed streaks \citep{khanna1998three}, our current results indicate that this convergence of high temperature fluid increases with increasing low-level surface heat flux from supercritical to ``max coherence state.'' Thus, the changing structure and coherence of updrafts with changing ABL stability state results from the combined increasing linkage between low-speed streaks and updrafts and increasing coherence of the low-speed streaks with the introduction of surface heating and the generation of buoyancy-driven vertical motions. 

It is surprising that the coherence length of the low-speed streaks should increase with the addition of surface heating to a previously neutral boundary layer. Given that the existence of streamwise-elongated coherent structures in lower-than-average streamwise turbulent velocity fluctuations is a direct result of the interaction of mean shear-rate and turbulence − even linear perturbation theory predicts elongation of turbulence structure by mean shear − one would anticipate that any addition of buoyancy-generated vertical motions would interfere with coherence of shear-driven low-speed streaks. Instead we find that the coherence length of low-speed streaks increases with the addition of surface heating above critical. Thus, the increasing coherence of streak structure work together with the increasing linkages between updrafts and streaks to increase coherence and strength of vertical fluctuations concentrated within thermal updrafts. The observation that the increases in streamwise and vertical coherence lengths of vertical velocity fluctuations lags the increase of streamwise coherence of streamwise velocity fluctuations with increased surface heating may be a manifestation of the physical separation in the vertical between shear-dominated surface layer dynamics below impacting buoyancy-dominated mixed layer dynamics above.

The strengthening of coherence with increased surface heating from the neutral ABL state ends as the ABL reaches the ``max coherence'' instability range $-z_i/L\sim 1.1 - 1.5$ where ABL turbulence coherence lengths peak and the ABL turbulence structure is characterized by maximally organized highly turbulent counter-rotating pairs of helical roll-like very-large-scale highly turbulent structures that fill the boundary layer, confined above by a soft capping inversion and below by a rigid rough ground surface. Streamwise coherence lengths of horizontal and vertical velocity fluctuations are at most $1.75z_i$ in the ``max coherence state,'' coherence lengths much shorter than the visual alignment of coherent integral-scale updraft structures in figures~\ref{fig:CriticalTrans-uisozibyL-108} and \ref{fig:CriticalTrans-uisozibyL-139}, and commonly observed in cloud formations visually from above, which can easily extend for tens of boundary layer depths, $z_i$. This super-integral-scale coherence within the streamwise helical structure of the turbulent velocity field is a characteristic of the maximal coherence state, a helical super-scale roll structure flanked on one side by strong coherent vertical updrafts and on the other side by much less coherent vertical downdrafts that cover a larger surface area than the concentrated updrafts. Similarly, the exceptionally concentrated coherent vertical updrafts in the mixed layer are exceptionally strongly tied to coherent low-speed streaks in the surface layer below with maximal streamwise coherence. Compared to instability states $-z_i/L$ above and below the max coherence instability range, large-scale rolls exist but are less coherent, in time as well as in space. In particular, in the max coherence state the thermal updrafts, and therefore the rolls, are exceptionally well formed, highly concentrated, and temporally rigid, both over the horizontal plane and in time, and with an exceptionally strong spatial correlation with exceptionally coherent low-speed streaks below. In contrast, at stability states somewhat outside the max coherence range $-z_i/L\sim 1.1 - 1.5$, the vertical updrafts experience cyclic periods of breakdown in coherence and reformation.

However, our research has also shown that the transition from the neutral to the maximal coherence state is not continuous as surface heating gradually increases from zero. The continual change in the interplay between buoyancy-generated motions in the mixed layer and shear-generated motions in the surface layer with increasing surface heat flux does not take place until the ABL has transitioned from a near-neutral to a supercritical stability state. The critical state is characterized by a bifurcation-like transition with a sudden dramatic change in ABL coherent structure. This critical transition occurs at extremely low levels of surface heat flux, only $1\%$ of the daytime maximum in surface heat flux, and correspondingly at the very low instability state parameter,$\left[ -z_i/L\right]_{crit}=0.40$. Only after the ABL has transitioned to the post-critical state does the coupling between low-speed streak coherence below, and vertical velocity thermal-updraft coherence above occur, with the systematic changes that lead to the max coherence state. The physical mechanisms underlying this transition from a shear-dominated neutral boundary layer to one in which shear and buoyancy interact in such as way as to enhance shear-generated streamwise coherence of horizontal velocity fluctuations, are not understood. Particularly fascinating is the observation from figure~\ref{fig:Lxx1vszAtzibyL} that the critical transition is associated with a sudden and dramatic change in the coherence structure of the ABL whereby streamwise coherence length of streamwise velocity fluctuations shifts from maximal in the surface layer subcritical to dramatically maximal in the mixed layer post critical.

\section*{Acknowledgements}\label{appOvershoot}
The research has been supported by the Department of Energy EERE office under grant DE-EE0005481. Computer resources were provided by the National Science Foundation XSEDE program at the National Institute for Computational Sciences in Tennessee and the Pittsburgh Computing Center under grant number TG-ATM0007.

\bibliographystyle{jfm}
\bibliography{ref}

\end{document}